\documentclass[aps, prc, showpacs, groupedaddress,amssymb,12pt]{revtex4-2} 

\usepackage{amsmath}  % needed for \tfrac, \bmatrix, etc.
\usepackage{amsfonts} % needed for bold Greek, Fraktur, and blackboard bold
\usepackage{graphicx} % needed for figures
\usepackage{bm}
\usepackage{dcolumn}

\begin{document}

\title{Towards Nonperturbative Solution of Quantum Dynamics : A Hamiltonian Mean Field Approximation Scheme with Perturbation Theory for Arbitrry Strength of Interaction }

\author{B.P.Mahapatra}
\email{bimal58.mahapatra@gmail.com, } % optional
\affiliation{ Professor ( retired ), School of Physical Sciences, Sambalpur University, India} 
\date{\today}

\begin{abstract}
   We introduce a non-perturbative general approximation scheme (NGAS) that can handle interactions of any strength in quantum theory. This approach starts with an input Hamiltonian that can be solved exactly. The interaction effects are then built into this Hamiltonian through a nonlinear feedback enforced by self-consistency conditions. While the method itself is nonperturbative, it can be systematically improved using a  new perturbation method called mean field perturbation theory (MFPT) which does not involve power-series expansion in any small parameter. We put this scheme to the test on one-dimensional anharmonic interactions using the harmonic approximation. The results are consistently accurate across various cases including: quartic, sextic, and octic anharmonic oscillators, as well as, the quartic double-well oscillator (QDWO)—even when the coupling strength varies widely. The method’s flexibility is demonstrated when we swap the input Hamiltonian for an infinite-square-well Hamiltonian and still achieve comparable accuracy. When applied to the ${\lambda}{\phi}^4$ quantum field theory, this approach aligns with the Gaussian-effective potential method under the harmonic approximation. Beyond that, it reveals the condensate structure of the effective vacuum and highlights the instability of the perturbative ground state. Notably, our ground-state energy results for the QDWO stand in stark contrast to those from standard perturbation theory, where Borel summation fails regardless of coupling strength.

\end{abstract}


\maketitle % title page is now complete

%% \usepackage{epsfig}
%
%\newcommand{\singlespaced}{\renewcommand{\baselinestretch}{1}\normalfont}
%\newtheorem{theorem}{Theorem}[section]
%%\newtheorem{theorem}{Theorem}[Chapter]
%\newtheorem{corollary}[theorem]{Corollary}
%%\newtheorem*{main}{Main~Theorem}
%\newtheorem{lemma}[theorem]{Lemma}
%\newtheorem{proposition}[theorem]{Proposition}
%\theoremstyle{definition}
%\newtheorem{definition}[theorem]{Definition}
%\theoremstyle{remark}
%\newtheorem{remark}[theorem]{Remark}
%\newtheorem{example}[theorem]{Example}
%%\numberwithin{equation}{chapter}
%\newcommand{\f}{\mathfrak}
%\newcommand{\mb}{\mathbb}
%\newcommand{\mr}{\mathrm}
%\newcommand{\mf}{\mathbf}
%\newcommand{\mc}{\mathcal}
%\newcommand{\e}{\emph}
%\newcommand{\vp}{\varphi}
%\newcommand{\Diff}{\textrm{Diff}}
%\newcommand{\Norm}{\textrm{Norm}}
%\newcommand{\Hom}{\textrm{Hom}}
%\newcommand{\ch}{\textrm{char}}
%\newcommand{\lcm}{\textrm{lcm}}
%%\newcommand{\ca}{\mathcal}
%\newcommand{\wt}{\widetilde}
%%\newcommand{\ol}{\overline}
%
%%\doublespaced

%  
%    \end{center}
%\begin{flushleft}
\newpage
\tableofcontents
\newpage
    \section { Introduction : The Role of Approximation in Quantum Theory (QT)}
   % Introduction: The Role of Approximation in Quantum Theory
    
    Quantum theory (QT) is widely recognized as a robust framework that elucidates the fundamental behavior of the physical universe, encompassing phenomena from the minutest particles to the expansive scales of the cosmos [1,2]. Despite its extensive applicability, exact analytic solutions for interacting quantum systems remain scarce. While certain special cases have been exactly solved and documented [3], these are exceptions rather than the rule. It is understood [4,5] that exact solvability often hinges on the factorization properties of the corresponding quantum Hamiltonian. In recent years, advances in supersymmetric quantum mechanics (SUSYQM) [6] have significantly broadened the catalog of exactly solvable potentials within non-relativistic QT. Given the limited range of exactly solvable models, approximation methods are indispensable for extracting meaningful insights from quantum systems—a necessity recognized soon after the birth of QT itself [7]. Over decades, numerous approximation schemes have been developed, some tracing back to QT’s earliest days [3,7]. The literature on such methods is vast and continually expanding, reflecting ongoing efforts to discover a broadly applicable, "ideal" approximation scheme suitable for the diverse array of interacting quantum systems. To set the stage for the investigations detailed in this report, it is crucial to review and contextualize existing approximation approaches. This overview is presented in the following section.

     \section{Overview and Categorization of Existing Approximation Methods} The principal approximation techniques in quantum theory (QT) that have demonstrated significant utility and are frequently cited in textbooks [3] include: (a) \textit{Perturbation Theory}, (b) \textit{Variational Method}, and (c) \textit{Semi-classical} (WKBJ) approximations. Numerous other methods exist, some of which are variations or extensions of these foundational approaches. A comprehensive discussion of these can be found in reference [8]. For completeness, we briefly summarize them here:
    
    %\end{flushleft}

    \subsection{ Perturbation theory (PT)} 
    Perturbation theory (PT) is among the earliest approximation methods recorded in the literature. In the context of classical physics, the method was applied by Lord-Rayleigh [7]  as early as 1873. In QT the application of the method is as old as the theory itself, see, e.g. ref.[7].Since the salient features of this method  are well discussed  in texts (e.g.[3]) only a brief discussion follows.
    
    The standard formulation of PT (SFPT) involves a power-series expansion of observables in some small parameter. Thus, e.g. if $E(g)$ is an observable of the system, then its dependence on the small-parameter denoted here as `$g$', is through the following power-series:
    \begin{equation}
    E(g)= \mathcal{E}_0+ \sum\limits_{k=1}^{\infty}{g^{k} \mathcal{E}_k\equiv \mathcal{E}_0(g)+ \Delta{\mathcal{E}(g)}}
    %\label{\textbf{1}}
    \end{equation}
    where the first term is exactly analytically computable. The parameter $g$ is usually chosen to be the \textit{coupling-strength} of interaction defined in a Lagrangian/Hamiltonian frame work. Thus, e.g. if $H$ is the system-Hamiltonian then it can be written in SFPT as: 
    \begin{equation}
    H = H_{s} + gH_{I},
    %\label{\textbf{2}}
    \end{equation}
    where $g$ is the coupling-strength of interaction, $ H_{s}$ is the exactly solvable unperturbed Hamiltonian and $gH_{I}$ is the ( usually non-linear) interaction part of the Hamiltonian. This latter interaction-term prevents exact analytic solution for the eigen-spectrum of $H$. 
    Interpreting the eigenvalues of $H$ through the expansion in equation (1), $E(g)$ is the energy of the full interacting system, $\mathcal{E}_0$ is the energy without interaction, and each $\mathcal{E}_k$ is the perturbative correction of order $k$ to the unperturbed energy. These corrections can, in principle, be computed to any desired order. For perturbation theory to be meaningful, the corrections must remain smaller than the leading term, which restricts $g$ to small values, typically $|g| < 1$. Additionally, the series in equation (1) should either converge or behave asymptotically so that truncating after a finite number of terms provides a good approximation of the true value. The speed of convergence or the decay rate of the asymptotic terms is therefore vital for the method’s practical success.
    
    \textit{\textbf{Strengths and Drawbacks of Perturbation Theory}}
    
     Perturbation theory is favored for many practical and theoretical reasons [9].
    
     \textit{Strengths:} 
     
     Two key advantages explain its popularity: (i) It allows systematic improvement, providing corrections order-by-order to a leading-order result that can be computed exactly. This capability for controlled refinement is its main strength. (ii) It is broadly applicable, as long as one can identify an exactly solvable unperturbed system and a small coupling parameter suitable for power-series expansion. This universality covers a wide array of quantum problems, including many-body systems and quantum field theory.

   \textit{ Limitations of SFPT:}

    SFPT is subject to several significant limitations:

    (i) Inability to address "non-perturbative" phenomena: SFPT employs a power-series expansion (refer to eqn.(1)), which becomes ineffective when the interaction strength $g$ is substantial, specifically when $|g|>1$. This scenario frequently arises in contexts involving strong interactions and bound states. SFPT encounters difficulties with processes that exhibit complex dependencies on $g$ from the outset. Numerous physical phenomena, such as tunneling, decay, phase transitions, critical phenomena, and collective behaviors like superconductivity and superfluidity, fall into this category. This represents a substantial challenge for SFPT. (ii) Instability of the "perturbative vacuum": The "perturbative vacuum/ground state" is derived from the non-interacting theory. Typically, the actual (interacting) vacuum/ground state is inaccessible from the perturbative one upon the introduction of interaction [8,10]. Furthermore, the perturbative ground state is unstable due to the significantly lower energy of the ground state in the interacting theory. (iii) Discrepancy with established analytic properties: The analytic properties of an observable, such as energy, in the initial complex-$g$-plane have been determined for certain systems [11], and they do not align with those in eqn.(1). (iv) Practical Challenges: In addition to theoretical issues, SFPT encounters practical challenges, such as computing corrections beyond the initial few orders, which involve an (infinite) summation over intermediate states. This issue has been addressed using specialized methods, such as the Dalgrano-Lewis method [12] and the hyper-virial theorem [9]. Additional limitations of PT are discussed in subsequent sections concerning specific systems.
         
    We conclude this subsection by noting that the challenges associated with SFPT have spurred the development of several alternative approaches, some of which are discussed below.
    
     \subsection{ Variational Methods (VM)}
        The Variational-Method of Approximation (VMA) is a powerful method  in obtaining the approximate  eigenvalues and eigenstates of observables in QT. Like perturbation theory (PT), the application of the method has a long history- initial application of VMA being ascribed to Lord Rayleigh [13] and to W.Ritz [14].  The VMA becomes especially suitable when perturbation theory (PT) fails or becomes inapplicable to the considered problem.  VMA is also used to test results based upon PT and for analysis of stability of the system under consideration.
        
        The essential ingredient of the method is based upon the variational theorem (Theorem-I), which provides an {\it upper bound} for the ground state energy of the system :
        \begin{equation} 
        \frac{<~\psi~|~H~|\psi~>}{< ~\psi|\psi~>}~\equiv~E~[\psi]~\geq~E_{0}~,
        %\label{\textbf{3}}  
        \end{equation}
        where, $~|\psi >~$ is any {\it arbitrary}, normalizable state known as the ``trial-state";  {\it H} is the  Hamiltonian, and $~E_{0}~$ is the ground-state energy of the system. ( In the above equation, we have used the standard Dirac-notation for expectation values and also denoted by $E[\psi]$, the energy- {\it functional}). The {\it proof} of the theorem is based upon straight forward application of eigenfunction  expansion method in  QT and can be found in any standard texts [3].
        
        For accuracy of estimation it is necessary to obtain an approximation to the  \textit{least upper bound} by judicious choice of the trial-state. This is provided by the Ritz-method [14], which consists in specifying the {\it trial} -state $~|~\psi~>~$ as a {\it function} of one or more free-parameters : $\{\alpha_{i}\}$ and then {\it minimizing}  the energy functional $~E[\psi~]$ with respect to these parameters :

    \begin{equation}
    \partial{E}[\psi]/\partial\alpha_{i}~=~0~;~~\partial^{2}E[\psi]/\partial
    \alpha_{i}^{2}~>~0.
    %\label{\textbf{4}} 
    \end{equation}
    
       Thus, the Ritz method enables the determination of a least upper bound (LUB) for a given choice of the trial-state, which represents the closest approximation to the ground-state energy for that choice.
        Thus, the Ritz method enables the determination of a least upper bound (LUB) for a given choice of the trial-state, which represents the closest approximation to the ground-state energy for that choice.

       Further relevant aspects of the VMA are discussed below:   
       
        (a) The equality sign in equation (3) is valid if and only if the trial state, $|~\psi~>$, coincides with the true ground state, $|~\psi_{0}~>$, of the system. (b) An error on the order of $O(\epsilon)$ in selecting the trial state results in an error on the order of $O(\epsilon^{2})$ in the computed energy value. This indicates that the system's energy is determined with greater accuracy by the Variational Method of Approximation (VMA) than the wave function. Consequently, the overall accuracy of the method is highly dependent on the choice of the trial state, which must be consistent with the system's physical boundary conditions. This necessitates considerable ingenuity and insight in selecting the trial state. (c) The generalization of the Rayleigh-Ritz method of the VMA to estimate or approximate higher excited states can be achieved through several methods. Some of these methods are outlined below: (i) A series of trial states, $~|~\psi_{\alpha}~>~;~\alpha~=~1,~2,~3......,$ approximating the higher excited states of the system, may be constructed with the requirement that these states are mutually orthogonal to each other and orthogonal to the variational ground state. This construction can be achieved using any suitable orthogonalization method, such as the Schmidt method [15]. Denote by $E_{\alpha}$ the energy functional for the $\alpha$-th state, minimized with respect to its free parameters and then arranged in a decreasing sequence, i.e.,

    \begin{equation}
    E_{\alpha}~\equiv~\Big[~\frac{<\psi_{\alpha}~|H|~\psi_{\alpha}~>}{<\psi_{\alpha}~|~\psi_{\alpha}~>}\Big]_{min}~;~\alpha~=~1,~2,~3,.........
    %\label{\textbf{5}} 
    \end{equation}
    The resulting sequence, $~ \mathcal{E}_0 ~<~E_{1}~<~E_{2}~<~E_{3}~.......,$ then represents the variational approximation to the energy of the higher excited states sequentially  in the order shown. 
    
    The other standard method to deal with the excited states within VMA is known as [16] the ``method of linear  variational approximation (LVA)''. This consists of the choice of the trial-state as a {\it linear} superposition of a suitably chosen set of eigenfunctions consistent with the boundary conditions and otherwise appropriate for the system, as described below :
    \begin{equation}
    |\psi>=\sum_{n}c_{n}~|u_{n}>~;\sum_{n}|u_{n}|^{2}~=~1;~~<u_{n}|u_{m}>=\delta_{nm}.
    %\label{\textbf{6}} 
    \end{equation}
    The energy functional as given in eqn.(3) computed with this trial-state leads to the following set of equations after minimization with respect to the coefficients $c_{n}$ (see, eqn.(6)) :
    \begin{equation}
    \sum_{n}c_{n}^{*}~(H_{nm}~-~E\delta_{nm})~=~0~;~~m~=~1,~2,~3,.....
    %\label{7} 
    \end{equation}
    where, $H_{nm}\equiv< u_{n} | H |  u_{m}>$.
    Condition for existence of non-trivial solutions of the above set of equation, is given by the vanishing of the secular determinant :
    \begin{equation}
    det~(~H_{nm}~-~E\delta_{nm}~)~=~0
    %\label{8} 
    \end{equation}
    For practical purpose, the determinant has to be truncated at some finite order `$k$'. The solution of eqn.(8) then provides, in general, $k$-roots for the energy $E$, {\it which} then correspond to the approximate energies of  the first `$~k~$' excited - states.
    
        For practical applications, it is necessary to truncate the determinant at a finite order $k$'. The solution of eqn.(8) typically yields $k$ roots for the energy $E$, which correspond to the approximate energies of the first $~k~$' excited states. It is apparent that this method becomes practically efficient when the truncation error, arising from the selected finite value of $k$, is minimized, and the spectrum stabilizes as $k$ increases. Consequently, the judicious selection of the basis states, $|~u_{n}>$, is essential in this approach.
        
{\it \textbf{Merits and Limitations of the Variational Method of Approximation}}

The merits and limitations of the VMA can be articulated as follows:

\textit{\textbf{Merits}} (i) \textit{Non-perturbative Nature} The method is effective for arbitrary coupling strengths of interaction and other non-perturbative phenomena with non-analytic dependence on the coupling parameter. This is arguably the primary advantage of the VMA. (ii) \textit{Accuracy} High precision can be achieved if the trial states are appropriately selected. Specifically, the diagonalization of the Hamiltonian in a suitably chosen basis, with parameters determined by variational minimization, can achieve high accuracy and stability by including a sufficient number of terms recursively. (iii) \textit{Universality} The method is applicable, in principle, to arbitrary Hamiltonian systems. (iv) \textit{Establishing Vacuum Stability} The stability of the ground state or vacuum of an interacting system can be established using the fundamental theorems of the VMA with a suitably chosen trial state. This method has been employed, for instance, to demonstrate the instability of the perturbative vacuum in anharmonic interactions [17], $\lambda \phi^4$ QFT [18], QCD [19], etc.

(v)\textit{Fail-safe method for guidance} The Variational Method Approach (VMA) offers guidance to the true ground state and the energy spectrum of a system when no other method is available or applicable. \bigskip 

\textit{\textbf{Limitations of VMA}} \bigskip 

(i)\textit{Not improvable order-by-order} In its original formulation, a primary limitation is that, unlike Perturbation Theory (PT), it does not allow for systematic improvement on an order-by-order basis. (ii)\textit{Non-uniqueness} There is no systematic or standard method for implementation, as the selection of trial states is largely arbitrary and relies on the ingenuity and insight of the researcher. (iii)\textit{Practical Difficulty in implementation} Practical challenges may arise in implementing the VMA when trial states involve multiple variational parameters. (iv)\textit{Not easily generalized to Quantum Field Theory (QFT)} In the context of QFT, the preferred method of investigation remains Standard Field Perturbation Theory (SFPT) due to its straightforward perturbative realization of the renormalization program, despite significant efforts within the VMA [18]. Several approaches have been proposed to address the limitations of the basic VMA. These include: (i) \textit{improving the input-trial state(s)} by incorporating additional free parameters or other modifications. A sample list can be found in references [20-22] for applications to standard examples of interacting systems with varying degrees of success. However, as previously noted, these attempts render the variational minimization of the energy functional more intractable. (ii)\textit{Tightening the variational estimates} by obtaining both upper and lower bounds, as seen in reference [22]. Lower bounds are derived using standard inequalities and other techniques. (iii)\textit{The method of minimum energy variance} (MEV): Instead of the Hamiltonian, $H$, itself, upper bounds can be derived, in principle, for any \textit{arbitrary} function $f(H)$ through the generalization of the standard Rayleigh-Ritz method. When the chosen function is considered as the \textit{variance} of the Hamiltonian, the method is referred to as the "method of minimum energy variance (MEV)" [23]. The MEV has been shown [23] to lead to improved estimates of the trial state as well as the energy eigenvalue. In particular, it has been demonstrated [23] that the estimates surpass the simple Gaussian approximation [17].

    \subsection{ Semi-Classical Methods (SCM)}
    The SCM also goes by the name of the original authors: Jeffery-Wentzel-Kramers-Brillouin (JWKB) approximation. The merits of the method are mainly  based upon the following features:
    
    \textit{\textbf{Merits}}
         (i) {\it model-independent and universal nature} : It can be applied to any {\it arbitrary, smooth potential}, which covers a large class of interacting systems. 
          (ii) \textit{The method is Non-perturbative}: It can be applied for arbitrary value of $g$ and also to cases with non-analytic dependence on the latter such as the\textit{ tunneling / barrier penetration }phenomena.  
          (iii)\textit{ Semi-classical method} : The "Semi-classical" nature of the approximation arises from the fact that the method is based upon the expansion of the ``action" of the system as a series involving powers of $\hbar$. Hence the leading-term being $\hbar$-independent, corresponds to the classical-action of the system, which may be easily computable.
        
           The formalism has been dealt at length in texts [24]. Therefore, it will not be repeated here. We note, however, the main features of the approximation scheme as applied to the discrete bound state problems in QT. 
    
    The relevant central formula for the above purpose is the so-called {\it JWKB quantization rule} :
    \begin{equation}
    \int^{b}_{a}~k(x)~dx~=~( n+\frac{1}{2})~\pi~,~n~=~0,~1,~2,~3......
     %\label{9}
    \end{equation}  
    where,$~k^{2}(x)~=~({2m}/{\hbar^{2}})~( E~-~V(x))~$; a ,b~ are a set of adjacent ``turning points'' obtained by solving the equation, $k(x)~=~0~$ and other terms have standard meaning. For those cases where the above integral can be evaluated, eqn.(9) can be expressed as :
    \begin{equation}
    f( E,\lambda,g...)~=~(n+\frac{1}{2})~\pi,
     %\label{10}
    \end{equation}  
    where, the l.h.s. of eqn.(10) represents the value of the integral as a function of the energy ($E$) and coupling-strengths $ ~(\lambda,g....)~$ occuring
    in the potential $V(x)$.  
    The determination of energy $E$ as a function of the other parameters $(n,\lambda,g...)$ therefore, requires the {\it inversion } of eqn.(10). 
    
   \textit{ \textbf{Limitations}}: With respect to cases covered when eqn.(10) is applicable, the main limitations could be as described below:
      (i) \textit{Non-invertibility of eqn.(10)}:
       For most cases of interest , the inversion of eqn.(10) can not be achieved in closed, analytic form except in a few known cases ( which, also happen to be exactly solvable by standard methods). However, eqn.(10) can be inverted by numerical-methods to the desired order of accuracy. 
        (ii)\textit{The complexity of the procedure }:
            The complexity of the method increases for the cases, when the WKBJ-integral can not be obtained in closed form as well as, in the cases  involving\textit{ multiple (closely spaced) turning points} [25] and the cases requiring \textit{higher order of approximation} in the WKBJ-series,(e.g. to deal with \textit{rapidly varying potentials}) become increasingly difficult to implement.
       
    Nevertheless, several useful information /insight about the energy levels can be obtained in limiting cases of the small-coupling regime, the strong-coupling and/or the large - n limits. We refer to Garg [25] who discusses the issues, successes and lists earlier references.
        
    %\subsection{ Other Methods}
    
    \subsection{Other Methods} 
    The multitude of methods precludes their exhaustive inclusion in this concise review. Herein, we briefly delineate those methodologies that are directly pertinent to the approach presented in this report.
         
    
    {\bf Combination of Variational and Perturbation Techniques} 
    
    Given that the Variational Method Approach (VMA) and Perturbation Theory (PT) offer complementary advantages and limitations, it is unsurprising that the literature contains numerous attempts to amalgamate the core elements of both methods into a unified framework. The following discussion highlights some of these efforts. Several scholars have adopted this combined approach, including Halliday and Suranyi [26], W. Caswell [27], J. Killingbeck [28], Hsue and Chern [29], Feynman and Kleinert [30] and collaborators [31], Patnaik [32], Rath [33], among others. The linear delta-expansion method [34], the Gaussian effective potential (GEP) approach [35], the self-consistent field method [36] and its generalizations [37], methods based on cluster-expansion [38], coherent and squeezed states [39], and various other methods incorporate the VMA and the VPA in some capacity. A comprehensive review of each individual work mentioned above is beyond the scope of this report. However, it is pertinent to highlight certain common features of the techniques employed by these authors and the consequent achievements attained.
    
    In most of the approaches listed above, the fundamental Hamiltonian of the system is modified by the addition and subtraction of terms involving certain \textit{additional} parameters. Perturbation techniques can then be applied with a redefinition of the unperturbed Hamiltonian and the perturbation correction. The arbitrary parameters are subsequently fixed order-by-order in the resulting perturbation series, either by the variational minimization of energy or by imposing other constraints such as the "principle of minimum sensitivity" (PMS) [40]. The resulting sequence of corrections is often claimed [26,27,32,33] to be convergent. Thus, the problem of convergence in naive perturbation theory, as well as the absence of a built-in mechanism for systematic improvement in the VMA, are largely overcome in these hybrid methods, which may be termed the {\it variation-perturbation method} (VPM). Apart from the aforementioned common underlying feature, each approach differs in detail, which may be found in the individual references cited above.

    {\bf  The operator Method of Approximation (OMA)}
    
    The OMA has been pioneered  by Feranchuk and Komarov [41]. The application of the OMA  to the anharmonic interaction  and to several other problems has been described in ref.[42] , which also provides a guide to earlier works of the authors. The basic idea of this method is to formulate the problem in the Fock-space of operators, instead of working with the co-ordinates and momenta. In doing so, an arbitrary parameter $\omega$ is introduced into the theory, which is then fixed by standard variational minimization of the energy-functional order-by-order in a modified PT. Details can be found in ref.[41] . The main limitation of the model appears to be the implementation of the scheme e.g. for non-polynomial interactions, interaction representing analytic functions expressible in infinite series, problems in higher dimensions, quantum field theory etc. In such cases, {\it additional} assumptions/methods/skills have to be employed [42].
        
    {\bf Approximation methods based on quantum-canonical transformation (QCT)}
    
    This is a powerful non-perturbative method, which has been successfully applied in QT, particularly in many-body systems exhibiting collective- and co-operative phenomena, e.g. superconductivity and super-fluidity [43-45]. The method had been primarily expounded by Bogoliubov [46] and  hence more  familiarly  known  as  the ``Bogoliubov-Transformation''. These transformations connect the Hillbert-space of the {\it interacting} system to that of the interaction-free case, while {\it preserving} the canonical structure of the { \it basic ( equal-time) commutation rules} . The method is particularly useful in variational studies by proposing the ansatz ( involving the variational - parameters) for the interacting vacuum state (IVS) generated through QCT. After minimization of the energy-functional, the approximation can be tested to dynamically establish the {\it stable} ground state of the system.   The formalism has been employed in  refs.[47,48] for the QAHO and the DWO problems. In the context of $~\lambda \phi^{4}~$ quantum-field theory, QCT has been employed in refs.[49,50,51]. QCT also provides important insight into  the structure of the interacting vacuum state [49,50,52,53]. 
   
     The \textit{limitations }[54] of the method appear to be those of variational methods as discussed earlier.
    
    {\bf  Approximation methods based on super symmetric quantum mechanics}
    
    The method of super symmetric quantum mechanics (SUSYQM) has been dealt in texts and several review articles. As a representative text [6], may be consulted, which provides guidance to earlier literature.
    
    This has been successfully applied to improve and extend the scope of known approximation methods, e.g. perturbation theory, variational method and JWKB-approximation scheme. In the context of non-relativistic QT, several exact results follow in SUSYQM valid for its partner potential [6]  such as the property of iso-spectrality, level degeneracy and positivity. These exact results provide important constraints in testing  various  approximation schemes.
    
    SUSY-improved perturbation theory starts from an initial guess of the ground state wave function, which can be based, for example, on a realistic variational {\it ansatz}. The {\it super-potential} corresponding to this trial-ground state wave function can then be constructed by computing the logarithmic derivative of the latter. The unperturbed Hamiltonian is then chosen to be the one obtained from the super potential and the perturbation correction is taken to be the difference of the original Hamiltonian and the unperturbed one. The development of the  RS-perturbation theory then becomes straight forward. For illustration of this method, ref.[6] can be consulted.
    
    The SUSY- based  JWKB approximation provides the following modification of the quantization-rule :
    \begin{equation}
    \int^{b}_{a}\sqrt{2m~[ E^{(1)}_{n}~-~W^{2}(x)]}~dx~=~n\pi\hbar~;~n~=~0,~1,~2,
    ~3,...
    %\label{11}
    \end{equation}    
    where, W is the {\it super-potential} and $E^{(1)}_{n}$ is the n-th energy-level of the $H_{1}$, which is one of the partner-Hamiltonians. Similar expression for the energy levels of the other partner-Hamiltonian, $H_{2}$ is given by,  
    \begin{eqnarray}
    \int^{b}_{a}\sqrt{2m~[ E^{(2)}_{n}~-~W^{2}(x)]}~dx~=~(n~+~1)\pi\hbar~;\nonumber\\~n~=~0,~1,~2,~3,...
     %\label{12}
    \end{eqnarray}    
    It may be seen from the above equations that the exact result on the iso-spectrality / level-degeneracy of the partner potentials, is respected in the SUSY-JWKB quantization condition. Moreover, the formulae are valid for {\it all} values of `$n$'  rather than for large-$n$ as in the case of the conventional JWKB-formulation. The SUSY-JWKB formula is demonstrated [6] to yield results with improved accuracy for many known cases.
    
    In the context of the QAHO/DWO, SUSYQM- methods have been investigated by several authors. A sample list is provided in refs.[55-65].
    
    \newpage  
    
    {\bf Approximation Schemes Based upon the Path-Integral Formulation of QT}
    
    The Path-Integral (PI) formulation of Quantum Theory [66-68] has traditionally served as a significant foundation for formulating and refining approximation methods. It is noteworthy that the Feynman-diagram technique, arguably the most popular computational method in perturbation theory, originated from the PI approach [69]. The approximation methods can be formulated either in the real-time formulation [66,67] or as the imaginary time, Euclidean formulation [70]. The latter is particularly well-suited for extending applications to quantum statistics [69-71]. A variational method was proposed in the Euclidean formalism of the PI in ref.[71] to approximate the partition function of the QAHO and thereby estimate the ground state energy. Applications to quantum field theory and statistical physics/critical phenomena are extensively addressed in the text [70]. In the context of the QAHO problem, the stationary-phase approximation was employed [72] in the PI formulation to deduce the inapplicability of the SFPT due to the presence of an essential singularity at the origin of the coupling strength plane. In ref.[73], variational lower and upper bounds on the energy of the QAHO have been obtained using inequalities for the partition function. The semi-classical approximation of partition functions for one-dimensional potentials using the PI formalism has been described in ref.[74].
    
   Various approximation schemes, including "variational perturbation theory" (VPT) [75,78], semi-classical methods [70,78], stationary-phase approximation [71,78], loop-wise expansion [78], large-order estimates of perturbation theory [70,78], and variational estimation using inequalities [71,78], among others [78], have been developed based on the PI-formulation. Notably, multi-instanton effects for potentials with degenerate ground states are detailed in ref. [76]. In the realm of scalar field theory, ref. [77] discusses non-standard expansion techniques for the generating functional within the PI formulation. Additional applications in field theory and statistical physics are documented in ref. [70,78]. 
    
    The primary advantages of this approach are its broad applicability [70,78] and its non-perturbative nature. However, a significant limitation of the PI-formulation is the mathematical complexity involved in evaluating multi-dimensional integrals. Even relatively simple, exactly solvable potentials, such as the hydrogen atom problem, necessitate complex techniques [78] for implementation. The inclusion of spin, formulation for fermions, and constrained systems present particular challenges. Consequently, in many instances, it is preferable to employ the standard operator-based formalism for implementing approximation methods. Furthermore, closed-form analytic expressions for the PI can only be derived for a limited number of problems with interaction, where the former can be approximated by a Gaussian around a saddle-point or stationary-phase point of the classical action and its fluctuations.
        
    {\bf  The self-consistent schemes of approximation}
    
    The self-consistent schemes of approximation are prevalent in the context of many-body systems and are commonly referred to as the "mean-field approximation" and the "Hartree-approximation," including various generalizations such as the "Hartree-Fock" method and the "Hartree-Fock-Bogoliubov" method. The fundamental concept of this scheme involves approximating the given potential with an exactly solvable one-body potential. If the original Hamiltonian is denoted as {\it H} and the approximating Hamiltonian as $H_{0}$, then a straightforward method to ensure self-consistency in the approximation is to impose the following constraint:
    \begin{equation}
    <n|H|n>~=~<n|H_{0}|n>~,
    %\label{13}
    \end{equation}
    where, the states $|n>$ are the eigenstates of $H_{0}$.
    Practically, the approximating Hamiltonian may be chosen with initially unknown, adjustable parameters, which are subsequently determined through the constraint, eqn.(13), and other criteria such as the variational principle and/or additional simplifying conditions. The non-linear feedback characteristic of the self-consistency condition can thus be readily ensured. Consequently, the ostensibly simple equation, eqn.(13), has the potential to incorporate interaction effects non-perturbatively while preserving the non-linearity of the original Hamiltonian. This method of implementation often yields results that are reasonably accurate even at the leading order of approximation, which can be further refined through a standard recursive procedure, as exemplified by the Hartree-Scheme. 
    
    In the context of $~\lambda \phi^{4}~$ quantum field theory and the QAHO problem, such a scheme was considered in ref.[79], where, however, the expectation value (see, eqn.(13)) was restricted to the ground state only in implementing the self-consistency requirement. This limitation has been addressed by proposing a "generalized Hartree-method" in ref.[52]. The 'mean-field' approach has also been explored in ref.[80]. 
    
    The primary limitations of the method appear to be the lack of uniqueness in the initial choice of the approximating Hamiltonian and the rate of convergence of the recursive procedure for subsequent improvement. 
    
    This concludes our survey of the various approximation methods in QT that are relevant in the context of the present report. It should be noted that we have described only those schemes with general applicability. However, there are several other methods that are system-specific. These latter often lead to greater accuracy but only at the cost of losing universality of application. Some such schemes will be described in later Sections when we consider specific applications.

     \section{ In Quest of an ``Ideal" Approximation Scheme}
    As detailed in the previous\textbf{ Section-II} , ref.[8-80] constitute only a partial list of  the main approximation methods /schemes for interacting quantum systems and the list is ever growing. This implies, among other things, that we are far from achieving an \textit{ideal} approximation scheme to deal with various cases of interacting quantum systems. In the quest of an ideal approximation scheme, the following requirements could perhaps form at least a subset of the list of the \textit{desirable} criteria characterizing it:\\
    
    \textit{ \textbf{A List of Desirable Criteria for an Ideal Scheme of Approximation} }:

    (i)\textit{General Applicability and the Universality of the Scheme}:
    This criterion necessitates that the scheme be applicable to a wide range of quantum systems and accommodate interactions of arbitrary strength. For instance, it should be relevant to quantum mechanics, many-body systems, and quantum field theory in both non-relativistic and relativistic contexts. This stands in contrast to "system-specific methods," which are tailored to particular systems and often prioritize high accuracy. Furthermore, the \textit{scheme should be applicable for arbitrary strength of interaction}, thereby covering both perturbative and non-perturbative domains as per standard nomenclature.\\

     (ii)\textit{Simplicity of Implementation}:\\
       The formalism's simplicity and ease of implementation or computation are evidently desirable attributes of an ideal approximation scheme.\\
       
        (iii)\textit{Achievement of Reasonable Accuracy}:\\
       In alignment with criterion (i), it is desirable for the scheme to achieve reasonable accuracy, ideally within a few percent of exact or numerical results, for general applications. However, it would be advantageous if greater accuracy could be attained while maintaining the scheme's general applicability.\\ 
        (iv)\textit{Provision for Systematical Improvement in an Ordered Manner}:\\
        The scheme should allow for systematic improvement in an ordered manner, progressing beyond a leading order (LO) result.\\
        (v)\textit{Achieving Rapidity of Convergence of the Perturbative Corrections}:\\
       In any perturbative framework, the sequence of higher-order corrections should either converge rapidly for a convergent sequence or decrease sufficiently in magnitude as an \textit{asymptotic} sequence, to ensure practical utility. 
    
    As previously mentioned, this set of criteria may constitute part of a more comprehensive set of requirements for evaluating an approximation scheme as 'ideal.' In the subsequent \textbf{Section}, we will assess these criteria against existing approximation schemes in quantum theory. 
    
    \section{ A critique of the various extant schemes of approximation}
    
   The preceding survey (\textbf{Section-II}), though concise, aims to be representative and reveals that despite significant advancements in addressing various challenges within interacting quantum systems, an 'ideal' approximation method that fulfills the outlined requirements has yet to be realized. It appears that none of the current schemes simultaneously meet the desired criteria. Consequently, there remains substantial potential to develop an ideal scheme based on the specified criteria. This potential constitutes the primary impetus for the investigations detailed in this report. We will further elaborate on this aspect in the following \textbf{Section}.
   
   \section{Motivation, Objectives, and Scope of the Present Report}
          
   This report is dedicated to the pursuit of the aforementioned ambitious objective: to develop an "ideal" approximation scheme that adheres to the stated objective of broad applicability to quantum mechanics and quantum field theory (QFT). The proposed scheme should meet the criteria of simplicity in its formulation, maintain self-consistency, and exhibit a non-perturbative nature. Additionally, it should allow for systematic enhancement through an improved perturbation theory, ultimately achieving a reasonable degree of accuracy.  
   
    To maintain focus on implementation, this study confines its scope to the quantum physics of self-interacting Bosonic systems. The systems under consideration include the one-dimensional quartic anharmonic oscillator (QAHO), quartic double-well oscillator (QDWO), sextic anharmonic oscillator (SAHO), sextic double-well oscillators (SDWO), octic anharmonic oscillator (OAHO), and the $\lambda \phi^{4}$ quantum field theory. These are all analyzed within the context of physical dimensions and specifically within the massive symmetric phase.   These investigations are contained in  subsequent \textbf{Sections}  and organized as described below: 

    \subsection{Organization of the subsequent sections}
      
      In \textbf{Section-VI}, we present the general formulation of the proposed approximation scheme, hereafter referred to as the \textbf{``Non-perturbative General Approximation Scheme" (NGAS)}. This section emphasizes the non-perturbative nature, self-consistency, and broad applicability of the method. \textbf{Section-VII} addresses the application of this scheme to a class of an-harmonic interactions (AHI) at the leading order (LO). Within this class, we sequentially examine the cases of the QAHO, QDWO, SAHO, SDWO, and OAHO. The formulation's generality is demonstrated through the uniform treatment of all these cases within the \textit{harmonic approximation}. We calculate the energy spectrum and compare the leading order (LO) results with those derived from previous approaches and exact numerical results, where available. This comparison illustrates how this straightforward approach reproduces, at the leading order (LO), results that are within a few percent of those obtained by earlier methods, which often involve different assumptions and sophisticated numerical analyses. In \textbf{Section-VIII}, we demonstrate the \textit{flexibility} of the current scheme by analyzing the AHI using the infinite square-well potential (ISWP) as the input approximation. The \textit{stability} and \textit{flexibility} of the scheme are confirmed by reproducing results of comparable accuracy and maintaining the general features of the scheme with this significantly different input.
      In \textbf{Section-IX}, we advance the formalism to encompass the $\lambda\phi^{4}$-quantum field theory in physical dimensions, concentrating on the massive symmetric phase. The process of \textit{renormalization} of the bare parameters is executed and demonstrated to align with the "Gaussian-effective potential" methodology. Importantly, the \textit{non-triviality} of the renormalized theory is highlighted. An analysis of the vacuum state's properties within the \textit{effective theory} at leading order (LO) reveals its \textit{non-trivial structure}. This approach is shown to be equivalent to the one employing Bogoliubov transformations. Moreover, it is demonstrated that the vacuum of the \textit{effective theory} is a \textit{condensate} of particle pairs from the free theory. Additionally, the \textit{instability of the perturbative vacuum} is confirmed by illustrating that the vacuum energy of the effective theory is considerably lower than that of the perturbative vacuum.
         In \textbf{Section-X}, we detail the \textit{analytic} computation of corrections to the leading-order (LO) results in NGAS, extending to arbitrary order through a novel perturbation theory (PT) framework. This innovative framework is termed "mean-field perturbation theory (MFPT)" in the subsequent discussion. To thoroughly evaluate the relative strengths of the standard formulation of perturbation theory (SFPT) and MFPT, it is crucial to first outline the fundamental characteristics of both approaches. These characteristics are explored in \textbf{Section-XI} and \textbf{Section-XII}, respectively, for SFPT and MFPT. A shared characteristic of both approaches is the \textit{divergence of the perturbation series (PS)}. Therefore, it is essential to clarify the process of interpreting the divergent PS by obtaining a finite "sum" for the former. This topic is examined in detail in \textbf{Section-XI}. The \textit{analytic} computation of perturbative corrections is achieved by recursive evaluation of the same using as "tools" , the Feynman-Hellman theorem (FHT) and the hyper-virial theorem (HVT) . In the context of anharmonic interactions  and using the standard formulation of perturbation theory(SFPT),  these tools were first used by Swenson and Danforth (SD) [81] and  it is described succinctly here in \textbf{Section-XI}. The generalization of the same formalism to MFPT is contained in \textbf{Section- XIII}.  In \textbf{Section-XIV}, we elucidate the application to anharmonic interactions by calculating the perturbation corrections, extending the application of the SD-method to the case of MFPT via the Feynman-Hellman Theorem (FHT) and the Virial Theorems (VT). The large-order behavior of perturbation corrections in MFPT confirms the \textit{asymptotic nature} of the series, as anticipated, yet it is demonstrated to be Borel-summable in all cases of the AHI examined in this report. The Borel summability of the divergent PS for the case of the QDWO in MFPT is emphasized in \textbf{Section-XIV}, as this result starkly contrasts with the situation in SFPT, where Borel summation is impeded by the "renormalon" singularity in the Borel plane. Additionally, this \textbf{Section-XIV} includes the characterization of the leading singularity in the Borel plane, inferred from the arge-order behavior of the perturbation series.
       In \textbf{Section-XV}, the computation of the total perturbative corrections (TPC) to the leading-order (LO) results is presented and tabulated for various cases of the AHI, utilizing two independent methodologies: (i) the method of optimal truncation (MOT) of the asymptotic series at the term of least magnitude, and (ii) Borel summation. In the latter approach, the Borel-Laplace integral is evaluated using the technique of conformal mapping for the analytic continuation of the Borel series beyond its circle of convergence, yielding highly accurate results for the ground state energy across a wide range of coupling strengths in each case of the AHI considered herein. In \textbf{Section-XVI}, a summary of the results is provided, with a comparison and contrast to findings from other studies in the literature. Potential further applications of NGAS to several other areas are discussed in \textbf{Section-XVII}.   
   % \chapter
%    \begin{center}
%    \textbf{CHAPTER-2}
%     \end{center}
    \section{A Non-perturbative General Approximation Scheme (NGAS) in Quantum Theory}
    
    %\section{A New General Approximation Scheme (NGAS) in Quantum Theory} 

    As has been highlighted in \textbf{Section-V}, the objective of the current investigations is to develop an ``ideal" approximation scheme with prescribed features outlined in \textbf{Section-III}.  For achieving  the above objective, we proceed [82] as follows:
     
    \subsection{ General Formulation for Hamiltonian systems }
    
   To ensure clarity and focus, this formulation is limited to one-dimensional self-interacting quantum systems with stationary, non-degenerate energy eigenvalues. We consider a generic Hamiltonian $H(g)$, which characterizes such a system, where $g$ denotes the coupling strength of the interaction. For simplicity in notation, the dependence of $H$ on other variables is omitted. The initial phase of the scheme's implementation requires the selection of an "Approximating Hamiltonian" (AH), represented by $H_0$. This AH must be well-suited to the system under study and align with essential system properties, including symmetries and the specified initial and boundary conditions. Further criteria for the AH are outlined below.   
   
   \subsubsection{Prescription of Conditions on the Approximating Hamiltonian (AH)} 
   
    The eigenvalue equation to be addressed for the system Hamiltonian is expressed as follows:
    \begin{equation*}
     H(g)|\psi_{n}(g)\rangle = E_{n}(g)|\psi_{n}(g)\rangle , 
     \end{equation*} 
     where the stationary energy, $E_{n}(g)$, of the quantum system is selected as the observable. The Hamiltonian, $H(g)$, can be decomposed into an {\it exactly solvable} dominant component $H_{0}$ and a {\it sub-dominant} perturbation, 
     \begin{equation*} 
     H(g) = H_{0}+H^{\prime} . 
     \end{equation*}
   The term 'approximating Hamiltonian' is represented as $H_{0}$. The 'New General Approximation Scheme (NGAS)' is implemented by imposing specific criteria on the AH:\textbf{ (i)} The AH, $H_{0}$, must be exactly and analytically solvable. This requirement implies that the eigenvalue equation for $H_{0}$, expressed as 
     \begin{equation} 
     H_{0}|\phi_{n}\rangle~=~{E}_{0}^{n}|{\phi}_{n}\rangle ; 
     \label{14}
      \end{equation}
 is exactly solvable. In other words, the eigenstates $|{\phi}_{n}\rangle$ and the eigenvalues ${E}_{0}^{n}$ are precisely known. Additionally, the eigenstates are normalized, as indicated by
         \begin{equation} 
       \langle\phi_{n}|\phi_{n}\rangle~=~1 
       \end{equation} 
   This stipulation, encapsulated in equation (14), is referred to as the "condition of exact solvability" (CES). Furthermore, it is required that \textbf{ (ii)} $H_0$ is constructed to depend on a set of {\sl free} (adjustable) parameters, $\{\alpha_{i}\}$, such that 
        \begin{equation*}
        H_{0} = H_{0} (\{\alpha_{i}\}) 
        \end{equation*} 
 In the subsequent step, the 'condition of equal quantum average' is applied, as detailed below.

   \subsubsection{ Condition of equal quantum-average (CEQA) }
   
   This condition is defined by the requirement that the AH must yield the same quantum average (QA) as the original system Hamiltonian, as expressed by:
    \begin{equation}  \langle\phi_{n}|H(g)|\phi_{n}\rangle~=~\langle\phi_{n}|H_{0}(\{\alpha_{i}\})|\phi_{n}\rangle,  
   %\label{16} 
    \end{equation} 
     for all physical $``n"$ and $``g"$, where $|\phi_{n}\rangle$ is defined by the eigenvalue equation for $H_{0}$, eqn.(14), with $``n"$ representing the spectral label. Hereafter, eqn.(16) will be referred to as the \textbf{{\it \textbf{``Constraint of Equal Quantum Average} \textbf{(CEQA)"}}}. The subsequent step involves the {\it optimization} of the approximation as delineated below:  
    
    \subsubsection{Determination of the free parameters in AH by variational-optimization}
    
    To determine the parameters \{ $~\alpha_{i}~$\}, we implement the CEQA alongwith the variational optimization of the AH with respect to the undetermined parameters $(\alpha_{i})$:
     \begin{equation}  
     \frac {\partial}{\partial{ \alpha_{i}}}  \langle H_{0}\rangle =  0
      %\label{17}  
      \end{equation} 
       where the notation is 
       \begin{equation}  
        \langle {A}\rangle ~ \equiv~\langle\phi_{n}|{ A} |\phi_{n}\rangle 
         %\label{18}  
     \end{equation} 
    We designate this condition, eqn.(17), as the ``\textbf{{\it\textbf{condition of optimality}} (CO)}''. This methodical approach not only conclusively determines the free parameters ${\alpha_{i}}$ but also implements the feedback of nonlinearity and $g$-dependence into the AH. This aspect is further discussed in the following subsection.                    
    
    \subsubsection{The non-linear feedback mechanism of the approximation}
    
     The concept of {\sl self-consistency} is embedded within the framework of eqs.(14-18). This approach effectively incorporates the nonlinear $g$-dependence of the system Hamiltonian, $H(g)$, into the approximating Hamiltonian, $H_{0}$, such that
    
  \begin{equation} 
  H_{0}(\{\alpha_{i}\}) \rightarrow H_{0}(g,n) 
  %\label{19} 
  \end{equation}
    As a result, this process establishes a mapping from the original (non-solvable) Hamiltonian $H(g)$ to an exactly solvable approximating Hamiltonian, $H_0$, while preserving the non-linearity of the original Hamiltonian in a self-consistent manner. 
    
  % \textbf{ Up to this point AI improved-Aug 08-2025}
    
  \subsubsection{Features of the Leading order (LO) results in NGAS}
    
    The steps outlined above form the fundamental ingredients of the proposed approximation scheme at the leading order (LO). \textit{The {\it exact} solution of the eigenvalue equation for $H_{0}(g,n)$, represented by $E_{0}^{n}(g)$, is naturally \textit{recognized as the leading-order (LO) result}} and is expected to significantly contribute to the true eigenvalue, $E_{n}(g)$. At this stage, the following observations regarding the approximation scheme are noteworthy:
     
     (i) It should be noted that in a restricted form, specifically when the quantum average in eqn.(16) is confined to the ground state only, the CEQA as expressed in eqn.(16) corresponds to the Hartree-approximation or mean field approximation in quantum field theory [36]. In this context, NGAS can be regarded as a {\it "generalized"} Hartree-approximation method [37].
    
    (ii) It is important to emphasize that even the LO results capture the dominant contribution of the (nonlinear) interaction through the self-consistency procedure of NGAS, despite the fact that one is always dealing with an exactly solvable Hamiltonian $H_{0}$. We regard this as a key feature of the approximation method.
    
     (iii) Systematic improvement of the LO-result can be achieved in NGAS through the development of an enhanced perturbative framework as follows. Consider the following equation derived from \textit{CEQA}, eqn.(16):
     \begin{equation} 
    \langle\phi_n|H'(g,n)|\phi_n\rangle=0, 
    %\label{20} 
    \end{equation}
     where 
    \begin{equation} 
    H'\equiv (H-H_0). 
    %\label{21} 
    \end{equation}
     It is further noted that

    \begin{equation}  
    \langle\phi_n|H'(g,n)|\phi_n\rangle \ll\langle\phi_n|H_0(g,n)|\phi_n\rangle  %\label{22} 
    \end{equation}
    
    for arbitrary values of $g$ and $n$, indicating that the above correction -  contribution remains sub-dominant (in the sense of \textit{QA}) \textit{regardless} of the coupling strength and excitation levels. Consequently, it is naturally suggested to develop a perturbation theory that treats $H_0$ as the unperturbed Hamiltonian and $H'$ as the perturbation. This novel perturbation theory is anticipated to exhibit characteristics that are significantly distinct from those in SFPT. The details are elaborated in \textbf{Section-X}.
    
    \subsection{Section  Summary}
   
   In this \textbf{Section-VI}, a novel general scheme of approximation (NGAS) is developed, which seeks to preserve the merits while addressing to overcome the challenges and limitations of the existing methods and schemes discussed in the preceding \textbf{Section-II}.The NGAS is characterized by the following features:
   (a) \textit{Universal applicability} to a general interacting system in Quantum Theory within the Hamiltonian formulation, regardless of the interaction strength. (b) By design, this scheme is shown to be potentially capable of providing \textit{accurate approximation for any {\it general} interaction of {\it arbitrary strength}} in quantum theory, even at the leading order (LO). (c) Consequently, the scheme is essentially \textit{non-perturbative} in nature at the LO. (d) Furthermore, this \textit{Section-VI} indicates how to enhance the accuracy of the LO results through the formulation of an \textit{improved perturbation theory} in a systematic manner within the scheme, while maintaining the aforementioned features of universality and non-perturbative character. (e) The fundamental input in NGAS is the selection of a suitable \textit{approximating-Hamiltonian (AH)} ${H_{}}$, which must be \textit{exactly} solvable and \textit{involves certain adjustable (variational) parameters, $\alpha_{i}$}. The scheme is implemented by imposing the constraint of \textbf{equal quantum-average (CEQA)}, ensuring that the quantum average (QA) of the original Hamiltonian $H$ is equal to that of the AH ${H_{}}$ with respect to any (arbitrary) eigenstate of the latter. The variational minimization of this QA with respect to the undetermined free parameters completely specifies the latter. This is referred to as the \textit{constraint of Optimality (CO)}. (f) Consistent with the basic requirements as discussed earlier, there exists considerable \textit{flexibility} in the choice of the AH which may be profitably used to improve the overall accuracy of the results . (g)The evaluation of the energy eigenvalue of the AH determines the leading order(LO) result in NGAS. (h) The potential development of an improved perturbation theory for arbitrary strength of interaction is contained in NGAS.
   
   In the following {\bf Sections}, we implement and illustrate the general approach described above by applying it to specific benchmark systems traditionally employed as theoretical laboratories to test approximation methods in quantum theory (QT).    
   
    \newpage
    
   % \chapter
%    \begin{center}
%        \textbf{CHAPTER-3}
%        \end{center}
            
   \section{ Application of NGAS in the Leading Order (LO) to Anharmonic Interactions (AHI)}
    
    Anharmonic Interactions (AHI) are characterized by potential functions that are polynomial additions to the harmonic interaction. Examples include quartic anharmonic and double-well potentials, as well as sextic, octic, and higher-order anharmonic potentials (we confine our discussion here to parity-symmetric potentials). These can be represented as:
    \begin{equation}
    V(x)=\sum\limits_{k=1}^ {K}a_k x^{2k}
    %\label{23}
    \end{equation}
    
    These interactions are fundamental in quantum theory due to their extensive applications (see, e.g., [31]) across various domains of physics, as well as their utility in serving as traditional testing grounds for various approximation schemes (see, e.g., [9]), given that exact or highly accurate numerical results are available (see, e.g., [8,9]) for comparison. Furthermore, a substantial body of literature is available for the comparison of different methods and schemes of approximation applied to the AHI. These are the primary reasons for the application of the NGAS to the AHI. In the following sections, we describe the application of the NGAS in LO approximation to the individual AHI systems.   

    \subsection{ The Harmonic Approximation}
    
    Given that the AHI are characterized by the modification of the harmonic oscillator potential through the incorporation of additional polynomial terms of higher degree, it is reasonable to select the approximating Hamiltonian (AH), $H_0$, as that of the harmonic oscillator (HO), albeit with a potential \textit{shift} in the field variable $x$ as well as in the energy. The frequency, $\omega$, of the HO becomes the other free parameter. The exact expression is provided in the subsections that follow. This choice then defines the \textit{Harmonic Approximation} for the different cases of the AHI.      
    
      However, as noted earlier, the NGAS is adaptable to the choice of the input AH, subject only to the defining properties of the latter. In a later section (\textbf{Section-VIII}), we will employ a different approximation, namely the \textit{Infinite-Square-Well Approximation}, to illustrate this flexibility.      
    
      In the subsequent subsections, we consider the different cases of the AHI in turn.

    \subsection{ The Quartic Anharmonic Oscillator (QAHO)}
    
    The quartic anharmonic oscillator (QAHO) represents the most fundamental system exhibiting self-interaction. This system has been the subject of extensive research, resulting in a substantial body of literature [8,9]. Its significance is underscored by its applications in various physical domains, including quantum field theory, condensed matter physics [83], statistical mechanics [84], non-linear systems [85], classical and quantum chaos [86], inflationary cosmology [87], lattice dynamics, and plasma oscillations, among others. Furthermore, the QAHO has functioned as a theoretical laboratory for examining the convergence of perturbation theory [88], the development of non-perturbative approximation methods [89], renormalization [90], vacuum structure [91], and stability analysis [92]. This quantum system serves as a foundational model for testing various aspects of approximation methods, as evidenced in numerous studies [9-11,17-18,21-23,26-42,47-60]. The Hamiltonian of the system is given by
     \begin{equation}
    H~=~\frac{1}{2}p^{2}~+~\frac{1}{2}~x^{2} + gx^{4},
    %\label{24}
    \end{equation}
   
   where $~ g~$ is a real and positive parameter (\textit{We employ dimensionless variables throughout by utilizing standard scaling methods}). To develop the NGAS for the QAHO, we adhere to the steps outlined in {\bf Section-VI}, as detailed below.   
   
    \bigskip
    
    {\bf (i) Choice of the Approximating Hamiltonian (AH) and the Solution for its spectrum}
   
     To derive the exact analytical solution of the spectrum of $~H_{0}~$, we proceed as follows. We adopt the "harmonic approximation" for the input approximating Hamiltonian (AH), defined by the equation   
    \begin{equation}   
    H_{0}~=~\frac{1}{2}p^{2}~+~\frac{1}{2}~\omega ^{2}x^{2} + h_{0}.   
    %\label{25}   
    \end{equation}   
    
   It is immediately apparent that the AH given by eqn.(25) corresponds to a "shifted" effective harmonic oscillator, where the energy is displaced by an amount: $h_{0}$. Furthermore, the parameter $ \omega $ is constrained by physical requirements to satisfy $~\omega > ~0$. For the implementation of CEQA, it is necessary to evaluate:   
   
    
   
   \begin{equation}   
    \langle H \rangle = \frac{1}{2}\langle p^{2}\rangle + \frac{1}{2} \langle x^{2} \rangle + g \langle x^{4} \rangle   
    %\label{26}   
   \end{equation}   
   
   where the notation $ \langle A \rangle \equiv \langle \phi_{n}|A|\phi_{n} \rangle $ is used for an operator A. Utilizing the definition of $|\phi_{n}\rangle $ as the eigenfunction of $ H_{0} $, the relevant operator averages in $\langle H \rangle $ can be calculated [8,17] using standard methods, such as the formalism of ladder operators for $ H_{0} $. These are given by the following equations:
   
   \begin{subequations}   
   \begin{align}      
   \langle x^{2} \rangle = (\xi /\omega)\\
   \langle p^{2}\rangle = \omega\xi ,\\   
   \langle x^{4} \rangle = 3 (1+4\xi^{2})/8\omega^{2} ,\\   
 \langle x^{6} \rangle = (5/8) (\xi/\omega^{3}) (5+4\xi^{2}),
   %\label{27}   
   \end{align}  
    \end{subequations}  
where ${\xi=(n+1/2)}$. By substituting the QA values from eqs.(27a-d), the QA of the original Hamiltonian $ H $ defined by eqn.(26) can now be evaluated as   
    \begin{eqnarray}   
    \langle \phi_{n} |H| \phi_{n}\rangle \equiv \langle \phi_{n} |H_{0} |\phi_{n} \rangle = \omega\xi/2 + (\xi/2\omega)+ (3g/8\omega^{2}) (1+4\xi^{2})\equiv E_{0}(g,\omega,\xi).   
    %\label{28}    
    \end{eqnarray}
       
    
    % \textbf{Up to this point AI improved on 09-08-2025}
    
    {\bf  (ii)  Determination of the free parameters by CEQA and CO}
    \bigskip
    
    Conducting the explicit variational minimization of $ \langle H_{0} \rangle $ with respect to $~'\omega ~'$ i.e., $ \dfrac {\partial \langle H_{0} \rangle } {\partial \omega } = 0 $, yields the following equation:
             
    \begin{equation}
     \omega^{3} - \omega - 6 g f(\xi) = 0
     % eqn(29)
     \end{equation}
     from which $\omega$ can be determined as the real, positive root. It is noteworthy that this equation has been derived by several authors [93], albeit from significantly different considerations. Similarly, by noting that $ h_{0} = E_{0}(g,\omega,\xi) - \omega\xi $, the expression for $ h_{0} $ is obtained and given by
    \begin{equation}
     h_{0} = \left(\frac {\xi} {4} \right) \left( \frac {1} {\omega} - \omega \right)
     % eqn(30)
     \end{equation}
     where $ f(\xi) \equiv \xi + \frac {1} {4\xi} $ and $ \xi = (n+\frac{1}{2})$; with the spectral index $ n$ taking values: $ n = 0,1,2,3,.....$. We refer to eqn.(29) as the \textit{``gap equation (GE)"} and eqn.(30) as the \textit{``energy-shift"}. As anticipated, these parameters thus acquire a functional dependence on $g$ and $\xi$:
     
     \begin{equation} ~~~~~~\omega\rightarrow\omega(g,\xi) ~~~~~~h_{0}\rightarrow h_{0}(g,\xi) 
    %eqn(31)
    \end{equation}
     \bigskip
    
     {\bf (iii) Determination of the Energy-Spectrum in the Leading Order (LO)} The solution of the gap equation (GE) eqn.(29) constitutes the key component in the calculation of the energy spectrum. From eqs.(28) and (31), it follows that the eigenvalues of $H_{0}$ also acquire the requisite $~g~,~\xi~$-dependence, and the leading order (LO) result for the energy is given by
    
    \begin{eqnarray}
    E_{0} (g,\xi) = (\xi/4)\left(3\omega + \frac {1} {\omega} \right) 
     %eqn(32)
     \end{eqnarray}
    
  where `$\omega $' is obtained as a solution of the $GE$ for the QAHO given by eqn.(29). It should be noted that the $GE$, eqn.(29), is in the form of a cubic equation of the type
   \begin{equation}
         x^{3} - 3Px - 2Q = 0;~ P, Q > 0
         %eqn(33)
    \end{equation}
    
  The real solution of this eqn.(33) is given by,
   \begin{equation}
     x = Q^{1/3}[ ( 1 + \sqrt {(1 - \frac{P^{3}}{Q^{2}})})^{1/3} + ( 1 - \sqrt {(1 - \frac{P^{3}}{Q^{2}})})^{1/3}]~~~~~
    %eqn(34)
    \end{equation}
    
   Then, by comparing the coefficients of eqs.(33) and (29), the required solution is obtained explicitly as
     \begin{equation}
    \omega=(3g f(\xi))^{1/3}[(1 + \sqrt {(1 - \rho)})^{1/3}+(1 - \sqrt {(1 - \rho)})^{1/3}],
    %eqn(35)
    \end{equation}
     where, $~\rho^{-1}~=~243g^{2}f^{2}(\xi)~$.
    
    It should be noted that the solution for `$~\omega~$' as given above exhibits the correct limiting behavior, $~\omega~ \longrightarrow~1$ for $~ g \longrightarrow 0~$, and further, it demonstrates the \textit{non-analytic dependence on the coupling '$~g~$' at the origin}, characteristic of the non-perturbative nature of the NGAS.
    
    At this juncture, several features pertaining to the leading order (LO) results based on eqs. (29) - (35) warrant attention: (a) As previously mentioned, equation (29) has been independently derived by several authors [93] from diverse independent considerations. (b) Furthermore, the rigorously established [11] analytic structure of $E(g)$ in the $g$-plane, along with the non-analytic dependence on the coupling strength '$g$', is inherently present [8,82] at the LO level through the solution of Eq. (29), as explicitly provided by Eq. (35). (c)The dependence of $\omega$ and $h_{0}$ on $\xi$ and $g$ implies that the AH, $H_{0}$, also depends on these parameters. The physical implication of this dependence is that the eigenfunctions of $H_{0}$, $|\phi_{n}\rangle$, are not mutually orthogonal: $\langle\phi_{m}|\phi_{n}\rangle \neq 0$ for $m \neq n$. As discussed later, this situation does not affect the development of perturbation corrections in the standard formulation of perturbation theory (SFPT), which can be generated by a "wave-function-independent methods," such as the Hyper Virial Theorem (HVT) and the Feynman-Hellman Theorem (FVT)-based formalism employed here (see, \textbf{Section-X}). (d) An additional significance of $\omega$ arises from the altered ground state structure [8,50,52] due to interaction, which differs non-trivially from the free field ground state. Moreover, the "dressed ground state" of the interacting system in the LO approximation possesses lower energy [8,50,52,82] compared to the "trivial" ground state of the noninteracting theory for any non-zero value of the coupling '$g$'. This result thus establishes the instability of the`free-field ground state’ in the presence of interaction. This aspect has been discussed in detail elsewhere [8,50,52,82]. (e) The $(g,\xi)$-dependence of $\omega$ and $h_{0}$ as determined in the LO does not undergo alteration in the computation of perturbation corrections at higher orders. We elaborate on this important feature later (\textbf{Section-X}).  (f) Furthermore, as previously stated, the accuracy of the energy spectrum obtained in the LO, Eq. (3.12), is quite significant, with deviations from the "exact" results being no more than a few percent [8,50,52,82] over the full range of $g$ and $n$. This result ensures that the dominant contribution indeed arises from the LO, as required in a perturbative framework. We also demonstrate the improvement of results obtained by the inclusion of higher-order corrections to the LO results, which is discussed in \textbf{Section-X}. We next apply the method to the case of the quartic-double well oscillator in the following subsection.
    
    \textbf{Up to this point on Aug 10-2025 (FN)}
    
    \subsection{The Quartic Double Well Oscillator (QDWO)}
    
    The quantum double-well oscillator (QDWO) is a system of significant theoretical and practical interest [94]. The Hamiltonian of this system is expressed as:   
    \begin{equation}   
    H~=~\frac{1}{2}p^{2}~-~\frac{1}{2}~x^{2}+g x^{4};~~ g~> 0     
     %\label{36}   
    \end{equation}    
    
     
    
    The negative sign of the $x^{2}$ term introduces a distinct physical scenario compared to the quantum anharmonic oscillator (QAHO), even in the classical limit. The classical potential, $~V_{c} \equiv - \frac{1}{2} x^{2}~+~g x^{4} ~$ displays the characteristic double-well shape with symmetric minima. These minima are located at $~\pm \frac{1}{2\sqrt g}~$, each with a depth of $~\frac{1}{16 g}~$. As $~g~$ decreases, the depth of the wells increases. The actual low-lying energy eigenstates of the system differ significantly from those predicted by the harmonic basis, which impedes the convergence of the resulting perturbation theory. Ideally, a natural solution would involve the simultaneous use of two harmonic basis centers around the minima at $~\pm \frac {1}{2\sqrt g}~$. However, implementing this approach requires the use of nonorthogonal states, which is cumbersome. Another challenge is that the theory is not defined for $~g \rightarrow 0~$, as the ground state does not exist in this limit due to the absence of a lower bound for $V_{c}$. Consequently, the simple harmonic oscillator (SHO) is not the free-field limit of the QDWO. Therefore, the standard formulation of perturbation theory (SFPT) is not applicable in this case. The perturbation expansion of the eigenvalues $~E_{n}( g)~$ in powers of $~g~$ is divergent [95-99] for all $~g~>~0~$. This divergence can be qualitatively understood by noting that the addition of the $~g x^{4}~$ term transforms a completely continuous eigenvalue spectrum of $~p^{2}~-~x^{2}~$ into a discrete spectrum bounded from below. A non-perturbative approach is thus necessary. Furthermore, it is noteworthy that the QDWO case is not Borel-summable [95-99] in SFPT for any value of the coupling strength. Recent advancements, such as the theory of resurgence and trans-series [95-97], distributional-Borel summation [98], and the generalized Borel-Padé method [99], have been developed to address this issue. In light of these considerations, the QDWO case assumes special relevance for investigation in the  non-perturbative general approximation scheme (NGAS).
    
           In contrast to the situation in SFPT, the NGAS can be effectively applied to the case of the QDWO, even in the LO, using a procedure analogous to that employed for the QAHO, over a considerably larger range of values for $~ n~$ and $~g~$. To develop the NGAS for QDWO, the "approximating Hamiltonian (AH)" for the system is again selected within the harmonic approximation as follows:

    \begin{equation} 
    H_{0} = (1/2) p^{2} + (1/2) {\omega}^{2} (x - \sigma)^{2} + h_{0} ,
      %\label{37}
    \end{equation} 
    
     but generalized to account for spontaneous symmetry breaking (SSB) through a nonzero vacuum expectation value for $ x $, denoted as $ \sigma $. Consequently, the various average values, analogous to eqs.(27a-c), are now expressed as:
    \begin{equation}
    \langle x^{2} \rangle = \sigma^{2} + (\xi/\omega),~~~~~~~
    \langle p^{2} \rangle = \omega\xi,
    % eqn(38)
    \end{equation}
    \begin{equation}
    \langle x^{3} \rangle = \sigma^{3} + 3\sigma(\xi/\omega),
    %eqn(39)
    \end{equation}
    \begin{equation}
    \langle x^{4} \rangle = \sigma^{4} + 6\sigma^{2} (\xi/\omega) + 3(1 + 4\xi^{2})/8\omega^{2}.
    %eqn(40)
    \end{equation}
     Equations (38-40) facilitate the evaluation of $\langle H \rangle $ in terms of the input parameters. These parameters: $ \omega$, $\sigma$, and $h_{0}$ are then determined analogously to the case of the AHO, in terms of $g$ and $\xi$. However, a distinctive feature in the case of the QDWO is the occurrence of a "quantum-phase transition (QPT)" [82] governed by a "critical coupling" $g_{c}(\xi)$, given by the expression:
         \begin{equation*}
      g_{c}(\xi) = \frac{(2/3)^{3/2}} {3(5\xi-(1/4\xi ))}
          \end{equation*}
     
  such that the "Spontaneously Symmetry Broken (SSB)" phase is realized with $\sigma \neq 0 $ for $ g \leq g_{c}(\xi)$, whereas the "symmetry-restored (SR) phase" is obtained with $\sigma = 0 $ when $ g > g_{c}(\xi)$. Numerically, we have: $g_{c} = 0.0907218$, for the ground state of the DWO. The transition across $g = g_{c}(\xi)$ is discontinuous, with the two phases governed by distinct expressions for $\omega$ and $E_{0}$, which are not analytically connected. It is, therefore, necessary to consider the two phases separately. However, due to the relatively small value of $ g_{c} $, the "SSB phase" exists only over a very limited range of $g$, i.e., when $ 0 \leq g_{c} \leq 0.0907218 $ for the ground state, and $ g_{c} $ takes even smaller values for the higher excited states, eventually vanishing for large $n$. Therefore, we confine our discussion to reporting the results for the "SR phase" only.
 In this phase, the condition $ g>g_{c}(\xi)$ is satisfied, with $\sigma = 0$, and $\omega$ fulfilling the gap equation:   
      \begin{equation}   
      \omega^{3} + \omega - 6g(\xi + \frac{1}{4\xi}) = 0    
      %(41)   
       \end{equation}   
       The energy levels in this phase are represented by the expression:    \begin{equation}   
       E_{0} = (\xi/4)(3\omega - (1/\omega))   
        %(42)    
       \end{equation}   
      and $h_{0} = E_{0} - \omega\xi$. In equation [42], $\omega$ is the solution of equation [41], which can be derived analogously and is expressed as:   
      \begin{equation}   
       \omega~=~(3 g f(\xi))^{1/3}[(~\sqrt{( 1 + \rho)} +1)^{1/3}~-~(~\sqrt {(1 + \rho)} -~1)^{1/3}]   
        %(43)   
      \end{equation}
where $~\rho^{-1}~=~243 g^{2}f^{2}(\xi)~$. 

    Similar to the case of the Quantum Anharmonic Oscillator (QAHO), the leading-order (LO) results for $E_{0}$ effectively capture the primary contribution in reproducing the energy spectrum [8,82] with a precision of a few percent - for the ground state energy, refer to \textbf{TABLE-I and -II in Section-X}.  These tables also illustrate the enhancement of accuracy of the LO results achieved through higher-order corrections in the ‘Mean Field Perturbation Theory (MFPT)’, as discussed in \textbf{Section-X}.
     
     In the next subsection we consider the case of the sextic-anharmonic oscillator in the NGAS.
     
    %\textbf{ Up to this point on 14-08-2025-FN}
     
    \subsection{ The Sextic Anharmonic Oscillator (SAHO)}
    The sextic anharmonic oscillator system is a prime example of higher anharmonicity and has been the subject of extensive research [100]. This system holds considerable interest and significance due to its applications in various branches of physics [100]. As a theoretical construct, it, along with its double-well variant, represents one of the simplest cases where supersymmetric quantum mechanics (SUSYQM) offers precise predictions [101] for energy levels based on specific Hamiltonian parameters. This allows for the evaluation of various models and approximation techniques against the exact analytical outcomes provided by SUSYQM. Additionally, the increased anharmonicity results in a more severe divergence of the SFPT at higher orders [102], characterized by $ E_{n} \sim \Gamma [n(m-1)] $ for $ gx^{2m}$ types of AHO. This feature underscores the sextic-AHO's importance in assessing convergent-approximation methods.  For these practical and theoretical reasons, the sextic-AHO serves as a distinctive testing platform for the NGAS, as detailed below. 
    
     The Hamiltonian for this system is defined as:  
        \begin{equation}
    H = \frac{1}{2} p^{2} + \frac{1}{2} x^{2} + gx^{2K},~~~~   (K=3)
    %(44)  
    \end{equation}
    The results for the SAHO can be obtained by employing a method similar to that used for the QAHO, utilizing the following input [82] for $\omega$, $h_{0}$, and $E_{0}$:  The value of $\omega$ is determined by the real, positive root of the 'gap-equation', 
     \begin{equation*} 
    \omega^{4} - \omega^{2} - (15g/4)(5 + 4\xi^{2}) = 0. 
    \end{equation*} 
    This leads to the derivation [82] of the following expression for the energy levels of the sextic AHO in the LO:
     \begin{equation*} 
    E_{0} = (\xi/3)(2\omega + (1/\omega)) ; ~~~~\\ h_{0} = (\xi/3)\left(\frac{1}{\omega}-\omega\right).
     \end{equation*}
     The LO results for the energy are obtained with high accuracy [82], within a few percent, across a broad range of $\xi$ and for any physical values of $ g\geq 0 $. The LO results along with perturbative improvement of accuracy through MFPT are discussed in \textbf{Section-X}. 
     \subsection{ The Octic- Anharmonic Oscillator (OAHO)}
    The octic-anharmonic oscillator, akin to its quartic and sextic analogs, is utilized in the modeling of molecular physics, lattice vibrations in solids, and quantum chemistry. Owing to its enhanced anharmonicity, this system also functions as a theoretical - laboratory for more stringent evaluations of non-perturbative approximation schemes in quantum theory, as the divergence of the SFPT at higher perturbation orders becomes increasingly pronounced [102] in this case. The system is examined here to assess the generality, uniform applicability, and reliability of NGAS in the context of higher anharmonicities in the AHI. For this purpose, the detailed description is as follows: 
    
    The Hamiltonian is expressed as
     \begin{equation} H~=~\frac{1}{2} p^{2}+\frac{1}{2}x^{2} + gx^{8}; ~ g ~ >0. %(45)
      \end{equation}
    In this context, we utilize the "approximating Hamiltonian (AH)" defined as follows: 
      \begin{equation} H_{0}~=~\frac{1}{2}p^{2}+\frac{1}{2}\omega^{2}(x - \sigma)^{2} + h_{0} 
      %(46) 
      \end{equation}.
  
   Equation [46] represents a "shifted" effective harmonic oscillator. The energy spectrum, consistent with previous instances, is given by 
   \begin{equation} 
   E_{n}^{(0)}~=~\omega~ \xi + h_{0}~, 
   \end{equation}
    where, as previously defined:~~~ $\xi~=~ ( n + 1/2 )~$ ;$~ n~ =~ 0,1,2,.....$.
   
    It is subsequently essential to determine the frequency $\omega$' and $h_{0}$'. To this end, we observe that the quantum average of equation (45) is expressed as: 
   \begin{equation} < H >~=~<H_{0}>~=~\frac{1}{2}< p^{2} >+\frac{1}{2}< x^{2} > + g < x^{8} > 
   %label eqn(48)
    \end{equation}
   
    
   
   where the quantum averages on the right-hand side of equation (48) can be computed using the standard properties of creation/annihilation operators as in prior cases (refer to equation (27)). By applying the variational minimization condition and considering $\sigma = 0$ to achieve a "physically acceptable" solution, we derive [82] the simplified gap equation, expressed as:
   \begin{equation}
    \omega^{5}-\omega^{3}-35 g h(\xi)=0
    %label eqn(49) 
   \end{equation} 
   where,~~~~~~~~~~~~~~ $h(\xi) = \xi^{3} + (7/2)\xi + (9/16\xi)$.
   \\ {\it Solution of the Gap Equation and Determination of the Energy Spectrum}\\
   The solution of equation (49) determines the frequency $\omega$' of the "shifted" harmonic oscillator. To obtain the energy levels, one substitutes $~\sigma~=~0$ and $\omega$' as the solution of equation (49). This leads, after some simplification, to the following simple formula [82]: 
   \begin{equation} 
   E_{0}=(\frac{\xi}{8})(5\omega+\frac{3}{\omega}).
    %(50) 
   \end{equation} 
   where `$\omega$' is obtained by solving equation (49) \textit{numerically}. 
   We compare the leading-order result in NGAS [82] with previous computations [103] over a wide range of values of $g$ and $n$. It is evident from this comparison [82] that the results obtained in the leading order of NGAS are already quite accurate over the full range of the parameters. \textit{This demonstrates the generality of the method and uniformity of the approximation with increasing anharmonicity. }
   
   As previously mentioned, a significant feature of NGAS is its flexibility in accommodating various choices for the input Hamiltonian, AH, to facilitate comparison and enhancement of the final results for the system-Hamiltonian. This feature will be demonstrated in the subsequent Section.

  % \textbf{Up to this point on 18-08-2025}
   
   \section{ Flexibility of Input in NGAS~-~The square-well approximation}
   
    In the preceding sections, the \textit{'simple harmonic-approximation'} was identified as the natural choice for the input/approximating Hamiltonian (AH) in NGAS when addressing various cases of AHI. In this \textbf{Section}, we explore the flexibility in selecting the AH in NGAS by considering the infinite-square-well (ISW) potential. A significant motivation for choosing the ISW potential as the input is its pedagogical value and the simplicity of this system. It is well-established that the ISW potential represents one of the simplest systems that allows for an exact analytical solution. Consequently, it is included in any standard introductory quantum mechanics course. Establishing an approximation that connects the AHO/DWO system to the ISW case may thus serve as an illustrative example of applying a standard textbook topic to advanced research. However, it must be acknowledged that the ISW approximation is perhaps the \textit{crudest} among the possible choices of AH. This is because the system described by the ISW potential propagates \emph{freely} between the infinite walls, in stark contrast to the actual situation for the AHO/DWO. \textit{Nevertheless, we deliberately choose it here to test the robustness and tolerance of NGAS to the crudest possible input approximation}, which could serve as an additional motivation. Furthermore, as a by-product of this study, it is possible to derive an approximation for the well-known case of the SHO in the limit where the anharmonic coupling strength is set to zero in the AHO Hamiltonian, thereby gaining further insight into the accuracy of NGAS. In the subsequent subsection, we demonstrate the method by applying it to the case of the quartic AHO, selecting the ISW Hamiltonian as the input. The LO results for the energy spectrum are obtained and compared with results from other calculations. We also emphasize the non-perturbative aspects and the analytic structure of energy as a function of the quartic coupling.

    \subsection{ The Quartic Anharmonic Oscillator in ISW-approximation}
    The quartic anharmonic oscillator (quartic-AHO) is characterized by the Hamiltonian previously introduced (refer to equation (24)) and is reiterated here for convenience with a different equation number:
   \begin{equation}
   {H}~=~{\frac{1}{2}~p^2}~+~{\frac{1}{2}~x^2}~+~{g{x^4}}
   %label{51}
   \end{equation}
   where $g>0$.
   
   The Hamiltonian for the infinite square well (ISW) potential is defined as follows:
    \begin{eqnarray}
   {H^{SW}_{0}~}=~{\frac{1}{2}~p^2}~+~{V_{SW}}
   %label{52}
   \end{eqnarray}
   where the potential $V_{SW}$ is specified by:
   \begin{eqnarray}
   {V_{SW}(x)}~=~\left\{\begin{array}{rc}
   \infty & \mbox{$|x|\ge a$}\\
   h & \mbox{$~~|x|<~a~$},\\
   \end{array}
   \right.
  %label{53}
    \end{eqnarray}
   The ISW potential is notably characterized by two free parameters: the width, denoted as $2a$, and the depth, denoted as $h$. Additionally, the potential is selected to be symmetric under spatial inversion, such that $V_{SW}(-x)= V_{SW}(x)$, to ensure consistency with the symmetry of the original Hamiltonian as specified in equation (51). The eigenvalue equation for $H_0^{SW}$ is presented as follows:
   \begin{eqnarray}
   H^{SW}_0~\phi^{SW}_n(x)~=~\mathcal{E}_n~\phi^{SW}_n(x),
   %Eqn(54)
   \end{eqnarray}
   This equation is readily solvable under the boundary condition:
   \begin{eqnarray}
   \phi^{SW}_n(x)~~=~  ~~~~~\textrm{for}~~ |x|\geq{a},
   %Eqn(55)
   \end{eqnarray}
   which ensures the physical requirement of absolute confinement of the system within the infinite potential barriers. For $|x|< a$, the normalized eigenfunctions, which vanish at the boundary of the potential well, are given by:
   \begin{eqnarray}
   \phi^{SW}_n(x)\equiv\phi^{(-)}_{n}(x)~=~\frac{1}{\sqrt{a}}~\sin\left(\frac{n\pi x}{2a}\right);~~\textrm{for}~~\nonumber\\ n=2,4,6,8...;
   %\label{56}
   \end{eqnarray}    and,   
    \begin{eqnarray}   
    \phi^{SW}_n(x)\equiv\phi^{(+)}_{n}(x)~=~\frac{1}{\sqrt{a}}~cos\left(\frac{n\pi x}{2a}\right);~~\textrm{for}\nonumber\\~~ n=1,3,5,7....   
     %\label{57}    
    \end{eqnarray}      
   
   In the above equations, the $(\pm)$ superscripts correspond to the even(odd)-parity solutions. The energy eigenvalues are straightforwardly obtained by solving the Schr\"{o}dinger equation and are given by   
   \begin{eqnarray}   
   {\mathcal{E}_{n}}~=~{h}~+~\frac{n^2\pi^2}{8a^2}   
   %\label{58}   
   \end{eqnarray}   
   The subsequent task involves determining the adjustable parameters, $\it{ a'}$ and $\it{ h'}$. The width parameter $`a'$ can be determined by the variational minimization of ${<H>}$. Here, we use the notation: ${<\hat {A}>}$ to define the quantum average/expectation value of the operator, ${\hat {A}}$ as given below:
   \begin{eqnarray}
   \langle {\hat A}\rangle ~\equiv~\int_{-a}^{+a}dx~{\phi_n^{\ast}(x)~\hat A~\phi_n(x)}
    %Eqn(59)
   \end{eqnarray}
   The evaluation of $<H>$ using the aforementioned definition, eqn.[59], is straightforward. It is noted that:
\begin{eqnarray} 
 \langle H\rangle~=~\langle\frac{1}{2}~p^2\rangle~+~\langle\frac{1}{2}~x^2\rangle~+~
 \langle g x^4\rangle,
   %\label{60}
\end{eqnarray}

    each term in eqn.(60) can be computed easily by exploiting parity-invariance, which forbids parity-changing transitions, i.e. $~<\phi_n^{+}|H|\phi_n^{-}>=0=<\phi_n^{-}|H|\phi_n^{+}>$.
    
     The result is given below:
    \begin{eqnarray}
    \langle H\rangle~=~ \left( \frac{n^2\pi^2}{8}\right)\left(\frac{1}{a^2}\right)~+~\left(\frac{1}{6}\right)c_n a^2~+ g a^4\left(\frac{1}{5}-\frac{4c_n}{n^2 \pi^2}\right)
    %\label{61}
    \end{eqnarray}
    where,
    \begin{eqnarray}
    c_n~\equiv~1-\left(\frac{6}{n^2 \pi^2}\right) ~~and~~{n= 1,2,3,4,...}.
    %Eqn(62)
    \end{eqnarray}
    The minimization of the expression for $<H>$ as given above, with respect to $u\equiv~\left(1/a^2\right)$ leads to the following equation:
    \begin{eqnarray}
    u^3-P\left(n\right)u-Q\left(g,n\right)~=~0,
    %Eqn(63)
    \end{eqnarray}
    where,
    \begin{eqnarray}
    P\left(n\right)~\equiv~\left(\frac{4}{3}\right)\left(\frac{c_n}{n^2 \pi^2}\right),
    %Eqn(64)
    \end{eqnarray}
    \begin{eqnarray}
    Q\left(g,n\right)~\equiv~\left(\frac{16 g}{n^2 \pi^2}\right){\left(\frac{1}{5}-\frac{4 c_n}{n^2 \pi^2}\right)}.
    %\label{65}
    \end{eqnarray}
    The real, positive root of eqn.(63) is required on physical grounds. This is given by,
    \begin{eqnarray}
    u=\left\lbrace {\left(\frac{8 g}{n^2\pi^2}\right)\left(\frac{1}{5}-\frac{4 c_n}{n^2\pi^2}\right)}\right\rbrace^\frac{1}{3}\left[\left\lbrace 1+\sqrt{1-\rho}\right\rbrace^\frac{1}{3}+\left\lbrace 1-\sqrt{1-\rho}
    \right\rbrace^\frac{1}{3}\right],
    %\label{66}
    \end{eqnarray}
    where
    \begin{eqnarray}
    \rho~\equiv~\left(\frac{4}{27}\right)\left(\frac{P^3}{Q^2}\right).
    %\label{67}
    \end{eqnarray}
    Substitution of eqn.(63) in eqn.(61) and following eqn.(16) and eqn.(17), one obtains the following simple expression for the energy eigen-values in the LO: 
    \begin{eqnarray}
    E_n^{LO}~=~\left(\frac{3n^2\pi^2}{16}\right)u~+~\left(\frac{c_n}{12}\right)\left(\frac{1}{u}\right),
    %\label{68}
    \end{eqnarray}
    where, $u$ is given by eqn.(66). The remaining parameter, $`h'$ can then  be determined by substitution of eqn.(68) in eqn.(58), given by:
    \begin{eqnarray}
    h~=~\left(\frac{n^2\pi^2}{16}\right)u~+~\left(\frac{c_n}{12}\right)\left(\frac{1}{u}\right)
    %\label{69}
    \end{eqnarray}
   
   At this juncture, several observations are warranted: \\ (i) It is noteworthy that the free parameters of the input-Hamiltonian, i.e., a'$ and $h'$ acquire functional dependence on g'$ and $n'$ through eqn.(66) and eqn.(69). This, in turn, implies that the input-Hamiltonian, $H_0^{SW}$ (see, eqn.(52)) also becomes a function of $g$ and $n$. An evident consequence is that the eigenfunctions of $H_0^{SW}{(g,n)}$ corresponding to different eigenvalues become non-orthogonal as the latter now depends on the excitation-label {'n'}, i.e.
    \begin{eqnarray}
    (\phi_m(x),\phi_n(x))~\equiv~\int_{-a}^{+a}dx~{\phi_m^{\ast}(x)\phi_n(x)}\neq 0\nonumber\\~for ~m\neq~n,
    %\label{72}
    \end{eqnarray}
    
    It follows from eqn. (66) and eqn. (68) that the energy in the LO is a non-analytic function of $g$ at the origin. Consequently, the results expressed in these equations are \emph{not} amenable to ordinary perturbation theory as a power-series expansion in $g$. In this context, the LO results are \emph{non-perturbative}.\\ (iii) The cube-root singularity at $g=0$ and the branch-point structure in the $ complex~g-plane$, as presented in eqn. (68), align with the rigorous derivation [11] of the analytic structure of the energy of the quartic-AHO in the coupling strength-plane, utilizing advanced tools of complex analysis. \textit{It is noteworthy that the correct analytic structure in $g$ emerges here as a straightforward consequence of the CEQA and the CO of NGAS} (see, eqn. (16) and eqn. (17)). (iv) Furthermore, it is crucial to emphasize that the observed features noted above under (i)-(iii) also apply in the analogous case of the simple harmonic approximation [82] applied to the same system of the QAHO/QDWO as described here in previous \textbf{Sections}. \textit{This fact demonstrates that these features are the \emph{inherent} properties of the scheme, NGAS, being \emph{independent} of the choice of the input Hamiltonian.}  
    
  % \textbf{ Up to this point on 19˘-08-2025 (FN) }  
  
   In \textbf{Table-I}, we present the leading-order (LO) results for the energy levels of the quartic anharmonic oscillator (AHO) at selected values of the oscillator-level index $n_s$ and the quartic coupling $g$. It is important to note that the infinite square well (ISW) level index and the oscillator-level index differ by one unit, i.e., $n_s = n-1$, for $n = 1, 2, 3, 4, \ldots$. Additionally, this table includes the energy levels $E_n^{(2)}$, which incorporate second-order corrections obtained through the 'improved perturbation theory' (IPT) (refer to \textbf{Subsection-D} below), as well as previous results from reference [105], denoted as $Exact$ in column 5, for comparison. The entries in the last column and the third column represent the relative percentage error with respect to the $Exact$ results.
  
   The tabulation reveals that the LO results are consistently accurate within approximately  $\sim (2-12)~\%$ of the standard results across the entire range of $n_s$ and $g$. Furthermore, it is evident that the accuracy of the LO approximation is significantly enhanced by the inclusion of the second-order correction in IPT.

    \begin{table}{\textbf{Table-I}}
        
        \begin{tabular}{c c c c c c c}
            $n_s$ & $g$ & $E_n^{LO}$ &$Error(LO)$ & $E_n^{(2)}$ & $Exact$ & $Error$ \\ 
            &  &  & ${(\%)}$ &     & ref.(8) & ${(\%)}$\\
            \hline  
            0 & 0.1 & 0.6312 & 12.886 & 0.5748 & 0.5591  & 2.793\\ 
            & 1.0 & 0.9033 & 12.386 & 0.8290 & 0.8038 & 3.136 \\ 
            & 10.0 & 1.6902 & 12.311 & 1.5546 & 1.5049 & 3.296\\ 
            & 100.0 & 3.5168 & 12.310 & 3.2359 & 3.1314 & 3.341\\ 
            \hline 
            1 & 0.1 & 1.9636 & 10.972  & 1.8058 & 1.7695 & 2.055\\ 
            & 1.0 & 3.0366& 10.900 & 2.7999  & 2.7379 & 2.267\\ 
            & 10.0 & 5.9051 & 10.911 & 5.4479 & 5.3216 & 2.374\\ 
            & 100.0 & 12.4162 & 10.965 & 11.4561 & 11.1872 & 2.403\\ 
            \hline  
            2 & 0.1  &  3.3470 &  6.640 & 3.0943  & 3.1386 & 1.412 \\ 
            & 1.0 & 5.5818  & 7.771  & 5.1796  & 5.1793 & 0.006 \\ 
            & 10.0 & 11.9046  &  8.151& 10.3893 & 10.3471 & 0.408 \\ 
            & 100.0 & 23.7124 & 8.242& 22.0163 & 21.9069 & 0.499\\ 
            \hline  
            4 & 0.1& 6.3332 & 1.812 & 5.8162 & 6.2203 & 6.496 \\ 
            & 1.0 & 11.2748 &  2.634& 10.3198 & 10.9636 & 5.872 \\ 
            & 10.0 & 23.1124  &  2.838& 21.1642 & 22.4088 & 5.556\\ 
            & 100.0 & 49.2384 & 3.139 & 45.0952 & 47.7072 & 5.475 \\ 
            \hline  
            10 & 0.1& 16.8320 & 2.997  & 18.5203 & 17.3519 & 6.733\\
            & 1.0 & 32.1164 & 2.479 & 35.3936 & 32.9326 & 7.471\\ 
            & 10.0 & 67.1872 &  2.350 & 73.8641 & 68.8036 & 7.355 \\ 
            & 100.0 & 143.8104 & 2.323 & 157.9952 & 147.2267& 7.312\\ 	
        \end{tabular}\\
        \caption{Table for results for the QAHO in the ISW-approximation. For the explanation of entries under different columns, see text}
        
       \end{table}

The energy eigenfunctions in the LO are as specified by equations (56) and (57), with the width parameter $a$ in these equations now exhibiting functional dependence on $g$ and $n$ in accordance with equation (66).
  
   The LO approximation for the simple harmonic oscillator (SHO) will be discussed in the subsequent subsection.
    
    \subsection{ The simple  harmonic-oscillator(SHO) in the ISW-approximation}
     The exact analytical solutions for the quantum Simple Harmonic Oscillator (SHO) are well-documented and can be found in any standard quantum mechanics textbook. Consequently, there is no necessity for approximation of this renowned example. Nevertheless, these 'exact' results serve as a benchmark for evaluating the accuracy and effectiveness of the ISW approximation in NGAS. It is for this reason that the SHO is examined here as a specific instance ($g= 0$) of the AHO. The Hamiltonian of the SHO is derived from that of the AHO, as presented in equation (51), by setting the quartic coupling $g$ to zero, and is expressed as:  
      \begin{eqnarray}
    {H^{SHO}}~=~{\frac{1}{2}~p^2}~+~{\frac{1}{2}~x^2.}
    %\label{71} 
    \end{eqnarray}
   The substitution: $g=0$ in Eqs.(60-63) lead to the following corresponding results for the  SHO:
    \begin{eqnarray}
    {u=\sqrt{P},}
    %\label{72}
    \end{eqnarray}
    \begin{eqnarray}
    E_n^{LO}|_{SHO}~=~\left(\frac{n^2\pi^2}{8}\right)u~+~\left(\frac{c_n}{6}\right)\left(\frac{1}{u}\right).
    %\label{73}
    \end{eqnarray}
    The `depth' parameter for the case of the SHO is given by,
    \begin{equation}
    h_{SHO}~=~\left(\frac{c_n}{6}\right)\left(\frac{1}{u}\right).
    %\label{74}
    \end{equation}
    The LO-approximation for the energy-levels following from Eqs.(72-73) are tabulated in \textbf{Table-II} for typical values of the level-index, $~n_s$  along with the results after inclusion of the 2nd-order correction in IPT  (discussed in \textbf{Subsection-D} ). The accuracy of the approximation for both the cases with respect to the \textit{exact} analytic result are also shown in the same Table. The entries in the 3rd- and the last column represent estimation of errors for the  results in the LO and with inclusion of 2nd-order correction respectively, as compared to the `exact' results in the 5th-column.\\
    
    \begin{table}{\textbf{Table-II}}
        
        \begin{tabular}{c c c c c c}
            $n_s$ & $E_n^{LO}$ & $Error-LO{(\%)}$ & $E_n^{(2)}$ & $Exact$ & $Error{(\%)}$ \\     
            \hline
            0 & 0.5678 & 13.56 & 0.5091 & 0.5 & 1.808 \\ 
            \hline
            1 & 1.6703 & 11.35 & 1.5239 & 1.5 & 1.578 \\ 
            \hline
            2 & 2.6272 & 5.05& 2.3888 & 2.5  & 4.462  \\ 
            \hline  
            5 & 5.3952 & 1.90 & 5.6456 & 5.5 & 2.609  \\ 
            \hline  
            10 & 9.9508 & 5.23 & 10.3945 & 10.5 & 1.155 \\
            \hline
            15 &  14.493 & 6.49 & 15.8155 & 15.5 & 1.260 \\
            
        \end{tabular} 
        \caption{The LO-results for the SHO in the ISW-approximation at sample-values of $~n_s~$. Other entries in the Table are explained in the text}
    \end{table} 
  
  % \bigskip
   
   In this context, it may be interesting to obtain an $asymptotic$ estimate for the accuracy of the LO-result given by eqn.(73) as compared to the $`exact'$ result. This is given by
    \begin{equation}
    \lim_{n_s\to\infty}\left(\dfrac{E_n^{LO}}{E_n^{exact}}\right)=\left(\dfrac{\pi}{2\sqrt{3}}\right)\simeq~0.9069,
    \end{equation} 
    which corresponds to an error of approximately $9.31\%$ . At finite values of $n_s$ the errors are of the same order of magnitude.However, the inclusion of the 2nd-order correction in IPT significantly improves the accuracy of approximation.
    
    The next Subsection is devoted to the discussion of the ISW-approximation for the quartic-DWO.
    \subsection{ The QDWO}
    The Hamiltonian for the QDWO is given by (refer to, eqn.(36))
    \begin{equation}   
    {H^{DWO}}~=~{\frac{1}{2}~p^2}~-~{\frac{1}{2}~x^2}~+~{g{x^4};~~~~~ g > 0}
    \end{equation}
   
   Apart from the various applications of the QDWO as discussed earlier, there is considerable theoretical/mathematical interest in the system. In particular, the instability at $g=0$ ( due to the non-existence of a physical ground state) prevents the application of the na\"{\i}ve perturbation theory. In some versions of modified-perturbation theory it has been established [95] that power series-expansion in  $g$ is not even Borel-summable. However, in the context of NGAS, the ISW-approximation is routinely applicable to the DWO-case as well, by merely a change of sign of the $x^{2}$:   $x^{2}\rightarrow~-x^{2}$ in the corresponding formulae for the AHO. In particular, the equation analogous to eqn.(63) for the QDWO now becomes:
    \begin{equation}
    u^3~+~P\left(n\right)u-Q\left(g,n\right)~=~0,
    %Eqn.(77)
    \end{equation}
    leading to the $physical$ solution (analogous to eqn.(66)) given by
    \begin{eqnarray}
    u=\left\lbrace{\left(\frac{8g}{n^2 \pi^2}\right)\left(\frac{1}{5}-\frac{4 c_n}{n^2 \pi^2}\right)}\right\rbrace^\frac{1}{3}~~~~~~~~\nonumber\\~\left[\left\lbrace \sqrt{\rho+1}+1\right\rbrace^\frac{1}{3}-~\left\lbrace \sqrt{\rho+1}-1\right\rbrace^\frac{1}{3}~\right]~~,
    %\label{78}
    \end{eqnarray}
    The results for the energy in LO are therefore given by:
    \begin{equation}
    E_n^{LO}|_{DWO}~=~\left(\frac{3n^2\pi^2}{16}\right)u~-~\left(\frac{c_n}{12}\right)\left(\frac{1}{u}\right),
    %Eqn.(79)
    \end{equation}
     However, several authors [105] have found it convenient to measure the energy of the DWO from the $\emph{bottom}$ of the (symmetric) double-well, which means that a term equal to $1/{16 g}$ be added to formula, eqn.(79). We denote this quantity as:~$E_n^{\textrm{ref}}\equiv~E_n^{LO}+\left(1/{16 g}\right)$.
    
   In \textbf{Table-III}, we present the results  for  $E_n^{\textrm{ref}}$ along with results corrected to include 2nd-order perturbation effects in IPT ( see, \textbf{Subsection-D} below) for a range of values of $g$ and $ n_s $. The relative accuracy of these results with respect to those from earlier computation are also given in this Table. Also shown, are corresponding results from ref.[105] denoted as $Exact$ under column-6.The entries in the last column correspond to $({\%})error$ of results under column-5 with respect to those under column-6. The entries under column-3 represent relative accuracy of the LO-results as compared with the $Exact$ results shown under column-6.\\    
    \begin{table}{\textbf{Table-III}}
        
        \begin{tabular}{c c c c c c c}
            $n_s$&$g$& $E_n^{LO}$ & $Err(LO){(\%)}$& $E_n^{(2)}$& $Exact(ref.8)$& $Err{(\%)}$\\     
            \hline
            0 & 0.1 & 0.5049 & 7.220 &  0.4726 & 0.4709 & 0.350\\ 
            & 1.0 & 0.6422 &  11.242& 0.5967 & 0.5773 & 3.358\\ 
            & 10.0 & 1.5468 & 12.266 & 1.4246 &  1.3778 & 3.398\\ 
            & 100.0 & 3.4480 & 12.313 & 3.1732 & 3.0701 & 3.359\\ 
            \hline
            1 & 0.1 & 0.8252 & 7.476 &  0.7857  & 0.7678& 2.333\\ 
            & 1.0 & 2.3101 & 10.898 & 2.1366 &  2.0830 &  2.573\\ 
            & 10.0 & 5.5457 & 11.009 & 5.1180 & 4.9957 & 2.449 \\ 
            & 100.0 & 12.2471 & 10.993 & 11.3007 & 11.0337 & 2.419 \\ 
            \hline
            2 & 0.1  &  1.7393 & 6.392 & 1.6632 & 1.6348  & 1.740  \\ 
            & 1.0 & 4.6254 & 8.741 & 4.2983 & 4.2536  & 1.052   \\ 
            & 10.0 & 10.7242 &  8.378 & 9.9565 & 9.8947 & 0.625 \\ 
            & 100.0 & 23.4937 & 8.292& 21.8101 & 21.6947 & 0.532 \\ 
            \hline  
            4 & 0.1&  3.8678 & 4.994& 3.6471 & 3.6836 & 0.991   \\ 
            & 1.0 & 9.9139 & 3.659& 9.0975 & 9.5641 & 4.878  \\ 
            & 10.0 & 22.4582 &  3.322& 20.5659 & 21.7365 & 5.385 \\ 
            & 100.0 & 48.9325 & 3.248 & 44.7980 & 47.3929 & 5.475 \\ 
            \hline  
            10 & 0.1& 12.2697 & 1.292 & 12.840 & 12.4303 & 3.731 \\
            & 1.0 & 29.7738 & 2.139& 32.5902 & 30.4248 & 7.117  \\ 
            & 10.0 & 66.0787 & 2.327& 72.6644 & 67.6167 & 7.465 \\ 
            & 100.0 & 143.2937 & 2.361& 157.7026 & 146.6738 & 7.519 \\ 
            
        \end{tabular}
        \caption{Results for the Energy-levels of the QDWO in ISW-approximation, for sample values of $~n_s~$ and$~g~$. For explanation of other entries in the Table, see text.} 
    \end{table} 
     The tabulated data indicate that the ISW-approximation consistently replicates the standard results with a deviation of approximately $\sim{(2-10)~\%}$ for the leading order (LO) case. Furthermore, when incorporating the second-order correction in the improved perturbation theory (IPT), the deviation is reduced to approximately $\sim{(0.5-7.5)~\%}$. This aspect is examined in detail in the subsequent subsection.
    
    \subsection{ Method for higher order corrections}
   
   In the context of NGAS, the 'improved perturbation theory (IPT)' represents an advancement of the standard Rayleigh-Schrödinger perturbation series (RSPS), wherein the $unperturbed~Hamiltonian$ is selected as the input-Hamiltonian, $H_0$, with its free parameters determined through PEQA and CO (refer to eqs. (16-17)). The $perturbation, H^{'}$ is subsequently defined by eqn. (21). The following properties of the IPT emerge from this configuration, distinguishing it from conventional perturbation theory in several respects:(i) By employing equations (20) and (59), it becomes apparent that the perturbation correction remains consistently subdominant (by design) to the unperturbed component in the following average sense: 
    \begin{equation}
       |\langle~H_0\rangle|~~>>~|\langle~H^{'}\rangle|\equiv~~0.
       %Eqn.(80)
       \end{equation}
   
    This characteristic is in contrast to the analogous situation in the conventional Rayleigh-Schrödinger Perturbation Theory (RSPT). For example, consider the case of the Anharmonic Oscillator (AHO), where the 'perturbation' eventually dominates the 'unperturbed' component of the Hamiltonian, irrespective of the magnitude of the quartic coupling.   (ii) Secondly, the first-order perturbation correction identically vanishes due to equation (20).   (iii) TheiImproved perturbation theory (IPT) is not restricted to small values of the coupling strength \(g\), as equation (80) is valid for arbitrary values of the latter.   (iv) There is no small parameter inherently associated with the perturbation; however, it remains small in the average sense as defined by equation (80). Therefore, to track order-by-order corrections in IPT, one can utilize the standard technique of introducing an arbitrary, real but finite parameter, \(\eta\), through the substitution: \(H^{'}\rightarrow~{\eta}H^{'}\), setting this parameter to unity post-computation.   (v) Property (iv) further implies the "universality" of IPT's application to arbitrary interactions, as the perturbation, \(H^{'}\), does not involve any parameter of the original Hamiltonian in contrast to the  expansion parameter for the RSPT.
   
   Due to property (ii), the initial non-trivial correction for the \(n^{th}\) energy level begins at the second order and is represented by the standard expression: 
   \begin{equation}
       \Delta{E_n^{(2)}}~=~\sum_{m\neq~n}\dfrac{\langle~n|H^{'}|~m\rangle\langle~m|H^{'}|~n\rangle}{E_n^{(0)}-E_m^{(0)}}
       %Eqn.(81)
       \end{equation}
   
     In eqn.(81), the notation used is:  
 \begin{equation}
\langle~n|H^{'}|~m\rangle~\equiv~\int_{-a(n,g)}^{+a(n,g)}dx~{\phi_n^{\ast}(x)~ H^{'}(n,g)~\phi_m(x)}~~
%Eqn.(82)
\end{equation}

where \(H^{'}\) is defined by eqn.(21), and the \((n,g)\)-dependence of the relevant quantities is displayed. As an example, the perturbation part of the Hamiltonian for the quartic-AHO case is given by     
 \begin{equation}     
 {H^{'}|_{AHO}}~=~{\frac{1}{2}~x^2}~+~{g{x^4}}-{h(n,g)},  
 %(83)
 \end{equation}     
 where \(h(n,g)\) is provided by eqn.(69).

 It is noteworthy that the RSPT can be derived [106] \emph{without} invoking the properties of the eigenfunctions of the unperturbed Hamiltonian. Consequently, the non-orthogonality properties as expressed in eqn.(70) do not impact the standard formulae of the RSPT.

As previously discussed and noted in \textbf{Tables-(I-III)}, the inclusion of the second-order correction to the energy levels, defined as: \( E_n^{(2)}\equiv~E_n^{LO}~+~\Delta{E_n^{(2)}}\), 
 significantly enhances the accuracy of the approximation across all cases considered here, namely the AHO, DWO, and the SHO. (Computation of higher-order corrections to energy levels can be conducted using standard techniques in a straightforward manner as discussed in \textbf{Section-X}). 

 Finally, we summarize and discuss the main results of this \textbf{Section} 
 as follows.
 \subsection{Section-VIII -  Summary}
    
    In summary, we have presented a simple yet accurate approximation within the framework of the NGAS for the quartic anharmonic and double-well oscillators, utilizing the elementary system of the infinite-square-well (ISW) as the approximating Hamiltonian. In the computation of the energy spectrum, uniform accuracy is achieved within a few percent of the exact numerical results \emph{even} at the leading order, for \emph{arbitrary} values of the quartic coupling, \(g\), and level index, \(n_s\), for \emph{all} the aforementioned systems. This situation contrasts with the results obtained using textbook methods of approximation, such as the naive perturbation method, variational calculations, and the WKB method. Furthermore, the formalism \emph{naturally} reproduces the correct analytic structure of the energy in the complex-\(g\) plane, otherwise established through rigorous mathematical analysis. To systematically improve the leading-order results, an enhanced perturbation theory is formulated in NGAS, which is \emph{not} restricted to small-coupling expansion and is shown to yield further significant improvement in accuracy with the inclusion of the second-order correction only. The results and the method are directly relevant in a pedagogic context due to the extreme simplicity of the scheme (NGAS) as well as the input ISW approximation. The method can be extended straightforwardly to address cases of higher anharmonicity.

   %\textbf{(Up to this point on 21-08-25)}
   
   \section{ Application of the NGAS in the Leading Order to $g\phi^{4}$ QFT in the Massive Symmetric Phase}
   
  In light of the successful application of the NGAS to the AHO and DWO, as detailed in preceding sections, it is logical to extend this formalism to the $~g\phi^{4}~$ quantum field theory and examine the implications of this approach. It is pertinent to note that the $~g\phi^{4}~$ field theory, both in physical and lower dimensions, represents a significant physical system with critical applications across various domains of physics, such as the standard model of particle physics (Higgs mechanism), cosmology [87], condensed matter physics, phase transitions, and critical phenomena [70], among others. Furthermore, this theory serves as a fundamental theoretical framework for evaluating various approximation schemes in quantum field theory (QFT). Consequently, it is essential to assess the current approximation scheme, NGAS, by applying it to $g \phi^{4}$ QFT. We undertake this task in the subsequent subsections, utilizing the harmonic approximation for the input/approximating Hamiltonian (AH):
    \subsection{ The Harmonic approximation}
    The Lagrangian for the system is given by   \begin{equation}
    {\cal L }= \frac{1}{2}(\partial_{\mu}\phi)(\partial^{\mu}\phi) - \frac {1}{2} m^{2}\phi^{2} - g \phi^{4}~,
    %Eqn.(84)
    \end{equation}
    where $~m^{2}~ >~ 0~$. This Lagrangian therefore describes the massive, symmetric version of the theory. The Hamiltonian density derived from the above Lagrangian is given by
    \begin{equation}
    {\cal H}~ =~ \frac {1}{2}(m^{2}\phi^{2} + \phi _{t}^{2}+\phi _{\alpha}^{2}) +g\phi^{4}
     %Eqn.(85)
    \end{equation}
    where  we have defined:
    \begin{equation*}
    \phi _{t}\equiv ~\partial \phi
        (\vec{x},t)/\partial t~
         and ~\phi _{\alpha}\equiv
        \partial \phi(\vec{x},t)/\partial x^{\alpha}~.
    \end{equation*}  
    \subsection{ The approximating Hamiltonian}
   To develop the NGAS for the aforementioned theory, we select an approximating Hamiltonian (AH) denoted by $~{\cal H}_{0}~$ such that, ideally, the defining constraints as specified by eqs.(16-17) are satisfied. However, unlike the scenario in quantum mechanics, the PEQA (eqn.(16)) is challenging to implement in QFT as it involves multi-particle states in evaluating the quantum averages (QA). Therefore, for simplicity, we relax the condition, eqn.(16), by restricting the QA to be evaluated in the "few-particle" states only. Specifically, we proceed as follows. As previously, the "Approximating Hamiltonian" (AH) $~{\cal H}_{0}~$ is represented by the following expression: 
   \begin{equation}
    {\cal H}_{0}~ = ~\frac{1}{2}M^{2}\xi^{2} + \frac{1}{2}\xi_{\alpha}^{2} +\frac{1}{2}\xi_{t}^{2} + h_{0},
     %Eqn.(86)
    \end{equation}
    where,  
    \begin{equation}
    \xi(\vec{x},t) \equiv \phi(\vec{x},t) - \sigma~,
    %Eqn.(87)
    \end{equation}
    In addition, $~\xi_{t} \equiv \partial\xi/\partial t~ ;~ \xi_{\alpha} \equiv \partial\xi/\partial x^{\alpha}$ etc.
    
    Eqn.(86) is immediately recognized as the Hamiltonian density of the hermitian scalar field $~ \xi(\vec{x},t)~$. The 'free parameters' introduced in the AH are: $M$, $\sigma$, and $h_0$, which are to be determined according to the prescribed procedure by applying eqn.(16-17). For this purpose, it is first necessary to obtain the spectrum of $~H_{0}~$, which is accomplished as follows. The diagonalization of the AH given by eqn.(86) is straightforward using the Fourier expansion in terms of creation and annihilation operators:
    \begin{equation}
    \xi(\vec{x},t)~ =~ \phi (\vec{x},t) - \sigma~ =~ \int \frac {d^{3}\vec{k}}{\Omega_{k}(M)}[b(\vec{k})e^{-ikx} + b^{\dagger}(\vec{k}) e^{ikx}]~,
    %Eqn.(88)
    \end{equation}
    where
    \begin{equation}
    \Omega _{k}(M)~ \equiv~ 2 (2\pi)^{3}\sqrt{|\underline{k}^{2}| + M^{2}}~\equiv~ 2 ~(2\pi)^{3}~\omega_{k}(M)~,
    %Eqn.(89)
    \end{equation}
    and  $~kx \equiv k^{0}t-\vec{k}.\vec{x}$, as usual.
    The operators$~b(\vec{k}),b^{\dagger}(\vec{k})$ satisfy the standard equal-time commutation relations (ETCR)~:
    \begin{equation}
    [b(\vec{k}),b^{\dagger}(\vec{q})]~ =~ \Omega _{k}(M)\delta ^{3}(\vec{k}-\vec{q}),
    %Eqn.(90)
    \end{equation}
    which is a consequence of the ETCR between the `field' $~\phi(\vec{x},t)~$ and its canonical conjugate momentum~: $~\pi(\vec {x},t)~ \equiv~ \partial{\cal L}/\partial\dot {\phi}~$, given by : 
    \begin{equation}
    [~\phi(\vec{x},t),~ \pi(\vec {y},t)~]~ =~i~\delta^{3}~(\vec{x} -\vec{y}).
     %Eqn.(91)
    \end{equation}
    The energy of the system described by $~H_{0}~$ is obtained by standard methods and given by : 
    \begin{equation}
    H_{0}\equiv\int d^{3}\vec{x}{\cal H}_{0}(\vec{x},t)=\frac{1}{2}\int \frac {d^{3}\vec{k}}{\Omega_{k}(M)}[b(\vec{k}) b^{\dagger}(\vec{k}) + b^{\dagger}(\vec{k})b(\vec{k})]
    +\int d^{3}\vec{x}~h_{0}
     %Eqn.(92)
    \end{equation}
    The  spectrum of the states  are analogously obtained and denoted by :$~|vac>,|\vec{p}>,~ |\vec{p_{1}},~\vec{p_{2}}>,...$ etc where the effective vacuum state $|vac ~>$ is defined by 
    \begin{equation}
    b(\vec{k})~|vac~>~ =~ 0~,
     %Eqn.(93)
    \end{equation}
    and the multi particle-states are generated  by multiple application of the creation-operator $~b^{\dagger}(\vec{p})~$ on $~|~vac~>~$:
    \begin{equation}
    b^{\dagger}(\vec{p})~|~vac~>~ =~ |~\vec{p}~>
     %Eqn.(94)
    \end{equation}
    \begin{equation}
    \frac {b^{\dagger}(\vec{p_{1}})~b^{\dagger}(\vec{p_{2}})}{\sqrt{2!}}~ =~ |~\vec{p_{1}},~\vec{p_{2}}~>,~etc.
     %Eqn.(95)
    \end{equation}

   It may be noted that,\\~~~ $H_{0}|vac>~ =~E_{0}~|~vac~>~; H_{0}|\vec{p}>~=~E_{1}(\vec{p})|\vec{p}>,....~$ etc, where $~E_{0},~E_{1}~$etc correspond to the energy of the corresponding states.

    The implementation of PEQA require the evaluation of the QA of monomials of the field $~\phi(\vec{x},t)~$ such as~: $<vac|~\phi^{n}(\vec{x},t)~|vac>~\equiv~<\phi^{n}(\vec{x},t)~>,$ $<\vec{p}|~\phi^{n}(\vec{x},t)~|\vec{p}>~$etc. We first turn to the  evaluation of $~<\phi^{n}(\vec{x},t)>~$. This is readily done using $~{translational-invariance}~$ of the vacuum-state~:$~|vac>~$ and the ETCR,~ as given by eqn.(4.7). Some  useful  results thus obtained are  given  below~:  
    \begin{equation}
    <\phi(\vec{x},t)>~=~\sigma~,
    %Eqn.(96) 
    \end{equation}
        \begin{equation}
    <\phi^{2}(\vec{x},t)> ~=~\sigma^{2}~+~ \int \frac {d^{3}\vec{k}}{\Omega_{k}(M)} 
    ~\equiv~ \sigma^{2}~+~I_{0}~,
    %Eqn.(97)
    \end{equation}
    \begin{equation}
    <\phi^{4}(\vec{x},t)> ~=~\sigma^{4}~+~6\sigma^{2}I_{0}~+~3I_{0}^{2}~,~etc.
    %Eqn.(98)
    \end{equation}
    Similarly,
    \begin{equation}
    <\phi_{\alpha}^{2}(\vec{x},t)> ~=~ \int \frac {d^{3}\vec{k}}{\Omega_{k}(M)}|\vec{k}|^{2},
    %Eqn.(99) 
    \end{equation}
    \begin{equation}
    <\phi_{t}^{2}(\vec{x},t)> ~=~ \int \frac {d^{3}\vec{k}}{\Omega_{k}(M)}~\omega_{k}^{2}(M)
    %Eqn.(100) 
    \end{equation}
    Using the above results and the PEQA, one obtains the following equation :
    \begin{equation}
    M^{2}(g,\sigma) \equiv  m^{2} + 12g\sigma^{2} + 12g I_{0}(M^{2})~,
    %Eqn.(101)
    \end{equation}
    i.e.,
    \begin{equation}
    M^{2}(g,\sigma) =  m^{2} + 12g\sigma^{2} + 6g \int\frac {d^{3}(\vec{k})}{(2\pi)	^{3}\sqrt{|\vec{k}|^{2}+M^{2}}}~,
    %Eqn.(102)
    \end{equation}
    
    Eqn. (102) can be interpreted as the generation of the 'mass-gap' (i.e., shift in the bare mass) due to interaction. In analogy with the terminology used earlier, we refer to Eqs. (101-102) as the "gap equation (GE)" of the theory. This equation plays a crucial role in the subsequent discussions. The physical implications of the theory in the leading order (LO), such as the spectrum, renormalization, stability properties, and the structure of the effective vacuum, are discussed in the following subsections.
               
        \subsection{The Effective Potential}
    The effective-potential (EP) can be  defined as :
    \begin{equation}
    U_{0}(\sigma)~\equiv~<vac|~{\cal H}_{0}~|vac>~, 
    \end{equation}
    %Eqn.(103) 
    such that 
    \begin{equation}
    \sigma~\equiv~<vac|~\phi(\vec{x},t)~|vac>.
    %Eqn.(104)
    \end{equation}
    It may be noted that the l.h.s. of eqn. (103) is $~\it{defined}~$ to be a function of $~\sigma~$ alone. This means that any other parameter occuring in $~{\cal H}_{0}~$ is to be fixed by variational-minimization of $~<{\cal H}_{0}>$. The procedure is made explicit below by working out the current example, with $~{\cal H}_{0}~$ defined in eqn.(86). To this end, we first calculate $<~{\cal H}_{0}~>~$ which  guarantees the following equation: 
    \begin{eqnarray}
    <~{\cal H}_{0}~>~\equiv~ <~{\cal H}~>
    ~=~\frac{1}{2}m^{2}<\phi^{2}>~+~\frac{1}{2}<\phi_{t}^{2}> 
        ~+~\frac{1}{2}<\phi_{\alpha}^{2}>~+~g<~\phi^{4}~>,
        %Eqn.(105)
    \end{eqnarray} 
    which works out to be : 
   \begin{eqnarray}
    <~{\cal H}_{0}~>~=~ \frac{1}{2}m^{2}(\sigma^{2}~+~I_{0})~+~ \frac{1}{2} \int\frac{d^{3}(\vec{k})}{\Omega_{k}(M)}~(~\omega^{2}_{k}(M^{2})~+~|\vec{k}|^{2}~)\nonumber\\
    +~ g(\sigma^{4}+6\sigma^{2}I_{0}+3I_{0}^{2}).~  
    %\label{106}        
    \end{eqnarray} 
    This can be rewritten as, ( using
    $~\omega^{2}_{k}(M^{2})~\equiv~|\vec{k}|^{2}~+~M^{2}$)
    \begin{eqnarray}
    <~{\cal H}_{0}~>~=~ I_{1}~-~\frac{1}{2}I_{0}~(~m^{2}~+~12g\sigma^{2}~+~
    12g I_{0}~)~~~~
    \nonumber\\
    +\frac{1}{2}m^{2}~~(~\sigma^{2}~+~I_{0}~) 
    ~+~~~~~\nonumber\\ g~(~\sigma^{4}~+~6\sigma^{2}I_{0}~+~3I_{0}^{2}~),~~~~~~~
    %\label{107}         
    \end{eqnarray} 
    where, we have defined~:
    \begin{eqnarray}
    I_{n}(x)~ \equiv~ \int~\frac {d^{3}\vec{k}}{\Omega_{k}(x)}~[~\omega_{k}^{2}(x)~]^{n},~~~~~\nonumber\\~~~~~~~~~~~~~~ n~ = ~0,~\pm ~1,~ \pm ~2,....
    %\label{108}
    \end{eqnarray}
    These integrals were first introduced  by Stevenson [18]. Eqn.(107) can be  simplified further  by using the``gap-equation'' as given by eqn.(101-102). One then obtains :
    \begin{equation}
    <~{\cal H}_{0}~>= I_{1}(M)~-~3g I_{0}^{2}(M)~+~\frac{1}{2}~m^{2}\sigma^{2}~+~g ~\sigma^{4}~~~
    %\label{109}
    \end{equation}   
    We thus derive the LO-effective potential of NGAS as given by 
    \begin{equation}
    U_{0}(\sigma)=\frac{1}{2}~m^{2}\sigma^{2}~+~g ~\sigma^{4}~+~I_{1}(M)~-~3 g I_{0}^{2}(M),~~~
    %\label{110}
    \end{equation}   
    where, {\it it is implicitly}  understood that  the ``gap-equation'', eqn.(101-102)  is  to  be {\it first}   solved to obtain $~ M^{2}~${\it as a  function of `$\sigma$'}.
    \subsection{Equivalence to the Gaussian-Effective Potential}
    In view of the results derived in the previous subsection , the following remarks/ observations are in order:
    (a)~~The results contained in eqs. (101-102 ) for the gap equation and the integrals defined by eqs. (108) were first derived by Stevenson [18] in the context of the``Gaussian effective potential (GEP)'' for the symmetric $~g\phi^{4}~$ theory and obtained by variational calculation using a Gaussian-trial wave-function. The reproduction of the results of the GEP in ref.[18] in the LO of NGAS demonstrates that  the \textit{GEP is contained in the NGAS as the leading order approximation.}
    (b)~~\textit{In view of the established equivalence  with the GEP, {\it all} the results obtained in the former approximation, are reproduced in the LO of the NGAS. In particular, the demonstration of {\it non-triviality} of the symmetric $~g\phi^{4}~$ theory in the GEP,~ being entirely based  upon the consequences of renormalization  is also reproduced in the LO of NGAS.}
     (c)  It must be emphasized, however, that the current scheme, NGAS is based upon { \it entirely~ different~ starting ~assumptions ~} and~ is much {\it more~ general}  than the GEP, which is obtained solely due to the choice of the  AH as given in eqn.(4.3) and that too, in the leading order.
   
     We elaborate in the following sub-section, some of the properties of the theory  concerning the stability  and non-triviality of the theory.  
   \subsection{Renormalization and realization of the 'non-triviality' of the theory in the leading order of NGAS}
    One can  carry out the renormalization program (in the LO) by noting that [18] :
    
    (i) the vacuum-state-configuration corresponds to the absolute (global) minimum of $~U_{0}(\sigma)~$, i.e. by solving :
    \begin{equation}
    \frac {dU_{0}}{d\sigma}|_{\sigma_{0}}~=~0~;~\frac{d^{2}U_{0}}
    {d\sigma^{2}}|_{\sigma_{0}}~>~0
    \end{equation}
    %(111)     
    (ii)~the $~\it{renormalized}~$ mass in LO is given by 
    \begin{equation}
    m_{R}^{2}\equiv\frac{d^{2}U_{0}}{d\sigma^{2}}|_{\sigma~=~\sigma_{0}}~,~~~~
    %(112)
    \end{equation}     
    (iii) ~the LO-renormalized coupling-strength is likewise defined to be: 
    \begin{equation}
    g_{R}~\equiv~\frac{1}{4!}\frac{d^{4}U_{0}}{d\sigma^{4}}|_{\sigma~=~\sigma_{0}}~,
     %(113)
    \end{equation}     
    where $~\sigma_{0}~$ corresponds to the vacuum-configuration as defined by eqn.(111). It is directly verified by minimization of $~U_{0}(\sigma)~$ ( see, eqn.(111)), that the {\it global} minimum  of the former occurs at $~\sigma_{0}~=~0~$, which is consistent with $\sigma = 0.$ as it should be since the symmetric case  is being considered. Next, evaluating eqn.(112) at $~\sigma_{0}~=~0~$ one gets the renormalized mass :
    \begin{equation}
    m_{R}^{2}~=~m^{2}~+~12g I_{0}(m_{R})~\equiv~M^{2}(~g,~
    \sigma^{2}~=~0),
    %Eqn.(114)
    \end{equation}
    Similarly,~ after a straight forward calculation, one obtains [18] the renormalized coupling as given by: 
    \begin{equation}
    g _{R}~ = ~g [\frac{1-12g I_{-1}(m_{R})}{1 + 6g I_{-1}	(m_{R})}],
    %Eqn.(115)
    \end{equation}  
    where again $~I_{-1}(x)~$ is defined as per the general definition given in eqn.(108).
    
     \textit{An important consequence of eqn.(114) is that $g$ must be chosen negative: $g < 0$, in order that the theory may be renormalized}. This is because   $I_{0}(m_{R})$ is divergent and the (unmeasurable) bare mass $`m'$ of the theory may be infinite but the physical (renormalized) mass $m_{R}$ has to be finite. This version of the theory has therefore, been designated in ref.[18] as  {\it  precarious} $~g\phi^{4}$ {\it  theory}.
    
     We next discuss some of the consequences  of the above \textit{non-perturbative} renormalization  scheme obtained in the LO of NGAS  leading to the $\it {stability}$ and $\it{non-triviality}$ of the theory. 
    \subsection{Emergence of the ``Non-Triviality" of the theory}
    In order to investigate the stability and non-triviality of the theory it is convenient to start with  eqs.(101-102), which involve the {\it divergent} integral due to the momentum-integration and which,  therefore, need a suitable  method  of subtraction.
    
    Using the {\it subtraction-procedure~ devised ~by ~Stevenson} [18], eqn.(115) can be inverted to express $~g~$  in terms of the {\it  observable} parameters, $~g_{R}~$  and $~m_{R}~$. This leads to {\it two} solutions for $~g~$ of which, the physical one is given by :  
    \begin{equation}
    g = [-1/6I_{-1}(m_{R})][1+1/2[g_{R}I_{-1}(m_{R}) + ....]
     %Eqn.(116)
    \end{equation}
    ( The {\it other} solution is $~g~ = ~-(1/2)g_{R}~ +~ O(1/I_{-1}(m_{R})~$. This solution can be shown to lead to  {\it instability}, since the minimum of the EP corresponding to this solution can be shown to lie (infinitely) higher than the minimum corresponding to eqn.(116)).
    
    It may be noted  at this point that eqn.(116) implies a viable, stable $~g \phi^{4}~$ theory~ when ~the ~(unobservable)~ bare-coupling~ becomes {\it negative~ but~ infinitesimal}.  
    
    Substituting for $~g~$ as given by eqn.(116), one can solve for the bare-mass, by inverting  eqn.(114), after carrying out the  subtraction as per the Stevenson-prescription. This leads to the following expression for the bare-mass : 
    \begin{eqnarray}
    m^{2} = {m_{R}}^{2} + 2I_{0}(m_{R})/I_{-1}(m_{R})\nonumber\\+\mbox{(sub-leading~ terms)}
    %(117)
    \end{eqnarray}
    With the aid of eqs.(116) and (117),~ the effective potential, as given by eqn.(110) can be recast in manifestly  renormalized form involving the $~\it{observable}~$ parameters: $~g_{R}~$ and $~m_{R}~$ only. The resulting expression is given by :   
    \begin{eqnarray}
    U_{0}(\sigma)~ =~ U_{min} + \frac {1}{4}t~{m_{R}}^{2}\sigma ^{2} - ({m_{R}}^{4}/128\pi
    ^{2})(t-1)^{2}\nonumber\\-({m_{R}}^{4}/64\pi^{2})(t-1)\eta,~~~~~~~
    %(118)
    \end{eqnarray}
    where 
    \begin{equation}
    t~=~M^{2}(\sigma)/{m_{R}}^{2}~;~~ \eta \equiv -4\pi^{2}/g_{R}~,
    %(119)
    \end{equation}
    and
    \begin{equation}
    U_{min} = I_{1}(m_{R}) - 3g I_{0}^{2}(m_{R}).
    \end{equation}
    Similarly, the renormalized  version of the ``gap-equation'' is given by :
    %(120)
    \begin{equation}
    (1-\eta)(t-1) - (16\pi^{2}/{m_{R}}^{2})\sigma^{2}~ =~ t~ln~t.
    %(121)
    \end{equation}
    It must be pointed out that, one has to $~\it{first}~$ solve the gap-equation, eqn.(121) to obtain  $ ~t~\equiv~t(\sigma)~$, which is then to be substituted  in eqn.(118) to infer the $~\sigma$-dependence of $~U_{0}(\sigma)~$.
    
    It may be noted that the gap-equation, eqn.(121) is a transcendental equation and its solution exists only when,
    \begin{equation}
    \sigma ^{2}~ \leq ~\sigma_{min}^{2}~\equiv~({m_{R}}^{2}/16\pi^{2})[e^{-\eta}~ +~\eta - 1]
     %(122) 
    \end{equation}
    
     The domain of validity for the effective potential (EP) is constrained by the range of $~\sigma^{2}~$ as determined by equation (122) for any given value of $\eta$. Specifically, in the regime of large coupling ($~\eta ~\rightarrow~0~$), the domain of the EP diminishes with decreasing $~\eta~$, as $~\sigma^{2} {min}~\rightarrow 0$ in this limit. This scenario corresponds to small oscillations around $~\sigma~\simeq~0~$, which is indicative of a pathological situation when $~|g{R}|~\rightarrow~\infty$. Conversely, in the small coupling regime ($~g_{R}~\rightarrow~0$), which corresponds to $~|\eta|>>~1$, the domain of the EP expands with increasing $~\eta~$. Thus, it can be summarized that the leading-order effective potential (LO-EP) of the symmetric $~g \phi ^{4}~$ theory is reasonable and well-behaved unless the renormalized coupling is excessively large. Consequently, one can conclude that a non-trivial and stable theory emerges in the leading order of NGAS provided the $~\it{physical}~$ coupling $~g_{R}~$ is $~\it{not}~$ unusually large. 
    
    Further commentary on the stability of the perturbative vacuum of the theory is as follows. To examine the stability issue, it is necessary to compute the effective potential based on the $~\it{perturbative}~$ vacuum (i.e., the vacuum of the free-field theory). This is achieved by setting $~M~\rightarrow~m~$. Thus, starting from equation (106) and letting $~M~\rightarrow~m~$, one obtains, after simplification, the following expression: 
        \begin{eqnarray}
    <{\cal H}_{0}>_{P}~ =~ \frac {1}{2}m^{2}\sigma^{2}~ +~ g \sigma ^{4}~+ ~\bar{I}_{1}~ +~\nonumber\\6g\sigma ^{2}\bar{I}_{0}~ +~ 3g {\bar{I}_{0}}^{2}
    %(123)
    \end{eqnarray}
    
   %\textbf{ (Up to this point on 21-08-25 FN )}
    
    In the above,$~<{\cal H}_{0}>_{P}~$  denotes $~<0|{\cal H}_{0}|0>~$ i.e. VEV of $~{\cal H}_{0}~$ in the $~\it{perturbative}~$ vacuum state ; and $~\bar{I}_{n}~\equiv~I_{n}(m^{2})$. By definition, the effective-potential based upon the perturbative vacuum 
    denoted by $~U^{P}(\sigma)~$ is identified with  $~<{\cal H}_{0}>_{P}~$. Hence one obtains :
    \begin{eqnarray}
    U^{P}(\sigma)~ =~ \frac {1}{2}m^{2}\sigma^{2}~ +~ g \sigma ^{4}~+ ~\nonumber\\~\bar{I}_{1}~ +~ 6g\sigma ^{2}\bar{I}_{0}~ +~ 3g {\bar{I}_{0}}^{2}
     %(124)
    \end{eqnarray}
    The renormalized-parameters following from the eqn.(124) are \\like-wise computed and denoted by
    \begin{equation}
    \bar{m}_{R}^{2}~ =~ d^{2}U^{P}/d\sigma^{2}|_{\sigma^{2} ~=~ 0}~;
    %(125)
    \end{equation}
    \begin{equation}
    \bar{g}_{R}~ =~ (1/4!)d^{4}U^{P}/d\sigma^{4}|_{\sigma^{2}~=~0}
     %(126)
    \end{equation}
    where $~\sigma^{2}~=~0~$ is again the location of the global-minimum of $~U^{P}(\sigma)~$, as can be readily verified. In this context, it must be emphasized  that the  integrals : $\bar{I}_{n}$ occurring in 
    eqn.(124) are $independent$ of $\sigma$. 
    
    ~~~Next, computing the derivatives of $~U^{P}(\sigma)~$ at the minimum, one obtains the following expressions for the renormalized  parameters based upon the perturbative vacuum: 
    \begin{equation}
    \bar{m}_{R}^{2}~ =~ m^{2}~ +~ 12g \bar{I}_{0}
     %(127)
    \end{equation}
    \begin{equation}
    \bar{g}_{R}~ = ~g
    %(128)
    \end{equation}
    
    The requirement for the finiteness of $~\bar{m}_{R}~$ and $~\bar{g}_{R}~$ necessitates that $g$ must be negative (refer to equation (127)), for otherwise $~\bar{m}_{R}^{2}~$ would be infinitely large since $~\bar{I}_{0}~$ is divergent and the bare (unobservable) mass, $m^{2}>0$. However, this would lead to instability since the effective potential $~U^{P}(\sigma)~$ will not have a lower bound. This is made manifest by explicitly writing the EP in terms of the renormalized parameters: 
    
    \begin{equation}
    U^{P}(\sigma) = \frac {1}{2}\bar{m}_{R}^{2}\sigma^{2} +\bar{g }_{R} \sigma ^{4} + 3\bar{g} _{R} {\bar{I}_{0}}^{2} + \bar{I}_{1}
     %(129)
    \end{equation}
    To prevent instability of the theory when renormalised about the perturbative vacuum it, therefore, becomes inescapable that 
    \begin{equation}
    ~\bar{g} _{R}~ =~ g~ =~0~,
     %(130)
    \end{equation}
    which is nothing but the $~\it{triviality~ scenario}$~!
    
A few remarks/observations are in order, in view of the above results: 
    
    (i)   We believe that the result in eqn.(130) constitutes perhaps, the most direct demonstration of triviality of symmetric $~g \phi^{4}~$ 
    theory in physical dimensions.  
    
    ( ii ) At the same time, the result, eqn.(130) also demonstrates that the conclusion of {\it triviality of the  theory is an artifact of  the	$na \ddot{i}ve$   perturbation theory built and renormalized around the free-field vacuum }. As demonstrated earlier, the theory renormalized about the NGAS-vacuum leads to a perfectly acceptable, stable and non-trivial $~g \phi^{4}$- theory ( see, eqs.(118-122) and discussions following ).     
    
    ( iii ) It may be further pointed out that the ground state of the trivial theory is still unstable as compared to that in the LO of NGAS, i.e.
    \begin{equation}
    U_{min}~<<~U_{min}^{P},
     %(131)
    \end{equation}
    which is readily established by referring to eqn.(129) 
    (with $~\sigma~=~0~$) and eqn.(120).
    
    This completes our results and discussions regarding the stability and the triviality of $~g \phi^{4}$- theory in the context of NGAS in the LO.
    
    In the next subsection, we discuss the structure of the interacting vacuum .
    
    \subsection{The Underlying Bogoliubov-Valatin Transformation}
    The actual/physical vacuum state in presence of interaction is approximated in the LO of NGAS by the state: $|~vac>$ which is the lowest energy state of $~H_{0}~$. The structure and properties of this state can be inferred from studying  the quantum-canonical transformation (i.e. Bogoliubov-Valatin transformation, ref.[46]) connecting the interacting vacuum state (IVS) with the free field vacuum ( FFV ) state.  
    
    For this purpose it is convenient to start from the Fourier-decomposition of the field $~\phi(\vec{x},t)~$ in terms of the $~\it{free~ field}~$ creation-and annihilation operators analogous to eqn.(88):
    \begin{eqnarray}
    \phi (\vec{x},t) = \sigma+ \int \frac {d^{3}\vec{k}}{\Omega_{k}(m)}[a
    (\vec{k})e^{-ikx} + a^{\dagger}(\vec{k}) e^{ikx}],~~~~~~~
     %(132)
    \end{eqnarray}
    where, now
    \begin{eqnarray}
    \Omega _{k}(m)~ \equiv~ 2 ~(2\pi)^{3}~\omega_{k}(m)~;\\~
    {k}^{0}~=~\omega_{k}(m)~\equiv~ \sqrt{|\vec{k}|^{2} ~+~m^{2}},
    %(133,134)
    \end{eqnarray}
    corresponding to the propagation of the free-field quanta satisfying the mass-shell condition : $~{k^{0}}^{2}~-~|\vec{k}|^{2}~=~m^{2}~$. The free-field  operators satisfy the standard commutation relations :
    \begin{equation}
    [a(\vec{k}),~a^{\dagger}(\vec{q})]~ =~ \Omega _{k}(m)~\delta ^{3}(\vec{k}-\vec{q}).
    %(135)
    \end{equation}
    Comparision of eqs.(90) and (135) implies that the modified operators :
    \begin{eqnarray}
    B(\vec{k}) ~\equiv~\frac{b(\vec{k})}{\sqrt{\Omega_{k}(M)}}
    \\
    A(\vec{k}) ~\equiv~\frac{a(\vec{k})}{\sqrt{\Omega_{k}(m)}}
    %(136,137)
    \end{eqnarray}
    satisfy identical commutation relations :
    \begin{equation}
    [B(\vec{k}),B^{\dagger}(\vec{q})]~ =~ \delta ^{3}(\vec{k}-\vec{q})~=~
    [A(\vec{k}),~A^{\dagger}(\vec{q}))].
    %(138)
    \end{equation}
    It  follows,  therefore, that the two sets  of operators must be connected through   Boguliobov-transformation ref.[46]  given by :
    \begin{subequations}
        \begin{align}
    B(\vec{k})~~ =~ cosh(\alpha_{ k})A(\vec{k}) ~-~ sinh(\alpha_{k})A^{\dagger}(-\vec{k})
    \\
    B^{\dagger}(\vec{k}) =~ cosh(\alpha_{k} )A^{\dagger}(\vec{k})~ - ~sinh(\alpha_{k})A(-\vec{k}) ,
    \end{align}
    %(139a-b)
    \end{subequations}

    whereas the inverse transformation is given by :
    \begin{subequations}
    \begin{align}    
    A(\vec{k})~ ~=~~ cosh(\alpha_{ k})B(\vec{k}) ~+~ sinh(\alpha_{k}) B^{\dagger}(-\vec{k})
    \\
    A^{\dagger}(\vec{k})~ =~ cosh(\alpha_{k} )B^{\dagger}(\vec{k})~ + ~sinh(\alpha_{k})B(-\vec{k}).
    \end{align}
    %(140 a-b) 
    \end{subequations}   
    In the above, $~\alpha_{k}~=~f(|\vec{k}|)~$ , is apriori an arbitrary real function of $~|\vec{k}|~$, i.e.
    \begin{equation}
    \alpha_{-\vec{k}}~=~\alpha_{\vec{k}}~=~\alpha_{\vec{k}}^{*} .
    %(141)
    \end{equation}
    However, eqs.( 90 ), (132) and (139) considered together
    further imply that 
    \begin{equation}
    exp~( 2\alpha_{k})
    ~=~\frac{\omega_{k}(M)}{\omega_{k}(m)}~=~\frac{\sqrt{|\underline{k}|^{2} + M^{2}}}{\sqrt{|\underline{k}|^{2} + m^{2}}}
    %(142)
    \end{equation}
    To show this, consider eqs.(90) and (132) at $~t~ =~ 0~$,
    which can be written as:
    \begin{subequations}
    \begin{align}    
    \phi (\vec{x},t)=\sigma+\int \frac {d^{3}\vec{k}}{\sqrt{\Omega_{k}(m)}}[~A(\vec{k})+ A^{\dagger}(-\vec{k})] e^{i\vec{k}.\vec{x}}
    \\
    =\sigma + \int \frac {d^{3}\vec{k}}{\sqrt{\Omega_{k}(M)}}[B(\vec{k})+ B^{\dagger}(-\vec{k})] e^{i\vec{k}.\vec{x}},
    \end{align}
    %(143a-b)
    \end{subequations}
    which implies  that :
    \begin{equation}
    \{B(\vec{k})+ B^{\dagger}(-\vec{k})\}=\sqrt{\frac{\Omega_{k}(M)}{\Omega_{k}(m)}}\{A(\vec{k})+ A^{\dagger}(-\vec{k})\}
    %(144)
    \end{equation}
    However, from eqn.(139) it follows that 
    \begin{equation}
    \{B(\vec{k})~ +~ B^{\dagger}(-\vec{k})\}~=~exp~(\alpha_{k})\{~A(\vec{k})~ +~ A^{\dagger}(-\vec{k})~\}
    %(145)
    \end{equation} 
    thus  leading to the desired  result, eqn.(142).
    
    The significant physical results  that follow from the above equations are discussed in the following subsection.
    \subsection{Structure and Stability of the Effective Vacuum}
    To obtain  the information regarding the  particle-content and other features of the IVS it is instructive to first compute the number-density of the free-field-quanta  residing in the IVS. To this end let us note that the free-field-number operator is given by the standard expression:
    \begin{eqnarray}
    N~\equiv~\int {\frac  {d^{3}\vec{k}}{\Omega_{k}(m)}a^{\dagger}(\vec{k})a(\vec
        {k})}
    \nonumber\\
    ~=~\int { d^{3}\vec{k}}~A^{\dagger}(\vec{k})~A(\vec{k})
    \end{eqnarray}  
    Hence the desired number density of the free-field quanta in the IVS defined as:
    \begin{equation}
    n(\vec{k})~\equiv ~<~vac~|~\frac {A^{\dagger}(\vec{k})A(\vec{k})}{v}~|~vac > ~,
    %(147)
    \end{equation}  
    where $~v ~\equiv~$ spatial-volume of quantization $~\equiv~\int d^{3}\vec{x}$ can be easily evaluated, using eqs.(140) and given by
    \begin{equation}
    n(\vec{k})~=~\frac {sinh^{2}(\alpha^{vac}(\vec{k}))}{(2\pi)^{3}}
    %(148)
    \end{equation} 
    where \textit{$~\alpha^{vac}(\vec{k})~$ is given by eqn.(142) evaluated for $~M~\equiv~ m_{R}~$=~ free-particle-mass renormalized~ about the IVS, $~| vac >$} ~(It may be recalled that $~M(\sigma~=~0)~=~m_{R}~$ and $~\sigma~=~0~$ define the IVS ). This leads finally to the
    expression :
    \begin{equation}
    n(\vec{k})~=~(\frac{1}{32\pi^{3}}) ~[~
    \frac{\omega_{k}(m)}{\omega_{k}(m_{R})}~+~\frac{\omega_{k}(m_{R})}{\omega_{k}(m)}~-~2~]~.
    %(149)
    \end{equation}    
    To extract further meaningful content from eqn.(149), we note that the bare-mass is divergent:$~(\frac {m}{m_{R}})~\sim~O(\frac{\Lambda}{\sqrt{ln~\Lambda}})$ where $~\Lambda~$ = momentum cut-off ( see, eqn.(117)). Since, \textit{according to the standard prescription of the renormalization procedure , the cut-off must be removed (i.e. $~\Lambda \rightarrow~\infty~$)$~\it{prior}~$  to the calculation of any physical quantity } ~of the theory , one obtains : 
    \begin{equation}
    \lim_{\Lambda \rightarrow \infty}~( \frac
    {n(\vec{k})}{n(\vec0)})~\equiv~\rho (\vec{k})~=~( 1~+~\frac {|\vec{k}|^{2}}{m_{R}^{2}})^{-\frac{1}{2}}~,
    %(150)
    \end{equation}       
    where $~n(\vec{0})~=~n(\vec{k})|_{max}~=~( \frac{1}{32\pi^{3}})( \frac 
    {m}{m_{R}})~$, is the maximum value of $~n(\vec{k })~$, occuring at $~\vec{k}~=~0$. 
    
    Equation (150){\it  provides direct physical content for the non-trivial} {\it structure of the IVS representing a condensate of off-shell correlated} {\it particle-pairs}. The situation is analogous to the structure of the physical vacuum state in case of the hard-sphere Bose-gas [43] and superfluidity . It is therefore , plausible that eqn.(150) might lead to interesting consequences for $~T~\neq~0~$, as happens in the case of the super-fluid and the hard-sphere Bose-gas.
    \subsection{Section-IX~:  Summary}
   We have demonstrated that application of NGAS to  the ${\lambda}{\phi}^4$ quantum field theory  under the harmonic approximation, reproduces the main results of the Gaussian-effective potential (GEP) method . In that sense, the scope of NGAS is broader since the GEP is contained as a particular realization of the latter. Beyond that, NGAS reveals the condensate structure of the effective vacuum and highlights the instability of the perturbative ground state. Besides,  the effective vacuum state in NGAS leads to a \textit{non-trivial and stable} theory with finite renormalized parameters whereas, the physics developed around the non-interacting (`perturbative') vacuum becomes either unstable or trivial.
  
    \bigskip
   % \chapter
   
  %\textbf{ (Up to this point on 22-08-2025)}
    
   % \section{ Perturbation theory for arbitrary coupling Strength in NGAS - motivation and scope}
    \section{Perturbation Theory for Arbitrary Coupling Strength in NGAS: Motivation and Scope} 
    In \textbf{Section-I}, we have examined the principal characteristics of the standard formulation of perturbation theory (SFPT) while reviewing various approximation methods, noting both its advantages and limitations. In this section, we introduce an enhanced perturbation theory, hereafter referred to as \textbf{"Mean Field Perturbation Theory" (MFPT)}, which is based on NGAS and aims to address the shortcomings of SFPT. (The terminology will be elucidated in subsequent sections.) Preliminary insights into achieving this objective were provided at the conclusion of \textbf{Section-VIII}, and these will be further developed in this \textbf{Section}. 
    
    Specifically, we will investigate how MFPT can maintain the nonlinearity and analytic properties of the system Hamiltonian, as well as other "non-perturbative" effects, which are not attainable in SFPT. However, certain features, tools, and techniques employed in SFPT may also be necessary for MFPT. In the following \textbf{Section}, we will consider those specific 'tools' and 'techniques' that may be pertinent for this purpose.

    \section{General features of the `standard' formulation of perturbation theory(SFPT)}
    
    As highlighted in \textbf{Section-II}, perturbation theory remains the preferred method of approximation for various practical and theoretical reasons [9]. However, SFPT is subject to several limitations, primarily due to its ${\it defining}$ characteristic that results for physical observables are expressed as a power-series expansion in the coupling strength (CS). To illustrate, consider the energy $E(g)$ as the observable, represented by the standard expansion given by eqn.(1), i.e.,
    \begin{equation*}
        E(g)= \mathcal{E}_0+ \sum\limits_{k=1}^{\infty}{g^{k} \mathcal{E}_k\equiv \mathcal{E}_0(g)+ \Delta{\mathcal{E}(g)}},
    \end{equation*}
     where the first term $\mathcal{E}_0$ represents the contribution known from the non-interacting theory, while $\mathcal{E}_k$ denotes the perturbative correction (PC) at the $k$-th order. For the PC to be meaningful, it is \textit{ab initio} necessary that the coupling strength, $g$, be sufficiently small, i.e., $~g\ll 1$.  Nevertheless, even when $~g~$ is appropriately small, it is generally observed [107] that SFPT results in ${\it divergent}$ but ${\it asymptotic}$ perturbation series (PS). Specifically, the PS exhibits [107] (generalized) factorial growth at large order, as described by the following generic form:
    \begin{equation}
    |\mathcal{E}_k|\sim c^k \Gamma(ak+b),for~k\gg~1,
    %(151)
    \end{equation}
     where $a,b,c$ are constants depending on the particular theory under consideration. Moreover, the \textit{asymptotic nature} of the PS is manifested through the following property [107,108]:
     \begin{equation}
     \lim_{g\rightarrow 0_+}(~E(g)-\sum\limits_{k=0}^{N}g^{k}\mathcal{E}_k~)/~g^{N+1}=\mathcal{E}_{N+1}.
    %(152)
     \end{equation}
     
   % \textbf{ (upto this point on 23-08-2025)}
     
     In this context, the systematic investigations of the large-order behavior of (renormalized) perturbation expansion were historically established through studies [88] of systems with anharmonic interactions (AHI) in quantum mechanics (QM) and subsequent investigations [107] in other systems. It is now established (see, e.g., [98,107,108]) that the generalized-factorial growth of perturbation correction is a general feature of perturbation theory for almost all cases of physical interest, and further, that this behavior is primarily due to the factorial growth of Feynman diagrams/intermediate states contributing at large orders, irrespective of the theory under consideration. Therefore,  It becomes foremost necessary, to   assign a meaning to the asymptotic-perturbation theory and investigate whether the latter can be \textit{``summed"} in some sense to yield a finite result. We will focus on two main methods to achieve this goal, namely the method of \textit{``Optimal Truncation of the Asymptotic Series"} and the \textit{``method of Borel-summation"}.These are discussed in some detail in the next subsection. 
     
      \subsection {Making sense of divergent, asymptotic series: Optimal truncation}
      
       This method utilizes the property~[107,108], that the initial terms of an AS continue to decrease in magnitude till the \textit{term of the  least magnitude}(TLM) is reached, beyond which the subsequent terms exhibit monotonic increase ( refer to the defining equation, eqn.(152)). Hence, the AS can be reasonably approximated by truncation at the TLM.
        \subsection{Making Sense of Divergent, Asymptotic Series: Borel-Summation}
           The method of Borel-summation is considered [107,108] the standard procedure for constructing the analytic function, $E(g)$ from its divergent, asymptotic series for $g\rightarrow{0_+}$. Indeed, the `sum' of the divergent, asymptotic PS is now customarily \textit{defined} [107,108] by its Borel-sum when the latter exists. However, Borel-summability holds under certain restrictive conditions imposed on the original PS . In particular, it is a \textit{necessary} condition [107,108] that the successive terms in eqn.(1) must \textit{alternate} in sign at large-order, $k\gg1$ to ensure Borel-summability. 
           
           This requirement exposes further limitations of SFPT , since it is found  that several  physically important theories, such as the case of the quartic double-well oscillator (QDWO) and other cases exhibiting degenerate vacua /ground states, \textit{fail} [95] this necessary condition for Borel-summability.
       
       Furthermore, \textit{in the context of SFPT, both the above methods are restricted to the perturbative domain characterized by small values ( compared to unity) of the coupling $g$.}
      
    In light of the deficiencies of SFPT discussed in \textbf{Section-II} and the preceding section, it is essential to investigate whether perturbation theory can be freed from the constraints of the small-coupling-power-series-expansion. The subsequent \textbf{Sections} are dedicated to this exploration in detail.
     
    % \textbf{( Up to this point on 24-08-2025)
      \section{General Features of the `Mean Field Perturbation Theory' (MFPT) in NGAS}
   One of the primary motivations for proposing the NGAS, as previously described, is the potential to develop an enhanced perturbation theory that is convergent or summable for \textit{all} permissible values of the coupling strength, $g$.
   
   Utilizing the standard Hamiltonian formalism, we can construct an improved perturbation theory wherein the system-Hamiltonian, $H(g)$, is divided into an {\it exactly solvable} dominant component $H_{0}$ and a {\it sub-dominant} perturbation $H^{\prime}$, such that
 \begin{equation}
   H(g) = H_{0}+H^{\prime}.
   %(153)
\end{equation}
   
   To establish a perturbative framework for arbitrary values of the coupling $g$, it may be {\it sufficient} to satisfy the following two conditions: (a) both $H_{0}$ and $H^{\prime}$ must depend {\it non-trivially} on the coupling strength $g$, i.e., $H_{0} = H_{0}(g)$ and $H^{\prime} = H^{\prime}(g)$, with the dominant $g$-dependence residing in $H_{0}(g)$, and (b) the contribution of the perturbing Hamiltonian $H^{\prime}(g)$ remains sub-dominant for any value of $g$.
   
   As previously discussed (\textbf{Section-VI}), a practical implementation of these conditions can be achieved [82,104] through the following steps: (i) select $H_0$ to depend on a suitable set of free parameters ${{\alpha_i}}$: $ H_0= H_0(\{\alpha_i\})$, (ii) impose the constraint that
   
   \begin{equation}
    \langle\phi_{n}|H(g)|\phi_{n}\rangle~=~\langle\phi_{n}|H_0(\{\alpha_i\})|\phi_{n}\rangle
   %(154)
   \end{equation}
   
   where $|\phi_{n}\rangle $ is defined by the eigenvalue equation for $H_0$ given by
   
   \begin{equation}
    H_0|\phi_{n}\rangle~=~E_0^{n}|\phi_{n}\rangle ,
   %(155)
   \end{equation}
   
   with $`` n"$ denoting the spectral label, (iii) finally, determine the parameters $\{\alpha_i\}$ by variational minimization of $\langle H(g)\rangle$ and subsequent imposition of the constraint given by eqn.(154).
   
   Through this straightforward procedure, the non-linear $g$-dependence of the system-Hamiltonian, $H(g)$, is incorporated into the reference Hamiltonian:
   
 \begin{equation}
    H_0(\{\alpha_i\})\rightarrow ~H_0(g)
     %(156)
    \end{equation}
   
   Solving the eigenvalue equation (eqn.155) for $H_0(g)$ then yields [82,104] the results for $E_0(g)$, which constitute the leading-order (LO) approximation. These LO results are found [82,104] to be highly accurate across all spectral labels `$n$' and for any physical values of $g$. It is noteworthy that this procedure is non-perturbative, self-consistent, and preserves the analytic structure of $E(g)$ in the $g$-plane [82,104].
   
   The aforementioned procedure has been previously discussed in \textbf{Section-VI} but is briefly recapitulated here for subsequent discussions. Having incorporated the major $g$-dependence at the LO, the task remains to develop a perturbation theory around the approximating Hamiltonian (AH), $H_0$, and systematically enhance the accuracy by computing order-by-order residual corrections, as described in the next subsection.
   
   \subsection{Definition and Implementation}
   
   A novel formulation of perturbation expansion [110,111], termed as "Mean-Field-Perturbation Theory (MFPT)," is \textit{naturally suggested} in light of the aforementioned considerations by defining the perturbation around the AH as:
   
    \begin{equation}    
    H^{\prime}(g,n)\equiv H(g)- H_{0}(g,n).     %(157)     
    \end{equation}    
   
    An immediate consequence of this prescription, eqn. (157), is the following result:     
   
   \begin{equation}    
   \langle \phi_{n}|H^{\prime}(g,n)|\phi_{n}\rangle = 0,    
    %(158)    
  \end{equation}     
   
   for all $``n"$ and $``g"$, by virtue of eqns. (154) and (156). This eqn. (158) ensures that the perturbation correction remains sub-dominant, in the sense of quantum average, for arbitrary values of $g'$ and $n'$ as required (see condition-(b) mentioned above, i.e.     
    \begin{equation}     
   \langle H^{\prime}(g)\rangle \ll \langle H_{0}(g)\rangle \equiv {E}^{n}_{0}(g).    
    %(159)     
\end{equation}    
   
    (Here, the quantum average of an operator A is defined as: $\langle A \rangle \equiv \langle \phi_{n}|A|\phi_{n}\rangle$). Furthermore, eqn. (158) has a direct consequence that the first-order perturbation correction in the Rayleigh-Schrödinger perturbation series (RSPS) {\it vanishes} identically for all $n'$ and $g'$:     
   
   \begin{equation}  
   {E}^{n}_{1} \equiv \langle\phi_{n}|H^{\prime}|\phi_{n}\rangle = 0,     %(160)    
    \end{equation}    
   
    It is shown later that these two general outcomes of eqn. (158) have significant implications for the properties of the perturbation series (PS) in MFPT, which is defined analogously to eqn. (1) as follows:     \begin{equation}     
   E(g) = E_{0}(g) + \sum_{k = 1}^{\infty} E_{k}(g) \equiv E_{0}(g) + \Delta E(g).    
    %(161)    
    \end{equation}     
   
   (Henceforth, we do not display the $n$-dependence of the various quantities, for notational convenience.)
   
          For the study of convergence properties of the above series, one needs to compute the energy corrections $E_{k}(g)$ to arbitrary order $k'$. However, since the above series is {\it not} a power series expansion in $g$, we resort to the well-known [109] \textit{recipe} of introducing an auxiliary, {\it dummy} parameter denoted as $\eta$, for generating a power series in this parameter (chosen real) and project out the $k'$-th order correction, $E_{k}(g)$ by the following procedure: consider an associated Hamiltonian (ASH), $\bar{H}$ given by     
   
   \begin{equation}    
    \bar{H}\equiv H_{0} + \eta H^{\prime},     %(162)    
    \end{equation}     
   
   and the corresponding eigenvalue equation:    
   
    \begin{equation}    
   ~~~~~~~~~~ \bar {H}(\eta,g)|\bar{\psi}(\eta,g)\rangle = \bar{E}(\eta,g)|\bar{\psi}(\eta,g)\rangle,    
     %(163)    
    \end{equation}     
   
   such that $\bar{E}(\eta,g)$ can be expanded as a {\it formal} power series in $\eta$ as follows:    
   
    \begin{equation}   
   ~~~~~~~~\bar{E}(\eta,g) = \sum_{k = 0}^{\infty}\eta^{k}E_{k}(g) %(164)     
   \end{equation}   
   
   The $`k'$-th perturbation correction $E_{k}(g)$, as referenced in equation (161), can be discerned from the preceding equation as the coefficient of $\eta^{k}$ prior to setting the limit $\eta \rightarrow 1$. It is important to note that this procedure serves merely as an intermediate book-keeping mechanism to isolate the $E_{k}(g)$ appearing in equation (161). Beyond this purpose, the formalism does not serve any additional function in this context. Specifically, the final results are independent of the dummy variable $\eta$ since, by design, $H(g) = \bar {H}(\eta = 1,g)$, $E(g)=\bar {E}(\eta = 1,g)$, and so forth. We compute the corrections $E_{k}(g)$ to an arbitrary order $`k'$ in MFPT, utilizing the recursion relations derived from the application of the "hyper-virial theorem (HVT)" and the "Feynman-Hellman theorem (FHT)" to the Hamiltonian specified in equation (163). This method is distinct from other approaches in that it generates exact values for perturbation corrections to any order, whereas other methods are prone to inevitable errors such as neglecting sub-asymptotic corrections and truncation errors. Additionally, the ease of implementing recursive evaluation through widely available symbolic computation is a practical advantage of this method. Here, we follow a procedure analogous to that in reference [9] used in the context of SFPT for the QAHO, with appropriate generalization for extending the application to MFPT as outlined in the subsequent section. 
   
  %\textbf{ (up to this point on 24-08-25 FN)}
 
    \section{Application of MFPT to Anharmonic -interactions(AHI) in the Harmonic Approximation}
    \subsection {Tools: The Feynman-Hellman Theorem (FHT) and the Hyper-Virial Theorem (HVT)}
    
    In many cases of physical interest, the concurrent application of the Feynman-Hellman Theorem and the hyper-virial theorem facilitates the computation of perturbation corrections $E_{k}$ without the necessity of summing infinite (sub)-series at each order. Furthermore, each term $E_{k}$ can be computed precisely (with zero error), as is inevitably the case due to finite truncation of the conventional perturbation series at each order. Moreover, precise computation of $E_{k}$ to an arbitrarily high value of `$k$' can be achieved exactly through system-specific recursion relations. For the AHI, the program was executed by Swenson and Danforth [81] and elaborated in texts such as by Fernandez [9]. We discuss these theorems in turn, as follows.
        \subsubsection{ The Hyper-virial Theorem(HVT)}.
    
     The content of the Hyper-Virial-Theorem (HVT) is given by the following equation , see, e.g.ref.[9]:
     \begin{equation}
    \langle\psi|\left[\hat{H},\hat {W}\right]|\psi\rangle~=~0,
     %eqn.(165)
    \end{equation}
    where,$ \hat{H}$ is the (hermitian) Hamiltonian of the system and $[\hat{A},\hat{B} ]$ stands for the \textit{commutator} of operators,   $ \hat{A}$ and $\hat{B}$  and $\hat {W} \equiv \hat {W}{(\hat{x}, \hat{p})}$ is any {\it arbitrary}, time independent operator function of $\hat x$ and $\hat p$ in the relevant Hilbert space. Here, we confine to the co-ordinate space- representation such that $\hat{p}~=~-i\frac{\partial}{\partial{x}}$  and  $\hat{x}~=~x$ are the momentum and position operators.
    
    Then the above eqn.(165) follows as a consequence of the appropriate energy-eigenvalue equation(and its hermitian conjugate) for the time-independent ( 'Stationary') situation, which can be verified directly using :  
        \begin{equation}
     \hat{H} |\psi\rangle~=~E |\psi\rangle ,  \langle\psi| \hat{H} ~=~ E \langle\psi|
     % eqn.(166) 
     \end{equation}

We consider next the \textit{Hellmann-Feynman Theorem}.
      
    \subsubsection{ The Hellmann-Feynman Theorem(HFT)}
    Consider a scenario where the Hamiltonian, $ {H}$, of a system is dependent on a parameter $\lambda$, in addition to the dynamic variables of the system under consideration. This situation is quite common, as $\lambda$ could represent the system's charge, mass, a coupling strength of interaction, or even a \textit{dummy parameter} introduced to define perturbation theory, among others. Consequently, the eigenvalues and eigenfunctions of $ {H}$ will also be dependent on this parameter. The Hellmann-Feynman Theorem (HFT) is applicable in this context and yields significant results in perturbation theories without necessitating the use of wave functions, as discussed in ref. [9]. The theorem posits that:
    
    \textit{The partial derivative of the total energy with respect to a parameter present in the Hamiltonian of the system is equivalent to the expectation value of the partial derivative of the Hamiltonian with respect to the same parameter}, as expressed by the following equation:
    
     \begin{equation}
             \langle\psi_{n}|\left(\frac{\partial H}{\partial\lambda}\right)|\psi_{n}\rangle~=~\frac{\partial E_{n}}{\partial\lambda} .
             %(167) 
             \end{equation}

     The Hellmann-Feynman-Theorem (HFT) applies to this situation and leads to profound results in perturbation theories without the use of wave-functions, see e.g. ref.[9] .The theorem states that: 
      \begin{equation}
   (H-E_{n})|\psi_{n}\rangle~=~0 ~=~\langle\psi_{n}| (H-E_{n}) ,
    %(168)
    \end{equation}
        which yields eqn.(167) on partial differentiation with respect to $\lambda$  if $|\psi_n\rangle$ is normalized to unity. 
        
        In this context, it is pertinent to note that: (i) there are generalizations of the (FHT) to its 'off-diagonal' version, as referenced in [106], and (ii) the off-diagonal FHT has significant applications, such as in reproducing the RS-Perturbation theory \textit{without the use of the unperturbed wave function}.

   \subsection{ Tools : The Method of Swenson and Danforth }
   The method of Swenson and Danforth [81] employs the hyper-virial and the Hellmann-Feynman theorems to derive the perturbation correction for anharmonic oscillators within the SFT framework. A comprehensive description of this method is provided in ref. [9], which we closely follow in this subsection to present an overview of the method.
 
 To start with, we consider the Hamiltonian for a non-relativistic, stationary system in one space-dimension  given by:
  ~~$\hat{H} = \hat{p}^{2}{/}{2m} + V(\hat x)  $,   
    and use the co-ordinate representation such that $\hat x  = x$ and  $ \hat p = -i \hbar\partial {/} \partial x $.
    
      In the next step,  we explore the consequence of the HVT by choosing a suitable linear operator, $\hat{W}$  depending up on $\hat x$  and $\hat p$ . Following ref. [9] ,  the following choice is made :
    \begin{equation}
    \hat {W}~=~f(x)\hat{p} + g(x),
    %(169)
    \end{equation}
    where, $ f(x)$ and $ g(x)$ are arbitrary, 'regular' (differentiable) functions to start with but allowing to be chosen conveniently for the system under study. Considering the one-dimensional Hamiltonian operator  as given above, i.e.   ~~$\hat{H} = \hat{p}^{2}{/}{2m} + V(\hat x)  $, ~~and using eqn.(169) in eqn.(165), one obtains:
   
    \begin{equation}
    \langle\left[H,fp+g\right]\rangle ~=~\langle\left[H,fp\right]\rangle
    + \langle\left[H,g\right]\rangle~=~ 0,
     %(170)
    \end{equation}
    
    where, $\langle~\hat A~\rangle$ denotes ~ the $~expectation~ value~ of ~the~ operator ~\hat A~ in ~  the~ state, ~\rvert\psi\rangle~$.
     This equation can be expressed as the sum of two terms i.e.  $~ A + B ~\equiv ~0$, where $~A =\langle\left [H,fp\right]\rangle~$ and $~B =\langle\left [H,g\right]\rangle~$. 
    
    Using  the basic commutation relation,~~ $[\hat x, \hat p] = -i \hbar$~~  and  the  result: $ [\hat p, f(x)] = i\hbar f^\prime $  (where $^\prime$ denotes derivative with respect to  $~x~$) and employing the standard manipulation techniques for $ commutators $, one obtains the following result for $A$:
    \begin{equation}
    A~=~\left(\frac{-\hbar^{2}}{2m}\right)\langle{f^{\prime\prime}}p\rangle-\left(\frac{i{\hbar}}{m}\right)\langle{f^\prime}p^{2}\rangle+(i{\hbar})\langle f{V^\prime}\rangle.
     %(171)
    \end{equation}
   Similarly, the $ B $ -  term can be evaluated as
    \begin{equation}
       B~=~\langle\left(\frac{-\hbar^{2}}{2m}\right){g^{\prime \prime}}\rangle
       - \left(\frac{i{\hbar}}{m}\right)\langle{g^\prime}p\rangle.
       %\label{172}
       \end{equation}
     The substitution of $ A$ and $B$ in the defining equation,
    \begin{equation}
    0 ~=~ A + B \equiv  \langle\left[H,W\right]\rangle 
    %(173)
    \end{equation}
    then yields the result
    \begin{equation}
    \left(\frac{-\hbar^{2}}{2m}\right)\langle{f^{\prime ^\prime}}p \rangle
    -\left(\frac {i{\hbar}}{m}\right)\langle{f^\prime}p^{2} \rangle
    +(i{\hbar})\langle f{V^\prime} \rangle
    +\left(\frac{-\hbar^{2}}{2m}\right)\langle{g^{\prime \prime}} \rangle
    -\left(\frac {i{\hbar}}{m}\right)\langle{g^\prime}p \rangle ~=~ 0  .
    %(174)
    \end{equation}
    Next, since $f$ and $g$ are \textit{arbitrary} regular functions,
    one can choose $g$ in terms of $f$ such that term linear in
    $\hat{p}$ vanishes.
    This is achieved by imposing the following constraint: 
    \begin{equation}
    \left(\frac{-\hbar^{2}}{2m}\right)\langle{f^{\prime \prime}}p \rangle
    + \left(\frac{-i{\hbar}}{m}\right)\langle{g^\prime}p \rangle~=~0.
    %(175)
    \end{equation}
    The constraint-eqn.(175) is thus satisfied by the following choice of the g-function:
    \begin{equation}
    g~=~\left(\frac{1}{2}\right) \left[p,f\right]~=~\left(\frac{i{\hbar}}{2}\right){f^\prime}.
    %(176)
    \end{equation}
    Substitution of eqn.(176) in eqn.(174) then leads to 
    \begin{equation}
    \left(\frac{-i{\hbar}}{m}\right)\langle{f^\prime}p^{2} \rangle 
    + (i {\hbar})\langle f{V^\prime}\rangle
    -(i{\hbar})\left(\frac{\hbar^{2}}{4m}\right)\langle{f^{\prime \prime \prime}}
    \rangle~=~0.
    %(177)
    \end{equation}
    The insertion of
    \begin{equation}
    \hat{p}^{2}|\psi\rangle~=~ \{2m(E-V)\}\psi\rangle,
    %(178)
    \end{equation}
    in eqn.(177) and simplifying, then yields:
    \begin{equation}
    2E \langle {f^\prime}\rangle - 2 \langle{f^\prime}V\rangle -\langle f{V^\prime}\rangle +\left(\frac {\hbar^{2}}{4m}\right)\langle {f^{\prime \prime \prime}}\rangle~=~0.
    %(179)
    \end{equation}
   
   In the above equation, the notation is as follows: $E$ represents the energy eigenvalue; $f = f(x)$ is an arbitrary differentiable function that can be conveniently chosen for the particular problem at hand, "prime(s)" denote differentiation, and $\langle A\rangle \equiv \langle\psi|A|\psi\rangle$ for an operator $A$, with $|\psi\rangle$ denoting the (normalized) eigenfunction of $H$.
   
    It is noteworthy that: (a) an alternative derivation of equation (179) can be achieved, e.g., [9] by using the Jacobi identity for commutators. (b) This equation, along with the use of the Hellmann-Feynman Theorem, will be shown (later in this \textbf{Section}) to lead to a recursive derivation of \textit{exact} perturbation corrections at \textit{arbitrary} order. (c) The yet-arbitrary function $f$ can be chosen according to the specific functional form for the potential function: $V(x)$. In the following subsection, we illustrate the SD-Method by applying it to the case of the general anharmonic interaction in one spatial dimension. 
          
   \textbf{ (Up to this point on 25 Aug 2025)}

    \subsection{The Swenson-Danforth Method (SDM) for the An-Harmonic Interactions (AHI)}
    Consider the Hamiltonian(dimensionless units used):
    \begin{equation}
    H~=~\frac{1}{2}p^{2} + V
    %(180)
    \end{equation}
    where
    \begin{equation}
    V~=~\frac{1}{2}x^{2} + \lambda x^{2K}, K= 1,2,3,....
     %(181) 
    \end{equation}
    Rewriting  eqn.(179) in these notations leads to:
    \begin{equation}
    2E \langle{f^\prime}\rangle - 2\langle{f^\prime}V \rangle - \langle f{V^\prime}\rangle +(\frac{1}{4}) \langle{f^{\prime\prime\prime}}\rangle~=~0.
     %(182)
    \end{equation}
    For the even an-harmonic oscillator as  defined in eqn.(181), it is suggested that we choose:
    \begin{equation}
    f(x)~=~x^{2j+1}
    %(183)
    \end{equation}
    where'$j$'= 0,1,2,3,4.....
    Computing ${f^\prime}$, ${V^\prime}$, ${f^{\prime \prime \prime}}$, etc, 
    substituting into eqn.(182) and simplifying , one obtains:
    \begin{eqnarray}
    2E(2j+1)\langle x^{2j} \rangle - 2(2j+1)\langle x^{2j}(\frac{1}{2}x^{2}
    + \lambda x^{2K})\rangle\nonumber\\
    -\langle x^{2j+1}(x+\lambda(2k)x^{2K-1})\rangle\nonumber\\ 
    + (\frac{1}{4})(2j+1)(2j)(2j-1) \langle x^{2j-2}\rangle &=& 0.~~~~~~~~
    %(184)
    \end{eqnarray}
    Denoting,
    \begin{equation}
    \langle x^{2j}\rangle \equiv X(j),
     %(185)
    \end{equation}
    and substituting eqn.(185) into eqn. (184),  yields
    \begin{eqnarray}
    2E(2j+1)X(j)-2(2j+1)(\frac{1}{2}X(j+1)\nonumber\\+\lambda X(j+K))- \{X(j+1)+\lambda(2k)X(j+K)\}\nonumber\\
    +(\frac{1}{4})(2j)(4j^{2}-1)X(j-1)&=&0.~~~~~~~
    %(186)
    \end{eqnarray}
    Rearranging eqn.(185) one finally gets  
    the desired recurrence relation for the expectation values of $X(j)$
    \begin{eqnarray}
    X(j+1)&=&\frac{1}{2(j+1)}\{2(2j+1)EX(j)\nonumber\\
    &&+\frac{j(4j^{2}-1)}{2}X(j-1)\nonumber\\
    &&-2\lambda(2j+K+1)X(j+K)\}
    %(187)
    \end{eqnarray}
    As a check of eqn.(187) , consider the {\it special case} when $\lambda=0$. 
    Then one recovers the $Hamiltonian$ of the simple-harmonic-oscillator (SHO), given by 
    \begin{eqnarray}
    H \rightarrow H_{SHO} \equiv \frac{1}{2}p^{2} + V_{0}~;~~
    V_{0}&=&\frac{1}{2}x^{2}~~~~~
    %\label{188}
    \end{eqnarray}
    Since this case is exactly solvable, we know the eigenvalues ,
    \begin{equation}
    E \rightarrow {\cal E} \equiv E_{SHO}~=~(n+\frac{1}{2})\equiv\xi 
    %\label{189}
    \end{equation}
    where , $n=0,1,2,3,......$.
    (Henceforth, we suppress the label $'n'$ in $ E$ and ${\cal E}$ for notational convenience.)
    Denote the eigenstates of $H_{SHO}$ by
    \begin{equation}
    |\phi_{n}\rangle\equiv~ {\lim}_{\lambda \rightarrow 0}~~|\psi_{n}\rangle.
    %\label{190}
    \end{equation}
    Also, let us denote by $Y(j)$:
    \begin{equation}
    Y(j) \equiv \langle\phi_{n}|x^{2j}|\phi_{n}\rangle
    %\label{191}
    \end{equation}
    Then from eqn.(187), substituting $\lambda=0$ and using eqns.(190-191) we get
    \begin{equation}
    Y(j+1)=\left\{(\frac{2j+1}{j+1})\right\}{\cal E}Y(j)
    +\left\{\frac{j(4j^{2}-1)}{4(j+1)}\right\}Y(j-1)~~~~
    %(192)
    \end{equation}
    Recursive solution for $Y(j)$; $j$=0,1,2,.... can be obtained from eqn.(192)in terms of the known eigenvalues ${\cal E}$ . Noting that: $Y(0)~=~1$ and $ Y(j)\equiv0$ for $j<0$ (by choice), we get
         
    \begin{eqnarray}
    Y(1)={\cal E}Y(0)\nonumber ={\cal E} \equiv \langle x^{2}\rangle_{SHO}, \nonumber\\
    Y(2)=(\frac{3}{2}){\cal E}Y(1)+ (\frac{3}{8})Y(0), \nonumber\\
    =(\frac{3}{2}){\cal E}^{2} + (\frac{3}{8}) \equiv
    \langle x^{4} \rangle_{SHO},\nonumber\\
    Y(3)=(\frac{5}{3}){\cal E}Y(2)+\frac{5}{2}Y(1)\equiv \langle x^{6}\rangle_{SHO}~~  ,
    %(193)
    \end{eqnarray}
    etc. , which are the {\it standard results} (see, eqns.(27a-27d)), obtained by other methods, e.g. the use of 'ladder- operators' for the SHO. 
    
         \bigskip
              
    Let us next consider the case when $\lambda \neq 0 $ in eqs.(180-181). Then, the Hamiltonian, (eqn.(180)) corresponds to the case of the anharmonic oscillator with anharmonicity = $2K$ (=4,6,8,....). Now of course, $E$ is {\it not} known analytically. However, if  the standard formulation of perturbation theory (SFPT) is applied (with $\lambda $ considered as the expansion parameter)  then one proceeds by expanding the observables appearing in the\textit{ `Master Formula' (eqn.(187)}  as follows: 
    \begin{eqnarray}
    E \equiv E(\lambda)=\sum_{p=0}^{\infty}E_{p}\lambda^{p} , ~~~
     X(j,\lambda) \equiv \sum_{i=0}^{\infty}\lambda^{i}X(j,i).
    %\label{194}
    \end{eqnarray}
     with $ E_{0} \equiv {\cal E}=\xi \equiv (n+\frac{1}{2})$ = SHO-energy levels. 
     
     The perturbation corrections, $E_{k}$ (k=1,2,3,....) can be computed using eqn.(187) together with the Hellmann-Feynman- Theorem , eqn.(167) as  given below: (This was originally done by Swenson-Danforth[81] . The derivation here is based on ref.[9].)
    To apply the Hellmann -Feynman-Theorem , eqn.(167) to the Hamiltonian given by eqn.(180-181) 
    the strategy is to evaluate $ \langle \frac{\partial H}{\partial\lambda}\rangle$ and $~\frac{\partial E}{\partial\lambda}$  each , as a power-series in$~\lambda~$ and then equate the resulting two series. This  yields the following equations :
       \begin{eqnarray}
      \langle \frac{\partial H}{\partial\lambda}\rangle~=~\langle x^{2K} \rangle\equiv X(K,\lambda)\nonumber\\
    ~=~\sum _{p=0}^{\infty}\lambda^{p}X(K,p).
    %(195)
    \end{eqnarray} 
    and
    \begin{equation}
    \frac{\partial E}{\partial\lambda}~=~\sum_{p=1}^{\infty}\lambda^{p-1}(pE_{p})
    %\label{196}
    \end{equation}
    Equating the two expressions, eqns.(195-196) as required by HFT (~eqn.(167)~) then leads to :
      \begin{equation}
    \sum_{p=1}^{\infty}\lambda^{p-1}\{X(K,p-1)-pE_{p}\}~=~0.
    %\label(197)
    \end{equation}
        Hence, equating the co-efficient of each power of $\lambda$ to zero in eqn.(197)  yields: 
    \begin{equation}
    E_{p}~=~(\frac{1}{p})X(K,p-1)
    %\label{198}
    \end{equation}
    This eqn.(198) thus relates the perturbation correction at order-$p$ in terms of the $X(K,p-1)$ , which can be evaluated recursively by the recurrence relation given by eqn.(187),again  by expanding the relevant observables in power-series in $\lambda$ as given by eqn.(194) and reproduced below:
    \begin{equation*}
   ~~~~ ~~~~~~~~~~~~ E \equiv E(\lambda)=\sum_{p=0}^{\infty}E_{p}\lambda^{p} ,  X(j,\lambda) \equiv \sum_{i=0}^{\infty}\lambda^{i}X(j,i).~~~~~~~~~~~~~~~~~~~~~~~~~~~~~~~~~~~~~~\mbox{(194)}
         \end{equation*}
   where $X(j,i)$ $ \sim $ Taylor-coefficients in power-series expansion of $ X(j,\lambda) $, i.e.
    \begin{equation}
    X(j,i)~=~(\frac{1}{i!})\frac{\partial^{i}}{\partial\lambda^{i}}
    X(j,\lambda)|_{\lambda = 0};
    %(199)
    \end{equation}
    Next, consider the power-series expansion for the product: $X(j,\lambda)E(\lambda)$ given by
    \begin{equation}
    X(j,\lambda)E(\lambda)~=~\sum_{i=0}^{\infty}f_{i}\lambda^{i}
     %(200)
    \end{equation}
    Then, from the rules of power-series multiplication , the coefficients $f_i$ get determined  in terms of $ E_{m} $ and $X(j,i)$ given by
    \begin{equation}
    f_{i}~=~\sum_{m=0}^{i}E_{m}X(j,i-m)
     %(201)
    \end{equation}
    
    After substitution of the above equations in eqn.(187), leads to
    \begin{eqnarray}
    \sum_{i=0}^{\infty}\lambda^{i}X(j+1,i)&=&\left(\frac{2j+1}{j+1}\right)
    \sum_{i=0}^{\infty}\lambda^{i}\{\sum_{m=0}^{i}E_{m}X(j,i-m)\}\nonumber\\
    && + \left\{\frac{j(4j^{2}-1)}{4(j+1)}\right\}\sum_{i=0}^{\infty}
    \lambda^{i}X(j-1,i)\nonumber\\
    && - \left\{\frac{2j+K+1}{j+1}\right\}\sum_{i=0}^{\infty}\lambda^{i+1}
    X(j+K,i)\nonumber\\
        %\label{202}
    \end{eqnarray}
    It can be checked that the $ \lambda$ independent terms in eqn.(202) reproduce eqns.(192-93) as required and these correspond to observables for the SHO.
      
    For, non-zero values of $\lambda $ i.e. for $i$=1,2,3,.... one gets from (eqn.202), the following recursion relation for $X(j+1,i)$ 
    by the following procedure:
    Rewrite (202) by changing $ i\rightarrow (i-1) $ in the last-term) leads to
    \begin{eqnarray}
    \sum_{i=1}^{\infty}\lambda^{i}\{X(j+1,i)-a(j)
    \sum_{m=0}^{i}E_{m}X(j,i-m)\nonumber\\
    - b(j)X(j-1,i) + c(j)X(j+K,i-1)\}&=&0,~~~~~~
    %(203)
    \end{eqnarray}
    Then equating coefficient of each power of $\lambda$ to zero, in (203) one arrives at:
    \begin{eqnarray}
    X(j+1,i)~=~a(j)\sum_{m=0}^{i}E_{m}X(j,i-m)\nonumber\\ +b(j)X(j-1,i)-c(j)X(j+K,i-1),
    %(204)
    \end{eqnarray}
    where we have denoted:
    \begin{eqnarray}
    a(j)&=&\left(\frac{2j+1}{j+1}\right),\nonumber\\
    b(j)&=&\left(\frac{j(4j^{2}-1)}{4(j+1)}\right),\nonumber\\
    c(j)&=&\left(\frac{2j+K+1}{j+1}\right),\nonumber\\
    %(205)
    \end{eqnarray}
    Finally, it is convenient to change $j\rightarrow (j-1)$ through out in eqn.(205) to get
    \begin{eqnarray}
    X(j,i)&=&\left\{(\frac{2j-1}{j})\right\}
    \sum_{m=0}^{i}E_{m}X(j-1,i-m)\nonumber\\
    && + \left\{\frac{(j-1)(4(j-1)^{2}-1)}{4j}\right\}X(j-2,i)~~~~~~~~~~~~\nonumber\\
    && -\left\{\frac{2j+K-1}{j}\right\}X(j+K-1,i-1).\nonumber\\
    %\label{206}
    \end{eqnarray}
%    Rewriting clearly eqn.(230), we get:
%    \begin{eqnarray}
%    X(j,i)&=&\left(\frac{2j-1}{j}\right)
%    \sum_{m=0}^{i}E_{m}X(j-1,i-m)\nonumber\\
%    && + \left\{\frac{(j-1)[4(j-1)^{2}-1]}{4j}\right\}X(j-2,i)~~~~~~~~~~~~~\nonumber\\
%    && - \left\{\frac{2j+K-1}{j}\right\}X(j+K-1,i-1)\nonumber\\
%    %\label{231}
%    \end{eqnarray}
%    It may be noted that the range of the variables in the recursion relation, eqn.(231) are given by: $i=1,2,3,....$; $K=2,3,4,....$ and $j=0,1,2,3,.....$.
    
   Using eqn.(206), along with the result from the FHT given by (see, eqn.(198)), i.e.
    \begin{equation}
   ~~~~~~~~~~~~~~~~~~~~~~~~~~~~~~~~~~~~ E_{p}~=~\left(\frac{1}{p}\right)X(K,p-1)\nonumber, ~~~~~~~~~~~~~~~~~~~~~~~~~~~~~~~~~~~~~~~~~~~~~~~~~~~~{(198)}
    %\label{none},
    \end{equation}
    one can completely solve the above recursion relations for $E_{p}$; $p$=1,2,3,....; provided, the following\textit{ boundary conditions} are imposed,
    \begin{equation}
    X(0,i)~=~\delta_{0,i}~~~~,~~~~~ X(j,i)~=~0~ for~ i<0.
    %(207)
   \end{equation}
        which follow trivially from the definitions of $X(0)=1$ and $i>0$ respectively.
               
    Regarding the recursion relation, equation (206), the following observations are pertinent: (a) The recursion terminates only when the \textit{second argument of the last term} in equation (206) becomes zero due to the boundary conditions, equation (207). This implies that for the evaluation of $E_{p}$, one needs to compute \textit{all} coefficients $X(j,i)$ for $i=1,2,3,\ldots,(p-1)$; and for each fixed $i$, $j=1,2,3,\ldots,(p-i)(K-1)+1$, such that the "maximum" value for `$j$' corresponds with the "minimum" value of `$i$'. This specifies the set of values of $i,j$ over which the recursion in equation (206) is to be carried out. (b) The array of $X(j,i)$ required to solve the problem for the computation of $E_{p}$ (i.e., $p$-th perturbation correction) can be generated by both numerical or symbolic implementation of computation. However, it soon becomes unwieldy in 'numerical' implementation when $p \gg 1$, due to large rounding-off errors and memory restrictions. In reference [9], symbolic implementation through "MAPLE" has been advocated, which we have followed and verified its suitability for $p \leq 200$, which suffices for most theoretical purposes (such as studies of \textit{large order behavior of SFPT}). (c) Sample results in SFPT for the QAHO were obtained by the recursion method, equation (206), which correctly reproduces [9] the results from asymptotic computation of Bender-Wu, Banks, Bender, and Wu [88,100] for establishing the large-order behavior, confirming the asymptotic nature and the predicted divergence in SFPT. (d) More importantly, the "Hypervirial-Feynman-Hellman-Theorem-(HVT-FHT)" method, followed here, provides \textit{exact} (i.e., no truncation) results at each order of perturbation theory.
    
    %(\textbf{Upto this point on 26ˆ-08-25 Fn})
    
     \section{Application of MFPT to Anharmonic Interactions} 
     
     In the preceding section, the method based on the Hypervirial and Feynman-Hellman Theorems (HVT-FHT) for calculating perturbation corrections within the "standard formulation of perturbation theory" (SFPT) was emphasized, demonstrating its superiority over conventional methods. The ease and simplicity of its implementation were established by reproducing results for perturbation corrections of arbitrary order in SFPT for various cases of anharmonic oscillators, as obtained by other approaches.
     In the subsequent subsections, we apply the HVT-FHT method within the framework of the "Mean-Field Approximation Scheme" and the "Mean Field Perturbation Theory" (MFPT) to the cases of the Quartic Anharmonic Oscillator (QAHO), Sextic Anharmonic Oscillator (SAHO), and Quartic Double Well Oscillator (QDWO), which constitute the primary objectives of this study, given the inadequacies of SFPT previously discussed (see Sections II, IV, and XI).
     \subsection{The Quartic- and Sextic Anharmonic Oscillators (AHO)}
    For both these cases,the system-Hamiltonian is given by:
    \begin{equation}
    H~=~\frac{1}{2}p^{2} + \frac{1}{2}x^{2} + gx^{2K},\mbox(K=2,3),
    %\label{208}
    \end{equation}
    corresponding to the quartic-AHO (QAHO) and the sextic-AHO (SAHO) respectively.To carry out the NGAS, we choose [82] the harmonic-approximation for the ``Approximating Hamiltonian" (AH), given by eqn.(25 but reproduced below for ready reference:
    \begin{equation*}
    H_0~=~\frac{1}{2}p^{2} + \frac{1}{2}\omega^{2}x^{2} + h_{0},~~~~~~~~~~~~~~~\mbox{(25)}
    %\label{25} 
    \end{equation*}
    which is the Hamiltonian of an energy-shifted-harmonic-oscillator involving two free-parameters:$h_0$ and $\omega$, corresponding to an `energy-shift' and the `frequency' respectively. These parameters are determined [82] in accordance to the procedure through the  steps(i)-(iii)as outlined in \textbf{Section}-VIA above. For the case of the QAHO, these are determined by equations(29-30) but reproduced here below for ready reference:
    \begin{equation*}
    ~\omega^{3}-\omega-6g(\xi+\frac{1}{4\xi})=0,~~~~~~\mbox{(29)}
    %(29)
    \end{equation*}
    and
    \begin{equation*} 
    ~h_0=\left(\frac{\xi}{4}\right)\left(\frac{1}{\omega}-\omega\right)~~~~~~~~~~\mbox{(30)}
    %(30)
    \end{equation*} 
    where, the notation is : $\xi\equiv(n+\frac{1}{2}); n=0,1,2,3,...$, being the spectral-level index.
    As derived earlier (in  \textbf{Section}-VI-A), these parameters acquire the required functional dependence on $g$ and $\xi$:$~\omega$=$\omega(g,\xi$); $~h_0=h_0(g,\xi)$. Similarly, the eigenvalue of the MFH, $E_{0}$ is obtained as a function of $g$ and $\xi$ and given by eqn.(32) (reproduced here, for ready reference)
    \begin{equation*} 
    ~E_0(g,\xi)=\left(\frac{\xi}{4}\right)\left(3\omega+\frac{1}{\omega}\right).~~~~\mbox{(32)}
    %(32)
    \end{equation*}
       (Let us note that because of the imposed constraint of CEQA ( see, \textbf{Section-VI}, eqn.(16)), one can also designate $H_{0}$ as the \textit{"Mean Field Hamiltonian (MFH)"}).   
      In \textbf{Section}-VIIB, it has been already discussed that the leading-order(LO) result given by eqn.(32), provides [82]  the dominant contribution to the energy-spectra over of the QAHO over the entire physical range of $n$ and $g$. Additionally, it reproduces the rigorous results [11] on the analytic properties of $E(g,\xi)$ in the complex-$g$ plane. 
      
      The next task is to develop the "mean field perturbation theory (MFPT)" for the QAHO by defining the "perturbation Hamiltonian" as generally given by eqn. (21). For the case of the QAHO, this yields:
      \begin{equation}
    H^{\prime}~\equiv~H-H_0=gx^{2K} - \frac{1}{2}(\omega^{2} - 1)x^{2}-h_{0}
    %\label{209} 
    \end{equation}
    To determine the perturbation-corrections to the LO-energy-spectrum,  we use the "Associated Hamiltonian (ASH)"  introduced earlier (See, eqn.(162)), which now reads as:
    \begin{eqnarray}
    \bar{H}=\frac{1}{2}p^{2} + \frac{1}{2}\omega^{2}x^{2} + h_{0}
    +\eta(g x^{2K} - \frac{1}{2}(\omega^{2} - 1)x^{2}-h_{0})\nonumber\\
    \equiv~H_{0} + \eta~ H^{\prime} .
    %(210)
    \end{eqnarray}
  
     The application of HVT  to $\bar{H}$ proceeds in analogous manner as outlined in the previous Section and leads to the following equation :
    \begin{align}
    2(2j+1)\bar{E}X(j)-2(j+1)\omega^{2}X(j+1)\nonumber\\
    -2(2j+1)h_{0}X(j)+\frac{1}{2}j(4j^2-1)X(j-1)\nonumber\\
    +\eta[2(j+1)\alpha X(j+1)+2(2j+1)h_{0}X(j)\nonumber\\
    -2g(2j+K+1)X(j+2)]=0
    %label{211}
    \end{align}
    In the eqn.(211) above, we have used the following notation: 
    $K$ is the anharmonicity-index,
    \begin{equation}
    X(j)\equiv\langle\bar{\psi}|x^{2j}|\bar{\psi}\rangle;
    %(212)
    \end{equation}
    $\alpha\equiv(\omega^2 - 1)$ and $\bar{\psi}$  ,  $\bar{E}$ have been defined earlier (see, eqn.(163) ). Next, using the power-series expansion in $\eta$ for the $\eta$-dependent quantities in eqn.(211) and equating the coefficients of each power of $\eta$ to zero,one arrives at the following recursion-relation:
    \begin{align}
    X(j,i)= a(j)X(j-1,i) + b(j)\sum_{m=0}^{i}E_{m}X(j-1,i-m)\nonumber\\
    + c(j)X(j-2,i)+d(j)X(j-1,i-1)\nonumber\\
    + e_{\omega}X(j,i-1)-f(j)X(j+K-1,i-1),\nonumber\\
    %\label{213}
    \end{align}
    where, $X(j,i)$ are defined by the relation: $X(j)=\sum\limits_{i=0}^{\infty}{\eta^i}X(j,i)$ and other quantities are given as follows:
    \begin{subequations}
    \begin{align}
    ~~~~a(j)=\left(\frac{-h_{0}}{\omega^{2}}\right)\left(\frac{2j-1}{j}\right),\\b(j)=\left(\frac{2j-1}{\omega^{2}j}\right),~~~~\\
    c(j)=\left(\frac{(j-1)\left(4(j-1)^{2}-1\right)}{4j\omega^{2}}\right),\\d(j)=\left(\frac{h_{0}}{\omega^{2}}\right)\left(\frac{2j-1}{j}\right) = -a(j),\\
    e_{\omega}=\frac{(\omega^{2}-1)}{\omega^{2}},\\~~~f(j)=\left(\frac{g}{\omega^{2}}\right) \left(\frac{2j-1+K}{j}\right).
    \end{align}
    %(214a-f)
    \end{subequations} 
    Similarly, following the analogous procedure, as in the previous Section, the Feynman-Hellman Theorem (FHT) provides the following additional relations between $X(j,i)$ and $E_p$:
        \begin{subequations} 
    \begin{align}
    E_{1}=gX(K,0)-\frac{1}{2}(\omega^2-1)X(1,0)-h_0~,~~~~\\
    pE_{p}=gX(K,p-1)-\frac{1}{2}(\omega^2-1)X(1,p-1); p=2,3,4,...
    %\label(215 a-b)
    \end{align}
    \end{subequations}
    The above equations imply that the evaluation of $E_p$ requires the computation of $X(j,i)$ for the values of $i$,$j$ lying in the following range(in steps of unity):$~0\leq~i~\leq (p-1)$, and$~1\leq~j~\leq (K-1)(p-i)+1$. Further note that, by definition, $X(0,i)=\delta_{0i}$, where$~\delta_{mn}$ denotes the standard Kr\"{o}necker-symbol and $X(j,i)=0$ for $j<0$. With the above inputs and boundary conditions, the perturbation corrections $E_p$ can, in principle, be computed \emph{exactly} to arbitrary high order $p$. 
    
    For the case of the QAHO, we note the following results for $E_p(g,\xi)$ at $g=1$ and $\xi=\frac{1}{2}$ (i.e. for the ground-state), for some low-lying values of $p$: $E_1=0$, $E_2= -\frac{3}{256}$,$ E_3= \frac{27}{4096}$, $E_4= -\frac{2373}{262144}$, $E_5= \frac{65457}{4194304}....$ etc. (Note that consistent with eqn.(20), we recover the result :$E_1=0$ in the limit: $\eta~\rightarrow~1~$, which is of special significance- see later. )
    
    {\bf Perturbation corrections for the  Sextic-Anharmonic Oscillator (SAHO)}
    
    Analogous results for the SAHO can be obtained by simply substituting $K=3$ in eqs.(213-215) and using the following input [82] for $\omega$, $h_0$ and $E_0$: $\omega$ is given by the real, positive root of the equation, (see, \textbf{Section-VIID})
    \begin{equation}
    \omega^4-\omega^2-(\frac{15}{4})(5+4\xi^2)=0;
    %(216)
    \end{equation} 
    \begin{equation} 
    E_0=(\frac{\xi}{3})(2\omega+\frac{1}{\omega}),~~~ h_0=(\frac{\xi}{3})(\frac{1}{\omega}-\omega).
    %(217)
    \end{equation} 
    Some sample-values of $E_p$ for the ground state i.e.$\xi=\frac{1}{2}$ and for $\omega=2$ are the following: $E_1=0$, $E_2= -\frac{49}{960}$, $E_3= \frac{671}{4608}$, $E_4= -\frac{53621891}{55296000}$, $E_5= \frac{2610955409}{265420800}....$ etc. As expected [100,102], the resultant asymptotic series exhibits \textit{more severe divergence} at large orders as compared to that of the QAHO. 
    
    In the next sub-section, we discuss the application of MFPT to the case of the quartic-double-well oscillator (QDWO).
    \subsection{The Quartic Double Well Oscillator (QDWO)}
    The QDWO represents perhaps the most straightforward example exhibiting ground-state degeneracy. Consequently, it has been instrumental in elucidating [82,95,98] the characteristics of the standard formulation of perturbation theory (SFPT). Notably, as previously mentioned, the successive terms in the perturbation series (PS) of SFPT for the energy eigenvalue are non-alternating in sign at higher orders, which impedes Borel-summability [82,95,98]. Several contemporary advancements, such as the theory of resurgence [95-99] and trans-series [97,98], are fundamentally motivated to overcome such typical issues of Borel-nonsummability encountered in SFPT when addressing systems with degenerate ground states. In light of this context, we find it particularly pertinent to examine the QDWO problem within the framework of MFPT. 
    
    The Hamiltonian in this instance is expressed as (see, \textbf{Section-VIIC}):  
    \begin{equation*}
    H~=~\frac{1}{2}p^{2} - \frac{1}{2}x^{2} + gx^4,~~~~~~~~~~~~~~\mbox{(36)}
    %\label{36}
    \end{equation*}
    The``Mean-Field Hamiltonian" (MFH)  is chosen in this case as:
    \begin{equation*}
    H_0~=~\frac{1}{2}p^{2}+  \frac{1}{2}\omega^{2}(x-\sigma)^2 + h_0. ~~~~~~\mbox{(37)}
    %\label{}(37)
    \end{equation*}
    This choice is motivated to account for the spontaneous symmetry breaking (SSB) through a non-zero vacuum-expectation-value (VEV) of the field operator $ \hat{x} $, denoted here by $\sigma$. The parameters $\omega$, $\sigma$, and $h_0$ are determined analogously to the case of the QAHO. The LO-results [82] (See, Section-VIIC) imply a "quantum-phase transition" (QPT) characterized by a critical coupling: $g_c(\xi)$ given by the expression: 
     \begin{equation*}
        g_{c}(\xi) = \frac{(2/3)^{3/2}} {3(5\xi-(1/4\xi ))}
        \end{equation*}
         such that the `SSB-phase' is realized with $\sigma\neq 0$ for $g\leq g_c(\xi)$,  whereas the `symmetry-restored(SR)-phase' is obtained with $\sigma=0$ when $g > g_c(\xi)$. The transition across $g=g_c(\xi)$ being discontinuous, the two phases are governed by distinct expressions for $\omega$ and $E_0$ ,which are \textit{not} connected analytically . We, therefore, consider the two-phases separately in the following:
    
    {\bf The SR-Phase of the QDWO}
    
    In this phase (See, \textbf{Section-VIIC}) we have: $ g > g_c(\xi)$; $\sigma=0$; $\omega$ satisfies the equation,
    \begin{equation*}
    ~\omega^{3}+\omega-6g(\xi+\frac{1}{4\xi})=0;~~~~~~~~~~~~~~~~~~~~~~\mbox{(41)}
    %(41)
    \end{equation*}
    The energy eigenvalues and the energy-shift parameter are given in LO as:
    \begin{equation*}
    E_0=(\xi/4)(3\omega-(1/\omega));~~~~~~~~~~~~~~~~~~~~~~~~~\mbox{(42)};
     %(42)
    \end{equation*} 
    and{   ~$ h_0= E_0-\omega\xi$ . }
   
    It has been demonstrated in Section-VII and ref. [82] that LO-results for $E_0$ as given above, provide an accurate estimate of the energy spectrum to within a few percent of the 'exact' result for arbitrary values of $g$ and $\xi$, thus constituting the major contribution to it. Further improvement over the LO-result can be attempted by the application of MFPT following the procedure analogous to that for the QAHO. For this purpose, the 'perturbation-Hamiltonian' and the 'associated-Hamiltonian' are now respectively given by: 
    \begin{equation} 
    H^{\prime}~=gx^{4} - \frac{1}{2}(\omega^{2}+1)x^{2}-h_{0} ,
    %(218)
    \end{equation} 
    \begin{equation}   
    \bar{H}=\frac{1}{2}p^{2} + \frac{1}{2}\omega^{2}x^{2} + h_{0}
    +\eta(g x^{4} - \frac{1}{2}(\omega^{2} + 1)x^{2}-h_{0}).
     %(219) 
    \end{equation}
   
    The application of the HVT to $\bar{H}$ leads to the recursion relation \emph{identical} to that of QAHO as given by eqn.(213-214)but with the following changes: the input-parameters, $\omega$,  and $E_0$ are as given above in eqs.(41-42); $K=2$, {   ~$ h_0= E_0-\omega\xi$  } and $e_{\omega}$  is now defined 
    as:$~e_{\omega}~=~{\omega{^2}}/{(\omega{^2}+1)}$. Similarly,relations analogous to eqn.(215a-b), follow from the application of FHT to this case:
    \begin{equation} 
    ~E_{1}=gX(2,0)-\frac{1}{2}(\omega^2+1)X(1,0)-h_0~,
     %(220) 
    \end{equation} 
    \begin{equation}
    ~pE_{p}=gX(2,p-1)-\frac{1}{2}(\omega^2+1)X(1,p-1);~~ p=2,3,4,....
    %(221)
    \end{equation}
    Sample-values for$~E_p$ evaluated at $\omega=1$ and $\xi=1/2$ are the following:~~$E_1=0$, $E_2= -\frac{1}{24}$, $E_3=\frac{1}{16}$, $E_4=-\frac{791}{3456}$, $~~~~E_5= \frac{7273}{6912}$....etc.
    
   % \textbf{Up to this point on 27 Aug 2025}
    
     We next consider the SSB-phase of the QDWO.
  
   {\bf The SSB-Phase of the QDWO}
  
    As stated earlier (\textbf{Section-VIIC}), a quantum- phase-transition(QPT) to this phase occurs when   $ g < g_c(\xi)$ manifesting in degenerate global-minima of the effective potential located at $x=\pm\sigma$, where
     {$~~~\sigma^2=(1-{12g{\xi}}/{\omega})/4g.$} 
    In this case, $\omega$ satisfies[82] the equation, 
    \begin{equation}
    ~~~~\omega^{3}-2\omega+6g(5\xi-\frac{1}{4\xi})=0.
     %(222)
    \end{equation}
    
   The LO-result for the energy-spectrum is given by [82] 
   
    ~~~~~~~~~~~~~~~~~~~~~$  E_{0}={(-1/16g)}+{(\xi/4)}{(3\omega+(2/\omega))}~$  \\
    
    and the energy-shift is given by: $~h_0= E_0-\omega\xi$. 
    Again, the LO-result for $E_0$ as given above, provides [82] accurate estimate of the energy-spectrum to within a few-percent of the `exact'-numerical result for arbitrary values of $ g $ and $\xi$.
    
     Further improvement in accuracy can be achieved by computing perturbation-corrections to the LO-result  via MFPT as described below: The perturbation-Hamiltonian and the auxiliary Hamiltonian (AH) are respectively given by,
      \begin{equation} 
    H^{\prime}~=gx^{4} - \frac{1}{2}(\omega^{2}+1)x^{2}+\omega^2x-\left(h_{0}+\frac{1}{2}\omega^2\sigma^2\right) , %(223)
   \end{equation}
    \begin{eqnarray} 
    \bar{H}=\frac{1}{2}p^{2} + \frac{1}{2}\omega^{2}\left(x-\sigma\right)^{2} + h_{0}~~~~~~~~~~~~~~~~~~~~~~~~~~~~\nonumber\\
    +\eta(g x^{4} - \frac{1}{2}(\omega^{2} + 1)x^{2}+\omega^2\sigma x-h_{0}  -\frac{1}{2}\omega^2\sigma^2). %(224)
    \end{eqnarray}
     Due to the presence of odd-powers of $x$ in the expression of $\bar{H}$ as given above, it is appropriate to choose:$f(x)=x^j, j=0,1,2,3,..$ for the application of HVT to $\bar{H}$. This leads to the following recursion relation [110,111]:
    \begin{align}
    X(j,i)= a(j)X(j-1,i) + b(j)\sum_{m=0}^{i}E_{m}X(j-2,i-m)\nonumber\\
    -\tilde{b}X(j-2,i)+ c(j)X(j-4,i)+d(j)X(j-1,i-1)\nonumber\\
    +\tilde{b}X(j-2,i-1)+ e_{\omega}X(j,i-1)-f(j)X(j+2,i-1),\nonumber\\
    %\label{225}
    \end{align}
    where, $X(j)\equiv\langle\bar{\psi}|x^{j}|\bar{\psi}\rangle$,$~X(j,i)$ are defined by the relation: $X(j)=\sum\limits_{i=0}^{\infty}{\eta^i}X(j,i)$\\ and other quantities are given as follows:
    \begin{subequations}
    \begin{align}
    a(j)=\sigma\left(\frac{2j-1}{j}\right),~~~~~~~~\\ b(j)=2 \left(\frac{j-1}{\omega^{2}j}\right),~~~~~~~~\\\tilde{b}(j)=(h_0+\frac{1}{2}\omega^2\sigma^2)b(j),~~~~~~~~\\
    c(j)=\frac{(j-1)(j-2)(j-3)}{\left({4j\omega^{2}}\right)},~~~~~~~\\d(j)= -a(j),~~~~~~~\\
    e_{\omega}=\frac{(\omega^{2}+1)}{\omega^{2}},~~~~~~~\\f(j)=\left(\frac{2g}{\omega^{2}}\right) \left(\frac{j-1+K}{j}\right).~~~~~~
    \end{align}
    %label(226)
     \end{subequations}    
    Application of FHT then relates $E_{p}$ to the $X(j,i)$ as follows:
    \begin{eqnarray}
    E_{1}~=~ g X(4,0)-\frac{1}{2}(\omega^{2}+1)X(2,0)+\nonumber\\ \omega^2\sigma X(1,0)-\left(h_{0}+\frac{1}{2}(\omega^{2}\sigma^2)\right),
    %\label{227}
    \end{eqnarray}
    and
     \begin{eqnarray}
    %\begin{equation}
    pE_{p}~=~ g X(4,p-1)-\frac {1}{2}(\omega^{2}+1)X(2,p-1)\nonumber\\+(\omega^{2}\sigma)X(1,p-1),~~
    %(228)
     \end{eqnarray}
    %\end{equation}
    where $p=2,3,4....$
      
      Note that the computation of $E_{p}$ using the eqns.(225-228) above, requires that the recursion in eqn.(225)  traverses within the following range of values of $i,j$ (in steps of unity):
    \begin{eqnarray}
    0\leq~i~\leq (p-1), ~ and~1\leq~j~\leq 2(p+1-i).
    %(229)
    \end{eqnarray}
        With the above inputs and boundary conditions, the perturbation corrections $E_p$ can, in principle, be computed \emph{exactly} to arbitrary high order $p$. We note the following results [110,111] for $E_p$ at  $\xi=\frac{1}{2}$ and $\omega = 1$ , which implies $g=(1/12)< g_c$ as required:\\
       
               $E_1=0$, $E_2= -\frac{17}{384}$, $E_3= \frac{83}{3072}$, $E_4= -\frac{69943}{884736}$, $E_5= \frac{464195}{2359296}....$ etc.\\
         
      Having thus achieved the objective of \textit{analytic} computation of the perturbation corrections $E_{p}$ in all the specified cases of the anharmonic-interaction, we consider the evaluation of the total~${\it `sum'}$ of these corrections in the next section. In that context, the following common features that emerge from the computations in \textit {all} the above cases may be noted:
    \textit{~~ $E_{1} = 0$, and the perturbation series have terms that \textit {alternate} in  sign.}
     These two features in MFPT are of special significance in obtaining the ``sum" of the perturbation series as described below.
    \section{Computation of the Total Perturbation Correction ( TPC )}
    We compute the TPC by two main methods: (a) the method of `optimal truncation' (MOT) of the original asymptotic perturbation series and (b) by 'Borel-summation'. These are sequentially described in the following sections. (For convenience and definiteness , we confine here to the computation for the \textit{ground state}.)
   \subsection { Method of Optimal Truncation (MOT)}
    This method utilizes the property~[108], that the initial terms of an \textit{asymptotic series}(AS) continue to decrease in magnitude till the ``term with the  least magnitude (TLM)" is reached, beyond which, the subsequent terms exhibit monotonic increase. Hence, the AS can be reasonably approximated by truncation at the TLM. 
    
    In the case of ``Mean Field Perturbation Theory (MFPT)",  in view of eqn.(161), we have:
    \begin{equation}
    \left(E(g)-\sum\limits_{k=0}^{N}{E}_k(g)\right) \leq {{E}_{N+1}(g)}.
    %(230)
    \end{equation} 
    Therefore, if $N_0(g)$ denotes the TLM  then the MOT leads to the following estimate: 
    \begin{equation}
    ~~E(g)~\simeq\sum\limits_{k=0}^{N_0(g)}{E}_k(g).
    %(231)
    \end{equation}
    \textit{This method (MOT) works for \textit{arbitrary} physical value of $g$ in MFPT}. In\textbf{ Table-I}, we present the results from this method for the case of the QAHO, QDWO and SAHO for sample-values of $g$. It can be seen from this Table that the TLM occurs at reasonably low values : $N_0(g)\sim$ (2-6).The primary reason for this occurrence can be traced back to the fact that $E_1(g)$ \textit{vanishes} for arbitrary $g$ and since the terms of the 'asymptotic perturbation series' must diverge at large orders,  the TLM is constrained to occur at fairly low-values, without compromising the accuracy , as exhibited in \textbf{Table-I}.
    
    In this \textbf{Table-I}, we have listed computed values for the 'total perturbation correction (TPC) ' by the method of 'optimal truncation(MOT)' of perturbation series(APS) in MFPT, for the \textit{ground-state energy} at different values of the coupling-strength $g$ (col.(1)). Results are shown (col.(4)) for the ground-state energy of QAHO, SAHO and QDWO and compared with accurate  numerical results of earlier calculations denoted as $`Exact' $ (col.(5)), from the given reference. The LO-results in MFPT are given in col.(2). In col.3, $N_0(g)$ denotes the "truncation-point" of the respective APS at the`term of least magnitude'(TLM). Accuracy as percentage-errors are depicted in the last-column relative to the`exact'-results.   
        For comparison with `standard' results, we have considered earlier computations in [112],[113] and [114] for the case of QAHO, SAHO and QDWO respectively. The relative-error incurred is seen to be within a few-percent uniformly over the considered range of $g$.
         
     In view of these results, the MOT in MFPT may be regarded as a `fail-safe' method in achieving reasonable accuracy for arbitrary values of $g$. This is because \textit{the MOT remains as the only practical method available when other methods , such as  Borel-summation etc fail or are inapplicable.}[115]  
     
     \begin{table}{\bf TABLE- I}\\
              \begin{tabular}{c c c c c c }
                 \hline
                  \hline        
                             & &  &$QAHO$& $ref.(112)$&\\
                  \hline   
                 $g$ & $E_0$ & $N_0(g)$ & $E_{MOT}$ & $Exact$ & $Error{(\%)}$ \\
                  
                 0.1 & 0.5603 & 6& 0.5593 & 0.5591 & 0.03 \\ 
                 1.0 & 0.8125 & 3& 0.8074 & 0.8038 & 0.44 \\ 
                 10.0 &1.5312 & 3& 1.5204 & 1.5050  & 1.02 \\ 
                 100.0 & 3.1924 & 3 & 3.1701 & 3.1314  & 1.23 \\
                 \hline 
                 & &  &$SAHO$& $ref.(113)$&\\
                 \hline     
                 0.1 & 0.5964 & 2 & 0.5787 & 0.5869 & 1.40 \\ 
                 1.0 & 0.8378 & 2 & 0.7694 & 0.8050 & 4.42 \\ 
                 50.0 & 1.9735 & 2 & 1.7241 &  1.8585 & 7.23  \\ 
                 200.0 & 2.7606 & 2  & 2.3986 &  2.5942 & 7.54 \\
                 \hline
                 & &  &$QDWO$ &$ref.(114)$&\\
                 \hline     
                 0.1 & 0.5496 & 2 & 0.4107 & 0.4709 & 12.78 \\ 
                 1.0 & 0.5989 & 3 & 0.5998 & 0.5773 & 3.91 \\ 
                 10.0 & 1.4097 & 3 & 1.4007 & 1.3778  & 1.66  \\ 
                 100.0 & 3.1338 & 3 & 3.1122 & 3.0701 & 1.37  \\
                 \hline\ 
             \end{tabular}\\
             
             \ {\textbf{TABLE-I} : See text below eqn.(231) for a description of entries in this Table.}\\	   
             \end{table}
             
             %\textbf{Up to this point on 28 Aug-2025}
             
     \subsection { Method of Borel Summation (MBS)}
    Consider the generic case ( see, eqn.(161)) when we have $E_j\sim~\Gamma(\alpha j+1)$ for $j\gg~1$. Then, it follows that 
    \begin{equation}
    \lim_{j\gg 1}b_j\equiv \dfrac{E_j}{\Gamma(\alpha j+1)}\rightarrow 0,~~\alpha>0.
    %(232)
    \end{equation}
    Using the integral representation of the Gamma function:
    \begin{equation}
    ~\Gamma(z)=~\int_0^{\infty}dt~exp(-t)t^{z-1},
    %(233)
    \end{equation}
    one can {\em formally} express the ``Total Perturbation Correction" (TPC)  $\Delta{E(g)}$  (see, eqn.(161)) as :
    \begin{align}
    \Delta{E(g)}\equiv\sum\limits_{k=1}^{\infty }E_k~=\sum\limits_{k=1}^{\infty } b_k \int_0^{\infty}dt~exp(-t)t^{\alpha_ k}\nonumber\\
    =\gamma\int_0^{\infty}du~exp(-u^{\gamma})u^{\gamma-1}B(u),
    %\label(234)
    \end{align}
    where, we have defined,$~u=t^{\alpha},\gamma=\frac{1}{\alpha}$ and $B(u)$ denotes the `Borel-Series(BS)', given by : 
    \begin{equation}
    B(u)\equiv\sum\limits_{j=1}^{\infty } b_ju^j. 
    %(235)
    \end{equation}
    It is important to note that, although by virtue of eqn.(232), $B(u)$ has a finite radius of convergence $r_c$, the Borel-Laplace-integral (BLI), eqn.(233), does not exist yet since the range of integration extends beyond $r_c$. The resolution to this issue is to perform the analytic continuation/extrapolation of the BS to the full path of integration of the BLI, i.e., to the entire positive real axis in the u-plane. Denoting such analytic continuation by $\tilde{B}(u)$ and substituting this for $B(u)$ in eqn.(234), the BLI can now be made to exist and hence can be used to define the TPC  as:
     \begin{equation}    
    \Delta{E(g)}\equiv\gamma\int_0^{\infty}du~exp(-u^{\gamma})u^{\gamma-1}\tilde{B}(u),
    %(236)    
    \end{equation}
    
    The feasibility of Borel-summation thus crucially depends upon the success of the analytic continuation/ extrapolation with the afore stated properties. In practice however, $\tilde{B}(u)$ can only be inferred approximately from the BS , if feasible, since only a \textit{finite} number of terms in eqn.(235) are computable/known. Under such circumstances, the \textit{method of conformal mapping} (MCM) has been preferably used [116] for implementing  analytic continuation in a model-independent way. We, therefore employ the same in the following. 
    
    \textit{The inputs, which are required for the implementation of MCM  are the location and nature of the singularity lying closest to the origin}, which we designate as the `leading singularity'(LS). By the `Darboux theorem' [117], the late-terms in the BS, eqn.(235), originate from the LS. Therefore, the nature and location of the LS may be inferred from the terms of the BS at large orders. This is readily demonstrated considering the following \textit{ansatz} of the LS: $\tilde{B}(u)\simeq\tilde{B}_0(u)\sim~(u+r_c)^p$. The `radius of convergence' $r_c$ and the `singularity-exponent'~$p$ can then be determined from a fast-converging set of equations [9] given below:
    \begin{equation}
            ~~~~ r_c(g)=\lim_ {j\gg1}~\dfrac{b_jb_{j-1}}{jb_j^2-(j+1)b_{j+1}b_{j-1}},
        %(237)
     \end{equation}
     \begin{equation}
      p(g)=\lim_ {j\gg1}~\dfrac{{jb_j^2}-(j^2-1)b_{j-1}b_{j+1}}{jb_j^2-(j+1)b_{j+1}b_{j-1}}.
     %(238)
    \end{equation}
    
    %\textbf{Upto this point on 28-08-25 FN}

   \textit{ We obtain the value:$~p(g)=~-0.5$, being independent of $g$ as expected} and the values for $r_c(g)$, as  shown in \textbf{Table-II}. Equipped with these information as inputs, consider [118] the conformal mapping , $z$  which maps the cut-u plan,$|arg(u)|<\pi$ into the interior of the unit-disk in the $z$-plane while preserving the origin:
    \begin{subequations}
        \begin{align}    
        z(u)=\dfrac{\sqrt{1+su}-1}{\sqrt{1+su}+1};~~s=\dfrac{1}{r_c};~~u=t^{\alpha},\\
        \mbox{and its inverse~~~~~~~~~~~~~~~~~~~}\nonumber\\ 	u(z)=\left(\frac{4}{s}\right)\dfrac{z}{(1-z)^2};~~|z|\leq 1.~~~~
        %(239a,b)   
        \end{align}
    \end{subequations}
    Note that the images of the points: $u=\pm\infty$ are at $z=1$ and further that $z(u)$ is a real-analytic-function of u. 
    
    The conformal transformation, as described in equation (239), possesses a significant property known as its 'optimality,' as demonstrated by the Ciulli-Fischer Theorem [118]. This theorem asserts that the truncation error is minimized when an arbitrary function $f(z)$, expressed as an infinite power series in $z$, is approximated by truncating the series to a finite number of terms, thereby ensuring an optimal rate of convergence. Consider the partial sum of the BS, as given in equation (235): 
         
     ~~~~~  ~~~~~~~~~~~~~~~ ~~~~~~~~~~~~~~~$B_N(u)\equiv\sum\limits_{k=1}^{N} b_ku^k$.
    
     Using the inverse-transformation, eqn.(239b),$B_N(u)$ can be consistently re-expanded [119,120] as follows, noting that $u^N\sim~ z^N+O(z^{N+1})$:
    \begin{subequations}
        %\label{240}    
        \begin{align}
        B_N(u)\equiv\sum\limits_{k=1}^{N} b_ku^k\rightarrow~\bar{B}_N(z)=\sum\limits_{k=1}^{N} B_kz^k,\\
        \mbox{where,}~~ B_k=\sum\limits_{n=1}^{k}b_n\dfrac{\left(n+k-1\right)!}{\left(k-n\right)!\left(2n-1\right)!},
        \end{align}
        %(240a,b)
    \end{subequations}
    where, we have denoted the analytic-continuation of the partial sum of the BS by:$~B_N(u)\rightarrow~\bar{B}_N(z)$.\\
    Using the above properties/results, the TPC can be evaluated by the change of the integration-variable to $z$ (see, eq(239b)).Then, by taking care of possible numerical-divergence at the upper limit, $~z=1$ and further, considering the limit of the sequence of partial-sums, the TPC is evaluated as follows ( we suppress the $g$-dependence of quantities for notational-clarity):
    \begin{subequations}
     \begin{align}    
        \Delta{E}=\lim_{N\gg1,~\epsilon\rightarrow0}\left(\Delta{E}\right)_N\equiv\int_0^{1-\epsilon}dz f_N(z),\\
        \mbox{~with~}f_N(z)=(\gamma\rho)~\dfrac{(1+z)}{(1-z)^3}\left(\dfrac{\rho~z}{(1-z)^2}\right)^{\gamma-1}\nonumber\\~exp\left[-\left(\dfrac{\rho~z}{(1-z)^2}\right)^{\gamma}~\right]\bar{B}_N(z),
        \end{align}
        %(241a,b)    
    \end{subequations}
    where,$~\rho=4r_c$ and $~\bar{B}_N(z)$ is defined by eqn.(240a). Note that term-by-term integration of the partial-sum,$~\bar{B}_N(z)$ can now be carried out in eqn.(241a), in contrast to the BS in eqn.(234) because of the incorporation of analyticity through the conformal-map $z$.
    
     \begin{table}{\bf TABLE II}
              
            \begin{tabular}{c c c c c c c c }
                \hline
                $g$&$r_c$&$N_c$& $\varDelta{E(g)}$&$E_0$&$E_{tot}(g)$& $Exact$ & $Er({\%})$ \\
                \hline
                QAHO& $(\alpha= 1)$& & &  & &$ref.(112)$&\\
                
                \hline     
                0.1 & 6.071 & 6 & -0.00116 & 0.5603 & 0.55914 & 0.5591 & 0.0083 \\ 
                1.0 & 2.667 & 7 & -0.00869 & 0.8125 & 0.80381& 0.8038 & 0.0003\\ 
                10.0 & 2.133 & 8 & -0.02619  & 1.5312 & 1.50501  & 1.5050 & 0.0005\\ 
                100.0 & 2.028 & 10 & -0.06101 & 3.1924 & 1.13139 & 3.1314 & 0.0001\\ 
                \hline
                SAHO & $(\alpha= 2)$& & & &  &$ref.(113)$ &\\
                \hline
                0.1 & 13.3 & 20& -0.0095 & 0.5964  & 0.5869 & 0.5869 & 0.001 \\ 
                1.0 & 8.56 & 20 & -0.0328& 0.8378  & 0.8050 & 0.8050 & 0.002\\ 
                10.0 & 7.14& 20 &-0.1149 & 1.9735  & 1.8586 & 1.8585  & 0.007\\ 
                100.0 & 7.02& 20 &-0.1662 & 2.7606 & 2.5944 & 2.5942 & 0.007 \\ 
                \hline
                QDWO & $(\alpha= 1)$& & & & &$ref.(114)$ &\\
                \hline
                0.5 & 1.191 & 20&-0.0232 & 0.4770 & 0.4538 & 0.4538 & 0.0027 \\ 
                1.0 & 1.455 & 11 & -0.0216 & 0.5989& 0.5773 & 0.5773 & 0.0006\\ 
                10.0 & 1.872 & 20 & -0.0320& 1.4098 & 1.3778 & 1.3778 & 0.0040\\ 
                100.0 & 1.972 & 18& -0.0637& 3.1338 & 3.0701& 3.0701 & 0.0005\\ 
                \hline\               
            \end{tabular}\\
            %	\label{$Table-II$}
            {TABLE-II: See text following eqn.(241b), for a description of the entries in this Table}
        \end{table}

    The results are presented in \textbf{Table-II}. There we have shown ( col.(4))  the Borel-sum for the total- perturbation-correction(TPC) computed using eqn.(241a-b) and  conformal-mapping, eqn.(239a) , for the \textit{ground state energy} of the QAHO, SAHO and the QDWO as a function of $g$ given in col.(1). For each case, the input-value of $\alpha$, (see eqn.(232)) is indicated. Entries in col.(2) are the input-value for the `radius of convergence'; $N_c$ in col.(3) denotes the number of terms retained in the Borel-series for convergence up to the specified accuracy; $E_0$ in col.(5) denotes the LO-contribution; $E_{tot}$ in col.(6) represents the `total corrected energy'; entries in col(7) labeled $Exact$, display the `standard'-numerical results from the indicated reference; and the last column specifies the ${\%}error$ with respect to the $Exact$-result. We have fixed the value: ~$\epsilon=0.001$ ,( see eqn.(241a)) in the computation of TPC in all-cases.
    
   It is noteworthy that the convergence of the partial-sum is quite fast- an accuracy of $\sim (99.99\%)$ is uniformly achieved  over the full explored range of $g$ and for all the  cases of anharmonic-interaction since  the cut-off $N_c$ on the number of terms retained in the partial-sum is as low as $\sim\left(10-20\right)$ (\textbf{Table-II}). Thus, the `$optimality$' and the `$accuracy$' of the 'method of conformal mapping' are firmly established in MFPT.
%     \begin{center}
%        \textbf{5:~ 
%                Summary}
  ~ %  \end{center} 
    \subsection{Section -  Summary}
    
    We have introduced a novel formulation of perturbation theory predicated on a mean-field Hamiltonian approximation, demonstrating its Borel-summability for \textit{arbitrary coupling strength} in the contexts of QAHO, SAHO, and the QDWO, by achieving highly precise results for the energy spectrum. This demonstration suggests that the distinction between perturbative and non-perturbative regimes may be an artifact of SFPT. 
    
    The results obtained for the QDWO are particularly noteworthy, as this system typically features degenerate ground states, which generally impede Borel-summability in SFPT.
    
    Given the simplicity and generality of the formulation demonstrated in this study, we are inclined to \textit{conjecture} that MFPT may be applicable to arbitrary interacting systems in quantum theory.     
   \bigskip
    \newpage
%    \begin{center}
%        \textbf{CHAPTER-6}
%            \end{center}   
    \section{Summary and Conclusions} 
   
   In this report, we introduce a non-perturbative approximation scheme in quantum theory, termed the "Non-perturbative General Approximation Scheme" (NGAS). This scheme is fundamentally applicable to general interacting systems described by a Hamiltonian and is characterized by its simplicity in formulation, non-perturbative nature, self-consistency, and flexibility to incorporate various choices of approximating inputs.
   
     A key feature of this scheme is the identification of a "mapping" that transforms the "interacting system" into an "exactly solvable" input model, while maintaining the primary effects of interaction through a self-consistent feedback mechanism that demands equal quantum averages of observables in both systems. This approximation method surpasses the standard formulation of perturbation theory (SFPT) and the variational approximation by overcoming their limitations: unlike the variational method, it is systematically improvable through the development of a new perturbation theory, namely, the "Mean Field Perturbation Theory" (MFPT). Notably, it allows for systematic improvement through a perturbation theory that is valid for arbitrary interaction strengths. Besides, in contrast to SFPT, MFPT preserves the analyticity and non-linearity of the original system order-by-order, while being Borel-summable for all permissible values of the coupling strength of interaction. Furthermore, MFPT demonstrates improved behavior of the divergent asymptotic perturbation series at large orders. The scheme replicates results obtained by several earlier methods, e.g., [112-114], while surpassing these methods in terms of wider applicability, systematic improvement, and better convergence. The other features of MFPT, which distinguish it from the conventional perturbation methods, include its freedom from power-series expansion in any \textit{small} parameter, including the coupling strength, and its applicability to arbitrary interaction strengths. Hence, by circumventing the limitations of SFPT in MFPT, it is established that the distinction between the 'perturbative' and 'non-perturbative' regimes vanishes, which suggests that this distinction may be a mere {\em artifact} of the former formulation, i.e., the standard formulation of perturbation theory (SFPT). 
     
      It is noteworthy that NGAS allows flexibility in the choice of the approximating Hamiltonian (AH).  Although the results presented here are based on the "harmonic-approximation," we have also demonstrated that comparable accuracy can be achieved using a much cruder approximation with the infinite- square-well-Hamiltonian as the input Hamiltonian within the framework of NGAS's general features.
     
     In the realm of quantum field theory, the scheme has been applied to the $g\phi^4$ theory, with the approximating Hamiltonian selected as that for the free-hermitian-scalar field, albeit with adjustable mass and shifted field. In this context, the standard results of the Gaussian approximation are replicated at the leading order, including the \textit{non-perturbative} renormalization of the mass and coupling strength. However, NGAS appears to offer a broader scope and deeper insight than the Gaussian approximation, as it provides a 'dynamical explanation' of the latter, emerging through the mechanism of altered vacuum structure introduced by the interaction. Furthermore, the scheme extends beyond the Gaussian approach by establishing new results, such as the calculation of the momentum distribution of the condensate structure function `$n(k)$'. It is noteworthy that by advancing beyond the leading order of NGAS, the results of the Gaussian approximation can be systematically refined, order-by-order. The resulting momentum distribution of the vacuum condensate structure function `$n(k)$' is particularly significant, as it exhibits the non-standard feature of a considerable spread in $~|k|~$ about the origin, scaled by the renormalized mass of the physical quanta. It is plausible to anticipate that this condensate structure of the physical vacuum persists at finite temperature, manifesting in the thermodynamic properties of the system with testable consequences. 
     
    % \textbf{Up to this on 29-08-2025}

   \section{Outlook}
     In this report we have confined the application of the scheme to some bench-mark systems in quantum mechanics and quantum field theory, namely the anharmonic interactions in one-dimension including  the quartic-, sextic- and octic-anharmonic oscillators , the quartic-double well-oscillators and to the $~g \phi^{4}~$ quantum field theory. The investigation can be extended in different directions, i.e. to include  finite temperature  field theory, quantum-statistics,  non-oscillator systems, super-symmetric cases and quantum field theories involving fermions and gauge-fields etc.
    
    Possible applications of the results derived here are envisaged in diverse area of current interest including critical phenomena (involving a scalar field as the order-parameter [84]), inflationary cosmology [87], finite temperature field  theory [121] , exploration of the vacuum structure [91]  of pure gluonic-QCD and  Higgs sector of the standard model ( by extending [90]  the analysis to  the spontaneously broken phase, which corresponds to the case of negative bare-mass $~m^{2}<0~$).
    
    One immediate task could be to apply MFPT to other systems known to be Borel non-summable in SFPT [94] to determine whether summability can be restored in such cases. Similarly, tunnel-splitting of energy levels [95] can be studied  after computation of perturbation correction to the excitation spectrum of the QDWO using more realistic [122] input - Hamiltonian, if necessary. Application to other Hamiltonian-systems are envisaged in a straight forward manner in view of the universal nature of the approximation scheme. 
    \bigskip
    \\{\textit{Acknowledgments}}\\
      Collaboration with Dr. Nabghan  Santi and Dr. Noubihary Pradhanon on  several aspects of the reasearch presented herein is acknowledged. The author acknowledges useful comments and suggestions from the anonymous referees of  the journals at references [48],[50],[82] and[110] . The computation and library facilities at the Institute of Physics, Bhubaneswar (IOPB) and NISER, India are gratefully acknowledged where one of the authors , (BPM) held positions of guest-scientist and visiting faculty respectively. The author is thankful to Prof. F.M.Fernandez for help in symbolic-computation using Maple and for helpful comments. Useful remarks from Profs. A. M. Srivastava, S. G. Mishra and P. Agrawall during seminar talks on the topic(s) of this investigation are gratefully acknowledged.

\end{document}